\documentclass[a4paper,12pt]{article}
\pdfoutput=1
\usepackage{epsfig}
\usepackage{amssymb}
\usepackage{amsfonts}
\usepackage{amsmath}
\usepackage{euscript}
\usepackage{verbatim}
\usepackage{latexsym}
\usepackage{graphicx}
\usepackage{caption}
\usepackage{float}
\usepackage{subcaption}
\usepackage{bbm}
\usepackage[normalem]{ulem}
\usepackage{soul}
\usepackage{listings}
\usepackage{esint}

\usepackage{booktabs}

\newif\ifdtup

\usepackage{esint}
\catcode`\@=11

\@addtoreset{equation}{section}

\def\@normalsize{\@setsize\normalsize{15pt}\xiipt\@xiipt
\abovedisplayskip 14pt plus3pt minus3pt%
\belowdisplayskip \abovedisplayskip
\abovedisplayshortskip \z@ plus3pt%
\belowdisplayshortskip 7pt plus3.5pt minus0pt}

\def\small{\@setsize\small{13.6pt}\xipt\@xipt
\abovedisplayskip 13pt plus3pt minus3pt%
\belowdisplayskip \abovedisplayskip
\abovedisplayshortskip \z@ plus3pt%
\belowdisplayshortskip 7pt plus3.5pt minus0pt
\def\@listi{\parsep 4.5pt plus 2pt minus 1pt
     \itemsep \parsep
     \topsep 9pt plus 3pt minus 3pt}}

\relax

\catcode`@=12

\catcode`\@=11

\def\section{\@startsection{section}{1}{\z@}{3.5ex plus 1ex minus
   .2ex}{2.3ex plus .2ex}{\large\bf}}

\def\SymBoxes#1#2#3#4{\newdimen\un@t \un@t#3%
\raisebox{#1}{\rule{#2\un@t}{#4}\hskip-#2\un@t% lower horizontal
\@tempdimb\un@t \advance\@tempdimb by-#4\@tempcntb#2\relax%
\@whilenum{\@tempcntb>0}\do{%                         % #2 vertical lines
\rule{#4}{\un@t}\hskip\@tempdimb \advance\@tempcntb by\m@ne}%
\hskip-#2\un@t \rule[\un@t]{#2\un@t}{#4}%
\rule[\un@t]{#4}{#4}\hskip-#4%             % upper horizontal line
\rule{#4}{\un@t}}\hskip-#4}                % rightest vertical line
\begin{document}
%\begin{letter}{~}

%%%%%%Define some new commands and  macros
\newcommand{\beq}{\begin{equation}}
\newcommand{\eeq}{\end{equation}}
\newcommand{\bea}{\begin{eqnarray}}
\newcommand{\eea}{\end{eqnarray}}
\newcommand{\beas}{\begin{eqnarray*}}
\newcommand{\eeas}{\end{eqnarray*}}
\newcommand{\defi}{\stackrel{\rm def}{=}}
\newcommand{\non}{\nonumber}
\newcommand{\bquo}{\begin{quote}}
\newcommand{\enqu}{\end{quote}}
%%%%%%%%%%%%%%%%
\renewcommand{\(}{\begin{equation}}
\renewcommand{\)}{\end{equation}}
%%%%%%%%%%%%%%%%%%%%%%%%%%%%%%%%%% definitions
\def \eqn#1#2{\begin{equation}#2\label{#1}\end{equation}}

\def\e{\epsilon}
\def\IZ{{\mathbb Z}}
\def\IR{{\mathbb R}}
\def\IC{{\mathbb C}}
\def\IQ{{\mathbb Q}}
\def\de{\partial}
\def\Tr{ \hbox{\rm Tr}}
\def\H{ \hbox{\rm H}}
\def\HE{ \hbox{$\rm H^{even}$}}
\def\HO{ \hbox{$\rm H^{odd}$}}
\def\K{ \hbox{\rm K}}
\def\Im{ \hbox{\rm Im}}
\def\Ker{ \hbox{\rm Ker}}
\def\const{\hbox {\rm const.}}
\def\o{\over}
\def\im{\hbox{\rm Im}}
\def\re{\hbox{\rm Re}}
\def\bra{\langle}\def\ket{\rangle}
\def\Arg{\hbox {\rm Arg}}
\def\Re{\hbox {\rm Re}}
\def\Im{\hbox {\rm Im}}
\def\exo{\hbox {\rm exp}}
\def\diag{\hbox{\rm diag}}
\def\longvert{{\rule[-2mm]{0.1mm}{7mm}}\,}
\def\a{\alpha}
\def\dag{{}^{\dagger}}
\def\tq{{\widetilde q}}
\def\p{{}^{\prime}}
\def\W{W}
\def\N{{\cal N}}
\def\hsp{,\hspace{.7cm}}

\def\br{\nonumber}
\def\IZ{{\mathbb Z}}
\def\IR{{\mathbb R}}
\def\IC{{\mathbb C}}
\def\IQ{{\mathbb Q}}
\def\IP{{\mathbb P}}
\def \eqn#1#2{\begin{equation}#2\label{#1}\end{equation}}

\newcommand{\C}{\ensuremath{\mathbb C}}
\newcommand{\Z}{\ensuremath{\mathbb Z}}
\newcommand{\R}{\ensuremath{\mathbb R}}
\newcommand{\rp}{\ensuremath{\mathbb {RP}}}
\newcommand{\cp}{\ensuremath{\mathbb {CP}}}
\newcommand{\vac}{\ensuremath{|0\rangle}}
\newcommand{\vact}{\ensuremath{|00\rangle}                    }
\newcommand{\oc}{\ensuremath{\overline{c}}}
\newcommand{\psizero}{\psi_{0}}
\newcommand{\phizero}{\phi_{0}}
\newcommand{\hzero}{h_{0}}
\newcommand{\psiin}{\psi_{\rh}}
\newcommand{\phiin}{\phi_{\rh}}
\newcommand{\hin}{h_{\rh}}
\newcommand{\rh}{r_{h}}
\newcommand{\rb}{r_{b}}
\newcommand{\psibnd}{\psi_{0}^{b}}
\newcommand{\psibndp}{\psi_{1}^{b}}
\newcommand{\phibnd}{\phi_{0}^{b}}
\newcommand{\phibndp}{\phi_{1}^{b}}
\newcommand{\gbnd}{g_{0}^{b}}
\newcommand{\hbnd}{h_{0}^{b}}
\newcommand{\zh}{z_{h}}
\newcommand{\zb}{z_{b}}
\newcommand{\man}{\mathcal{M}}
\newcommand{\hbr}{\bar{h}}
\newcommand{\tbr}{\bar{t}}

\begin{titlepage}
%\begin{flushright} CHEP XXXXX
%ULB-TH/09-10\\
%hep-th/yymmnnn\\ \end{flushright}
%\bigskip
\def\thefootnote{\fnsymbol{footnote}}

\begin{center}
{
{\Large
{\bf Normal Modes of the Stretched Horizon: }}\\
\vspace{0.1in}
\large{\bf A Bulk Mechanism for Black Hole Microstate Level Spacing  } 
}
\end{center}

\bigskip
\begin{center}
Chethan Krishnan$^a$\footnote{\texttt{chethan.krishnan@gmail.com}}, \ Pradipta S. Pathak$^a$\footnote{\texttt{pradiptap@iisc.ac.in}}, \ %Soham RAY$^a$\footnote{\texttt{raysoham14@gmail.com}}  
\vspace{0.1in}

\end{center}

\renewcommand{\thefootnote}{\arabic{footnote}}

\begin{center}
%\vspace{0.2cm}

$^a$ {Center for High Energy Physics,\\
Indian Institute of Science, Bangalore 560012, India}\\

\end{center}

\noindent
\begin{center} {\bf Abstract} \end{center}
In 1984,  't Hooft famously used a brickwall (aka stretched horizon) to compute black hole entropy up to a numerical pre-factor. This calculation is sometimes interpreted as due to the entanglement of the modes across the horizon, but more operationally, it is simply an indirect count of the semi-classical modes trapped between the stretched horizon and the angular momentum barrier. Because the calculation was indirect, it needed both the mass and the temperature of the black hole as inputs, to reproduce the area. A more conventional statistical mechanics calculation should be able to get the entropy, once the ensemble is specified (say via the energy, in a microcanonical setting). In this paper, we explicitly compute black hole normal modes in various examples, numerically as well as (in various regimes) analytically. The explicit knowledge of normal modes allows us to reproduce $both$ the Hawking temperature as well as the entropy, once the charges are specified, making this a conventional statistical mechanics calculation. A quasi-degeneracy in the angular quantum numbers is directly responsible for the area scaling of the entropy, and is the key distinction between the Planckian black body calculation (volume scaling) and the 't Hooftian calculation (area scaling). We discuss the (rotating) BTZ case in detail and match the thermodynamic quantities {\em  exactly}. Schwarzschild and Kerr normal modes are discussed in less detail using near-horizon approximations. Our calculations reveal a new hierarchy in the angular quantum numbers, which we speculate is related to string theory. 

%Normal modes of the stretched horizon may provide a universal $bulk$ mechanism for the level spacing of black hole microstates that is non-perturbative in the Planck length as defined by the stretched horizon.

%$O(1)$ numbers 

%The relationship between our explicit normal modes and the semi-classical  ``black hole atmosphere" behind the angular momentum barrier. 

%The normal modes of the brickwall (aka stretched horizon) were $implicitly$ used by 't Hooft in 1984, to account for black hole entropy up to an $O(1)$ number. 

%gives us a ``hands on" understanding of why 't Hooft's calculation was successful. It also

\vspace{1.6 cm}
\vfill

\end{titlepage}

\tableofcontents

\setcounter{footnote}{0}

%%%%%%%%%%%%%%%%%%%%%%%%%%%%%%%%%%%%%%%%%%%%%%%%%%%%%%%%%%%%%%%%%%%%%%%%%%%%%%%%%%%%%%%%%%%%%%
%%%%%%%%%%%%%%%%%%%%%%%%%%%%%%%%%%%%%%%%%%%%%%%%%%%%%%%%%%%%%%%%%%%%%%%%%%%%%%%%%%%%%%%%%%%%%%

\section{Introduction and Outlook}

Smooth horizons have entropy, but without an explicit state-counting explanation in the underlying bulk effective field theory (EFT). In some examples in string theory, we have however been able to count these states holographically via a dual quantum description \cite{Strominger-Vafa, Maldacena}. Despite this, little is known about the nature of black hole microstates in the {\em bulk}. Progress on this question is likely crucial, for resolving the information paradox(es). 

In the zero temperature BPS-protected cases, there has been significant progress on constructing supergravity solutions that are candidates for black hole microstates. These solutions form the foundation of the fuzzball program \cite{Mathur, BenaWarner}. Fuzzball solutions cap off before the horizon and are presumably proxies for the fully quantum/string descriptions of BPS microstates, but in supergravity. Their interpretation in the full string theory (especially at finite temperature) has not been very clear, but they are a suggestion in a stringy setting that black holes may not have interiors\footnote{In some yet-to-be-made-fully-precise sense.} -- an idea advocated by Mathur \cite{MathurLunin} and others \cite{Bena2, Bena3}. 

More recent developments \cite{Leutheusser1, Leutheusser2, Witten} suggest that in the large-$N$ limit above the Hawking-Page transition, the algebra of small fluctuations of a black hole turns into a type III algebra instead of a type I algebra. Type III algebras have non-trivial commutants, which is a strong suggestion that this transition is to be understood as the emergence of the black hole interior. This immediately raises the possibility that black holes at finite-$N$ may not have interiors\footnote{Note that this is to be understood in the sense of any thermodynamic phase transition. Magnets are the result of a large-volume transition, but magnetic domains of finite volume exist in the ``real'' world. The ideas is that they should be understood in the sense of a suitable scaling limit. This is yet to be completely elucidated in the black hole case, but see comments in \cite{CK-VM}.}.

The idea of black holes without interiors goes back to a 1984 paper of 't Hooft \cite{tHooft} (see also \cite{SusskindThorlaciusUglum}). As we will explain in great detail, 't Hooft's calculation can be understood as an indirect count of the semi-classical modes of a scalar field that are trapped behind the angular momentum barrier and a ``brick wall'' placed a Planck distance outside the horizon\footnote{To emphasize that the brickwall can result in dynamics, \cite{SusskindThorlaciusUglum} called it the stretched horizon. We will go back and forth between these two names.}. He was able to show that the area law for the entropy is reproduced (upto a numerical pre-factor), once both the Hawking temperature and the mass of the black hole were specified. 

't Hooft's calculation was clever -- he used a semi-classicality demand (that the modes are the trapped modes behind the barrier) and the associated Bohr-Sommerfeld quantization, to determine the entropy without actually determining modes. But there is a price to pay. He had to specify both the temperature and the energy (mass) of the system to obtain the entropy. In a conventional statistical mechanics calculation, only one of these should be necessary to determine the rest of the thermodynamic quantities. In this paper, we will not be as clever as 't Hooft, but we will be willing to work harder. We will compute the normal modes of the system with the stretched horizon explicitly in various examples, to various degrees of completion -- both numerically  and analytically. Using these explicit stretched horizon normal modes, we will be able to take a few steps beyond \cite{tHooft}:
\begin{itemize}
\item We show that {\em both} the temperature and the entropy are reproduced upon the specification of the black hole mass. The computation of normal modes is a difficult problem in general, because of various technical complications (starting from the lack of simple solutions of radial wave equations, even for the Schwarzschild black hole). But in the case of the BTZ black hole where they are most under control\footnote{The radial mode in BTZ is a hypergeometric function.}, we are able to reproduce {\em both} the temperature and the entropy {\em exactly}, and not merely up to $O(1)$ numbers.
\item Our calculation for BTZ works in a holomorphically factorized way, so we show that it holds for the rotating case as well, upon specifying the angular momentum as well of the BTZ black hole. (More precisely, we specify the $L_0$ and $\bar L_0$.)
\item In other cases like Schwarzschild and Kerr where our calculations are more crude, we are again able to fix the temperature and entropy, but up to $\mathcal{O}(1)$ numbers. We strongly suspect that these calculations can be improved, and we hope to come back to them in the future. 
\item Our formulation of 't Hooft's calculation reveals its parallels and distinctions from Planck's black body calculation. We can view the operational difference between the two as simply that in 't Hooft's calculation, the box in which the radiation lives, is not in flat space -- it is bounded on one end by the stretched horizon brick wall and on the other by the angular momentum barrier.
Knowledge of the normal modes gives us clear intuition about the reason why we get a volume scaling in the Planckian case, while we get an area law in the 't Hooftian case. The key observation is that there is an approximate degeneracy\footnote{Somewhat more precisely, it grows quasi-logarithmically. Explicit expressions are presented in various places in the text.} in the angular quantum number direction in 't Hooft's calculation, while it grows linearly in all spatial directions in Planck's. The linearity of the mode number growth due to the harmonic oscillator like spectrum is missing in {\em one} of the dimensions, ie., the angular   Casimir dimension, in the black hole case. This is the operational origin of the area law. 
\item We present a detailed study of the normal modes using multiple techniques -- including numerical, analytic and perturbative tools, and approximating the equations in various regimes. We believe that even though this paper is moderately big, we have not fully exploited these tools and more remains to be done.  
\item In the low-$\omega$ limit where they contribute to the black hole entropy, we present general analytic expressions for the normal modes and note their key features. We expect that understanding the origin and implications of these features will be enlightening for understanding quantum black holes.
\item In some recent papers \cite{Adepu, Sumit1, Sumit2} it has been noticed (largely numerically) that the one-particle spectral form factor (SFF) computed from the normal modes has a linear ramp. This is believed to be a quantum chaos diagnostic \cite{Cotler}. Using our analytic 
approximations for normal modes, we are able to cleanly identify the origin of the ramp in these normal modes. This connects with the ramp in the $\log$ spectrum noted in \cite{Sumit2}. 
\item We note that it is the same low-lying modes that are responsible for both the thermodynamics as well as the ramp. We can reproduce the thermodynamics even if we ignore the (weak) $J$-dependence of the spectrum, while the ramp emerges due to the $J$-dependence. Here we use $J$ to denote a suitable angular quantum number -- details vary slightly with specific black holes. 
\item Using what we learnt from these investigations, we write down a simple ``model" spectrum which can carry entropy while also exhibiting the ramp. This model spectrum can be viewed as the ``skeletal" normal mode structure that is responsible for the physics. A version of this spectrum without the $J$-dependence was known previously -- it was perhaps first written down by Solodukhin \cite{SoloFirst}. 
\item 't Hooft's modes are semi-classical because they are bound behind the angular momentum barrier. This classical trapping effectively imposes a cut-off in $J$ in his calculations. Our normal modes are true eigenmodes of the wave equation -- in other words, they are aware of tunneling through the angular momentum barrier. But we find that the demand of matching the entropy automatically puts a cut-off on them as well. Remarkably, this cut-off is hierarchically {\em below} the 't Hooft cut-off, by a factor of about $\sim \log S_{BH}$. Perhaps because of this, they are able to capture both the precise thermodynamics as well as the linear ramp. Equally interestingly, it turns out that 't Hooft's cut-off is at a value of $J$ at which our low-$\omega$ analytic approximation for the normal modes undergoes complete breakdown. Curiously, the reduction in the cut-off that we find, brings the modes into a regime where the low-lying analytic form is more reliable. It will be good to understand why this happens in such a convenient way, and what the significance of this refined $J$ cut-off is. We emphasize that with this reduced cut-off, not only are we able to get precise matches for entropy and temperature -- we are able to show that these are the modes responsible for the linearity of the ramp as well. It is tempting to think that this mode cut-off is related to a Hagedorn transition when viewed as a string mode \cite{Itzhaki, Mertens}.
\end{itemize}

It is known that bulk EFT can count microstates indirectly -- this relies on the black hole interior and ``bags of gold" being overcomplete \cite{Balasubramanian}. But while this gives us an implicit count of the microstates, it still leaves us wanting for an understanding of individual microstates themselves. This we believe will require some guess (or actual knowledge) about the UV complete theory. The stretched horizon is such a guess, if we believe that the UV complete description and the EFT description should effectively coincide, once we are more than a few Planck lengths away from the horizon. 

In the case of 2-charge D1-D5 black holes, it is known that the precise entropy \cite{Rychkov,CK-AR} can be reproduced from the phase space of horizonless fuzzball solutions. However in the more interesting 3-charge case, where the classical black holes have non-vanishing area, a full understanding of bulk microstates is lacking.  Even more puzzling is the case of black holes at finite temperature, which also require an explanation for their entropy and temperature in terms of microstates. Considering the fact that black holes are a {\em generic} prediction of general relativity, it is plausible that there may be generic ways of understanding these microstates in the bulk. Instead of viewing the stretched horizon as a phenomenological input (as is sometimes done), we view it as a way of describing bulk microstates of the underlying UV-complete theory, at small but finite $G_N$. The results of this paper strengthen the case that this generic {\em bulk} mechanism can capture the level spacing of microstates that is non-perturbative in $G_N$.

From the bulk EFT point of view, the stretched horizon is an ad-hoc boundary condition. The claim is that it is a regulator that captures a piece of the UV-complete description. The remarkable thing about this boundary condition is that it has enough magic to reproduce the microstate level spacing expected in black holes -- it reproduces both the temperature and the entropy from a conventional statistical mechanics perspective. What this shows, is that the distance from the horizon to the stretched horizon contains non-perturbative information in $G_N$\footnote{We will use the Planck length and the string length interchangeably in this paper because we are working with a single scalar field. But if there are many fields, it may be natural to have a hierarchy \cite{Nomura}. Ultimately, we will see that the relation between stretched horizon location and $J_{cut}$ is the crucial information, so some of these choices have some flexibility.}. 

In this paper, we have taken some old ideas, and tried to develop them systematically from a new perspective. The paper deals with various technicalities, but we have tried to outline the big picture in various places in the text. We will discuss another closely related reason to take the stretched horizon seriously, in a companion paper \cite{Vaibhav} -- we will show that the effective correlators of the stretched horizon are that of a smooth horizon.

\section{Schwarzschild Normal Modes}\label{SchSec}

We will start with 3+1 Schwarzschild, as per practice \cite{tHooft, Nomura}. But the lack of solvability of the wave equation means that in many ways this is {\em not} the simplest case study. We find BTZ to be the most tractable example, which we will study later. Writing the metric as
\begin{equation}  \label{SchwMetric}
ds^2 = -f(r)dt^2 + \frac{dr^2}{f(r)} + r^2  d\Omega_{2}^2  
\end{equation}
and separating variables in a massless scalar field as $\Phi(t, r, \theta, \phi) = e^{-i \omega t} R_{\omega l}(r) Y_{lm} (\theta,\phi)$, we get the radial wave equation in the form(s)
\bea  \label{radial-Sch}
&\frac{f(r)}{r^2} \partial_{r} (r^2 f(r) \partial_{r} R_{\omega,l}(r)) + \left(\omega ^2 - \frac{l(l+1)}{r^2}f(r)\right) = 0 \\
&f^{2}(r) R_{\omega l}^{''} (r) + \left[2\frac{f^{2}(r)}{r}+ f(r)f^{'}(r)\right] R_{\omega l}^{'} (r) + \left(\omega^2 - \frac{l(l+1)}{r^2} f(r)\right) R_{\omega l}(r) = 0
\eea
with the spherical harmonics satisfy their usual equation $\partial_{\Omega_2}^2 Y_{lm}(\theta, \phi) = - l(l+1) Y_{lm}(\theta,\phi)$. We can change to tortoise coordinates $r_*$ defined by $dr_* = \pm \frac{dr}{f(r)}$, 
\begin{equation}  \label{Sch-tort}
ds^2 = -f(r)(dt^2 + dr_*^2)+ r^2  d\Omega_{2}^2  
\end{equation}
The scalar field modes in the form 
\begin{equation}  \label{tort-separation}
\Phi_{t}(t, r_*, \theta, \phi) = e^{-i\omega t} \frac{R_{lm}(r_*)}{r} Y_{lm} (\theta, \phi)
\end{equation}
lead to a Schrödinger equation 
\begin{equation} \label{TortoiseSchrodinger}
-\frac{d^2}{dr_*^2} R_{lm}(r_*) + (V_l(r_*) - \omega^2) R_{lm}(r_*) = 0 
\end{equation}
where, the effective potential $V_l(r_*)$ is given by, 
\begin{equation} \label{tort-eff-pot}
V_l(r_*) = f(r)\big(\frac{f'(r)}{r} + \frac{l(l+1)}{r^2}\big).
\end{equation} 
%The tortoise coordinate radial wave equation can also be obtained by replacing del_r by del_r* using the relation between them in the first equation of eq(2.2) and from there r\Phi automatically comes up as the ansatz we propose for in tortoise coordinates, under the assumption R(r) = R(r*) which nomura has also used probably because this 1/r doesn't matter for him too. 

\subsection{The Effective Potential}

For Schwarzschild, $f(r) = 1-\frac{r_+}{r}$, where $r_+$ is the horizon radius related to mass $M$ of the black hole through $r_+ = 2 M l_p^2$. Here $l_p$ is the natural Planck length equal to $\sqrt{G_N}$ in natural units ($\hbar = c = 1$) in 3+1 dimensions.

The tortoise coordinate, with plus sign choice, is given by, $r_* = r + r_{+} {\rm ln}\big(\frac{r-r_+}{r_+}\big)$. For $r$ lying between, $r_{+} \leq r < \infty$, $r_*$ lies between $-\infty$ to $\infty$, where $-\infty$ is near the horizon and $+\infty$ is the asymptotic boundary. Now, in the same spirit as done by 't Hooft in \cite{tHooft}, we calculate Planck Length $l_{p}$ as the radial invariant distance starting at the horizon radius $r_+$ to $r_s\ \equiv r_{+}+x$, which is the brickwall location (in tortoise coordinates, the brickwall location is denoted $r_{*s}$):
\begin{equation} \label{l_p}
    l_{p} = \int_{r_+}^{r_{+}+x} dr \frac{1}{\sqrt{1 - \frac{r_{+}}{r}}}
\end{equation}
Since $x$ is small compared to $r_+$, it is approximately
\begin{equation} \label{l_p-approx}
     l_{p} \approx 2\sqrt{r_{+}x}
\end{equation}
We will not distinguish between the Planck length and string length in this paper, so will use the notations $l_s$ and $l_p$ interchangeably. But when there are a large number $N$ of fields in the effective field theory at the stretched horizon, it is useful to distinguish them \cite{Nomura} and their relative hierarchy is controlled by $N$. We will simply work with a single scalar field throughout the paper. With these definitions, we have
\begin{equation} \label{stretch-distance}
    r_s - r_+ \approx \frac{l^{2}_s}{4 r_+} \hspace{0.2cm} \leftrightarrow \hspace{0.2cm} r_{*s} - r_{+} \approx 2 r_{+} \log\left(\frac{l_s}{2 r_+}\right)
\end{equation}
and
\begin{equation} \label{l_s}
l_s = l_p = 2 r_{+}  e^{r_{*s}/(2 r_+)}  e^{-1/2}
\end{equation} 
The full effective potential is then given by
\begin{equation} \label{Sch-effective-pot}
V_l(r_*) = \left(1-\frac{r_+}{r}\right) \left(\frac{r_+}{r^3} + \frac{l(l+1)}{r^2}\right)
\end{equation}
In the near horizon limit, $f(r) \approx \frac{l_s^2}{4 r_+^2}$ , $f'(r) \approx \frac{1}{r_+}$ , $r_{*} \approx r _+ + r_+ \log \big(\frac{r -r_+}{r_+}\big)$ and also \eqref{stretch-distance}. So the near-horizon potential is  
\bea \label{NHP}
V^{nh}_l(r_{*}) = \left(r-r_{+}\right) \left(\frac{l(l+1)+1}{r^{3}_{+}}\right)
= \frac{l(l+1)+1}{r_+^2} e^{\frac{r_{*}-r_+}{r_+}}
\eea
and its value at the stretched horizon is
\bea \label{V-min-rs}
 V^{nh}_l(r_{*s}) = \frac{l_s^2}{4 r_+^4} (l(l+1)+1).
\eea
\begin{figure}
\centering
       \includegraphics[width=14 cm]{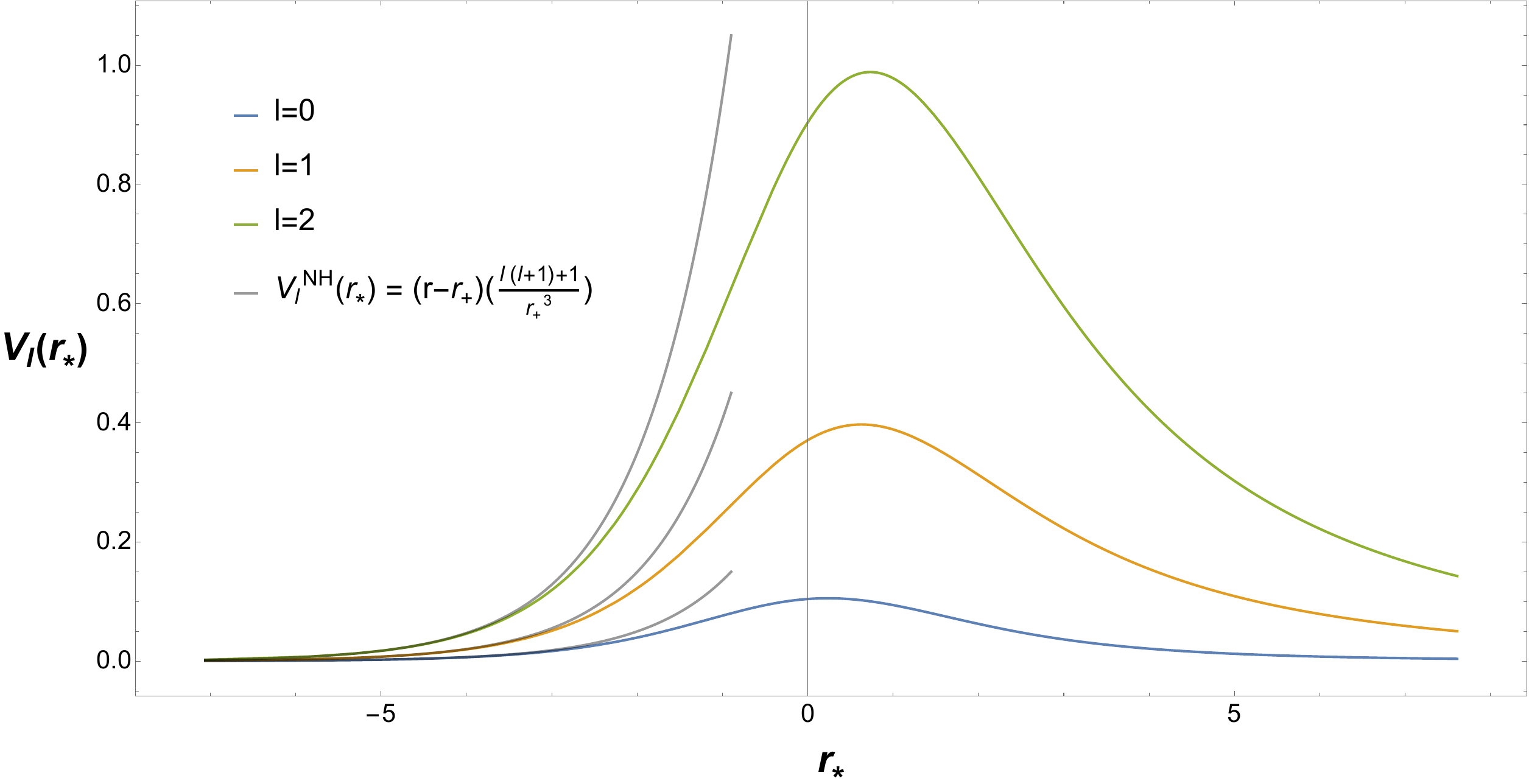}
       \caption{The massless scalar effective potential $V_l$ of  Schwarzschild for $l = 0, 1, 2$ as a function of tortoise coordinate $r_{*}$. The black lines represent  \eqref{NHP}. Near horizon modes are the ones from large and negative $r_*$ till the peak of potential and this is the region contributing to entropy of black hole. The plot is made with $r_+=1$.
}
\label{fig1}
\end{figure}
The solutions of the Schwarzschild radial equation are in terms of Heun functions which are not well studied functions compared to (say) hypergeometric or Bessel functions.  This makes Schwarzschild and Kerr cases, somewhat more complicated than (say) BTZ. If we are to study Schwarzschild directly, we may have to do even the first step (solving the radial equation) numerically, which is something we would like to avoid.  But we can make significant progress even in Schwarzschild, by making a near-horizon Rindler approximation. 

In the Rindler limit, the Schwarzschild radial equation can be written as 
\begin{equation} \label{Rindler-limit-eqn}
    R^{''}_{lm}(r_{*}) + \left(\omega^2 - \left(\frac{l(l+1)+1}{r^{2}_{+}}\right)e^{-1} e^{r_{*}/r_{+}}\right) R_{lm}(r_*) = 0 
\end{equation}
We will study the normal modes of the (Rindler wedge)$\times S^1$ in the next section as a continuation of some of our previous work \cite{Adepu, Sumit1}. The wave equation of that geometry is presented in \eqref{Rindler-S1-eqn}, and it maps to the wave equation above, once we make the identifications  
\begin{equation} \label{Schwarzschild-Rindler-map}
    a \equiv \frac{1}{2 r_+}, \hspace{0.3cm}, R \equiv r_+ \hspace{0.3cm}, r_* \equiv \xi \hspace{0.3cm},
      \frac{l(1+1/l+1/l^2)^{1/2}}{r_+} e^{-1/2} = \frac{J}{R}
\end{equation}
So we will develop the discussion of the normal modes in great detail in the Rindler setting, in the next section. This will go beyond the numerical discussion of the low-lying modes in \cite{Adepu, Sumit1}. Note that because we are concerned only with the radial equation for the most part, the dimensionality of the compact dimensions that we attach to Rindler will not be important. The only difference will be that $J$ can get replaced by the Casimir of the compact directions in higher dimensions, but for us this will only be a difference in nomenclature. Of course, if we want to compute the entropy and temperature of the black hole via normal modes and want to fix the numerical coefficients exactly, we will have to be more careful. But in this paper, we will only do that for the (rotating) BTZ case for which the normal modes are more directly tractable. 

Before turning to Rindler, we first develop some intuition 
by discussing some features of the effective potential.

\begin{itemize}
\item In the full Schwarzschild case, the effective potential has a peak for all values of $l$ (including $l=0$) and the peak gets steeper and steeper as $l$ increases. For any $l$, the low-lying $\omega$'s are bound (this range depends on $l$), but for sufficiently high $\omega$ the mode is above the barrier. This is not the case in Rindler, where the peak grows unboundedly for all values of $l$ (including $l=0$). All the near-horizon modes are trapped. 
\begin{figure}
\centering
\begin{subfigure}{.5\textwidth}
  \centering
  \includegraphics[width=0.9\linewidth]{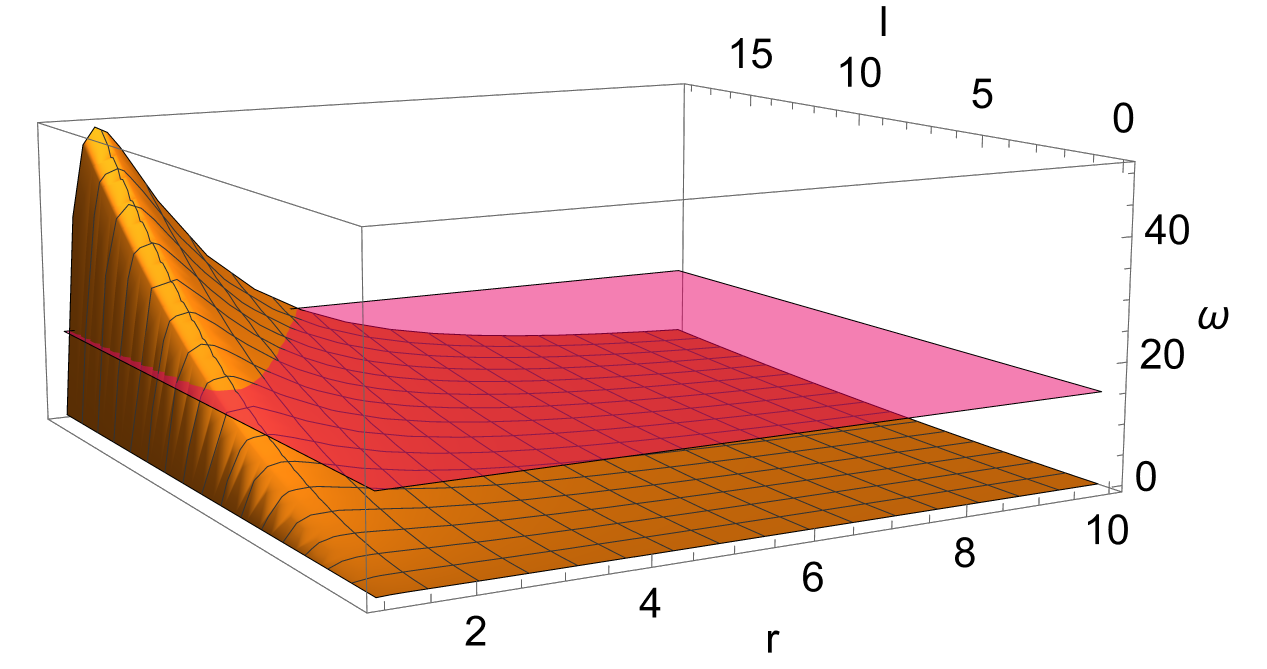}
  \caption{\textbf{Schwarschild potential}.}
  \label{flat-pot}
\end{subfigure}%
\begin{subfigure}{.5\textwidth}
  \centering
  \includegraphics[width=0.9\linewidth]{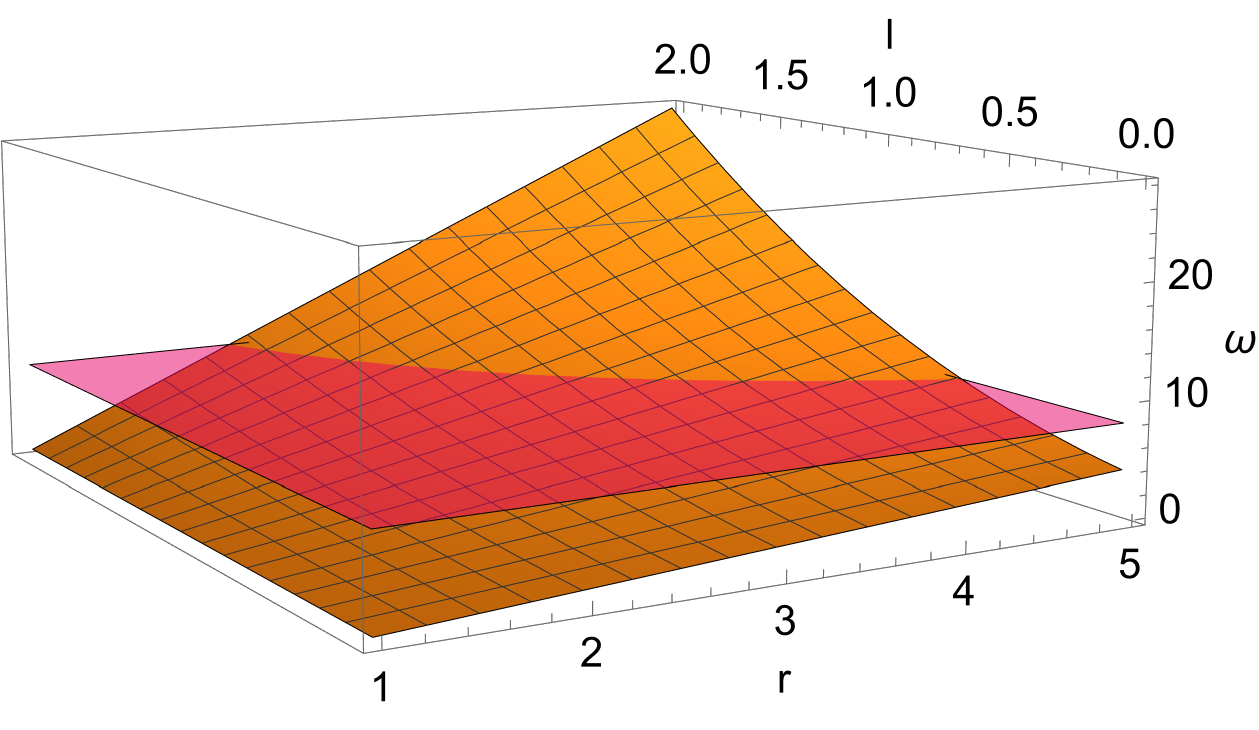}
  \caption{\textbf{Rindler potential}.}
  \label{rind-pot}
\end{subfigure}
\caption{3D Plots of Schwarschild and Rindler potentials plotted against $l$ and $r$. Constant $\omega$ planes are shown in pink.}
\label{fig2}
\end{figure}
\begin{figure}[h!]
       \centering
       \includegraphics[width=0.5\linewidth]{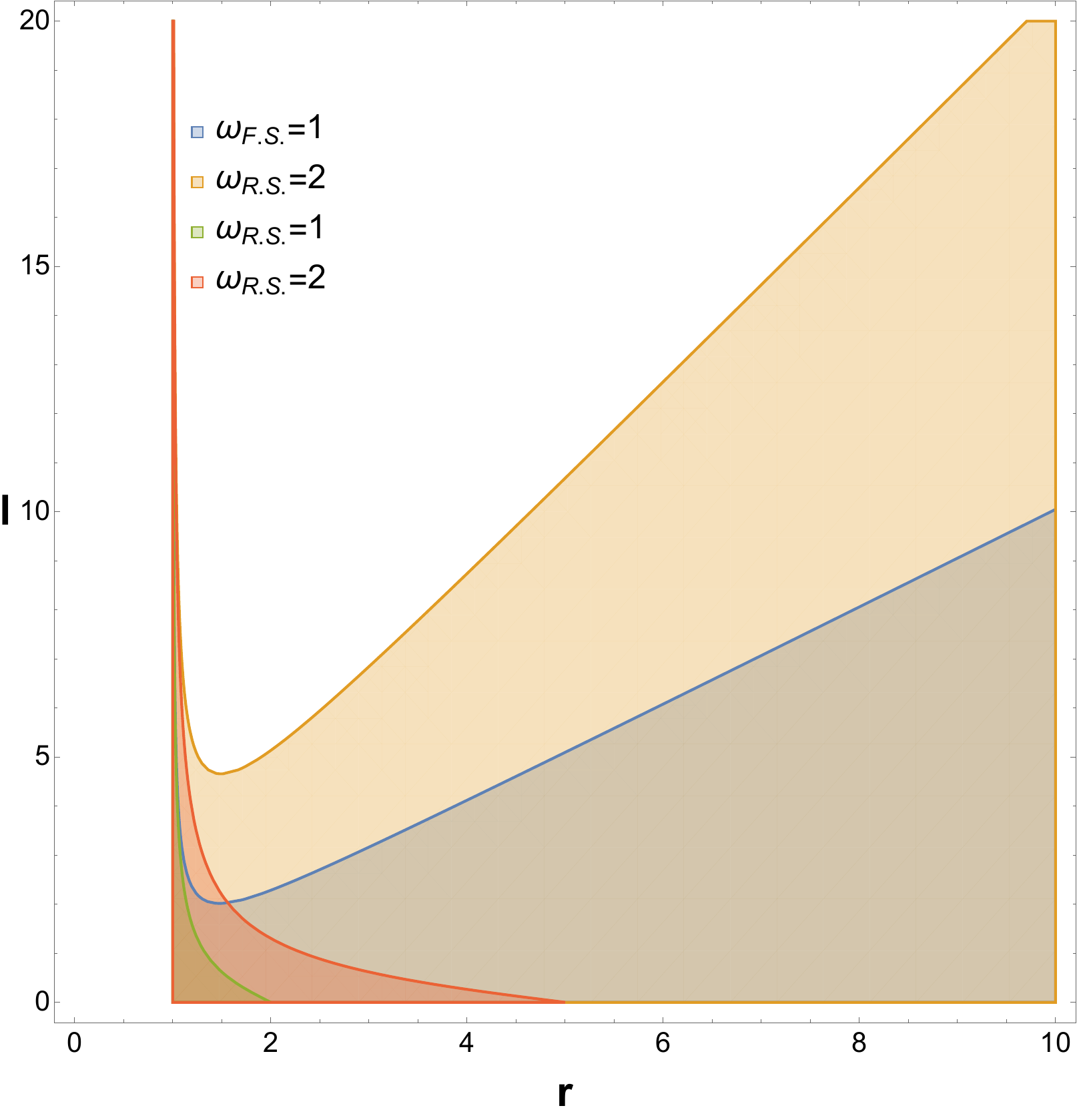}
       \caption{Shaded region represents  $\omega^2 - V_l(r_*) \geq 0$. Schwarschild case has a classical turnover point which causes some $l$'s to get unbound for a given $\omega$. Rindler case has monotonously dying down regions that prevent any mode from leaking out of the potential barrier. $\omega_{F.S.}$ stands for the full asymptotically flat black hole geometry and $\omega_{R.S.}$ for Rindler geometry. The choice is $r_+ = 1$ for these figures.}
       \label{pot-boundary}
\end{figure} 
    \item Fig \ref{pot-boundary} is a 2D visualisation of the 3D plots. The flat space potential has a turnover for a given $\omega$. This is due to the existence of a peak in the potential, and we call it a classical turnover point. So, for a given $\omega$, all the $ l$ modes lying below this turning point are free modes. Rindler does not have free modes. Also, as $\omega$ increases, we get more shaded region, because more states can be unbound. 
    \item The classical turning point tells us what is the minimum value of $l$ required for a given $\omega$ to be regarded as a bound state. The wavelength of Hawking radiation is of the order of the size of a black hole. Then for $\omega \sim \frac{1}{r_+}$, we see from Fig \ref{pot-boundary}, that at most the lowest-lying $l$-modes are unbound (say, $l=0$ and $l=1$).
    \item So for a given $\omega$, a rough idea about the near horizon bound states can be obtained from those $l$'s between the turnover point to the maximum allowed value ($l_{max}$), given by the equation $\omega^2 - V_l(r_*) = 0$ at a given radius $r$.  Since the peak of the potential is at about $r \sim \frac{3}{2}r_+$, the values of $r$ for which we have bound modes is $r_{s} \leq r \lesssim \frac{3}{2}r_{+}$. 
    \item Since we are interested in near horizon physics, we can make useful estimates of this $l_{max}$ using \eqref{V-min-rs} and a typical Hawking mode of energy $\omega \approx 1/r_{+}$. The result is  
   \bea
        l_{max}(l_{max}+1)+1 \approx \frac{4\omega^2 r^{4}_{+}}{l^2_{s}}, \implies 
        l_{max} \approx \frac{2 r_+}{l_s}
  \eea
This is a quantity that will play an important role in our discussions later, and is essentially the cut-off implicit in 't Hooft's calculation.
    \item Since only the lowest lying modes are affected by the distinction between Rindler and flat space Schwarzschild, we expect that the determinations of the entropy and temperature from normal modes should work even if we replace Schwarzschild with Rindler normal modes (up to perhaps $O(1)$ factors). This will indeed turn out to be the case. 
\end{itemize}

\section{Rindler as Poor Man's Schwarzschild}\label{RindlerSec}

It has been argued in \cite{Adepu, Sumit1} that to see signatures of quantum chaos in the normal modes, we need the angular quantum numbers $J$ associated to compact dimensions. So the Rindler metric we will work with is the simplest example with a compact extra dimension: 
\begin{equation}  \label{Rind-metric}
ds^2 = e^{2a\xi} \left(-d \eta^2 + d \xi^2 \right) + R^2 d\phi^2  
\end{equation}
Solving massless scalar field in this geometry with the ansatz $\Phi(\xi, \eta, \phi) = e^{-i\eta \omega} e^{i\phi J} \phi(\xi)$, we get the radial equation
\begin{equation}  \label{Rindler-S1-eqn}
\Phi_{lm}^{''}(\xi) +\big(\omega^2 - \frac{J^2}{R^2} e^{2a\xi}\big)\Phi_{lm}(\xi) = 0                 \\~\\
\end{equation}
This equation has two great virtues. Firstly, it allows exact solution in terms of Bessel functions of imaginary index. This was noted in \cite{Adepu, Sumit1}. Secondly, it is straightforward to solve for the normal modes directly using these Bessel functions using numerical tools like Mathematica. We will exploit this feature in this section to determine the exact normal modes to very large $J$. Note that neither of these simplifications happens in Schwarzschild. In BTZ, we can determine the exact solutions in terms of hypergeometric functions, but the second step is harder than in Rindler -- nonetheless we will be able to make progress in BTZ using perturbative methods in a later section.

Following \cite{Adepu} we first define variables $A = \omega/a$ and $y = e^{a\xi}(J/aR)$. Near the horizon\footnote{Near the horizon means $ \xi \rightarrow -\infty$ or $y \rightarrow 0$.} after imposing Dirichlet boundary conditions ($\Phi(\xi_o) = 0$) we require \cite{Adepu, Sumit1} 
\bea \label{exact-equation}
{\rm I}[-iA,y_o]-{\rm I}[iA,y_o] = 0  
\eea
where ${\rm I}$'s are certain Bessel functions and $y_o$ is the location of the stretched horizon. 
We will call this the {\em Exact Equation}. The exact normal modes can be found numerically for even very large values of $J$ by plotting the zeroes of the left hand side. This will be one of our results. 

We will also be interested in obtaining analytic expressions for the normal modes in the low-lying part of the spectrum. This is the part of the spectrum that contributes to the entropy and the quantum chaos signatures. The low-lying normal modes were obtained numerically in \cite{Adepu, Sumit1}. But for obtaining our results about black hole thermodynamics, it is very useful to have analytic expressions. This is part of our goal here. For doing this, we start with the near-horizon equations obtained in \cite{Adepu, Sumit1} which take the form
\begin{equation} \label{cos-sin}
\begin{split}
\cos(\alpha)+\cos(\theta) &= 0\\
\sin(\alpha)+\sin(\theta) &=0
\end{split}
\end{equation}
and $\alpha$ and $\theta$ are defined as  
\begin{equation} \label{alpha-theta-def}
\begin{split}
\alpha &= {\rm Arg} \left[\left(\frac{J}{aR}\right)^{-2iA}  \frac{\Gamma(iA)}{\Gamma(-iA)}\right]\\
\theta &= {\rm Arg}\left[\left(\frac{e^{a\xi_o}}{2}\right)^{2iA}\right]
\end{split}
\end{equation}
The two equations in \eqref{cos-sin} are in fact a single equation and we can write it as
\begin{equation} \label{phase-eq}
\cos\left({\rm Arg}\left[\Gamma\left(i \frac{\omega}{a}\right)\right] - \frac{\omega}{a} {\rm log}\left(\frac{e^{a\xi_o} J}{2aR}\right)\right) = 0
\end{equation}
which we call the {\em Phase Equation}. This equation is essentially as hard/easy to solve as the Exact Equation\footnote{In fact, the solutions of the Phase Equation are in excellent agreement with those of the Exact Equation as well, in the regimes of $J$ where $J \lesssim J_{max}$. We will define $J_{max}$ momentarily.} so one might wonder what we have gained by writing it.

\subsection{Analytic Low-Lying Spectrum}

The answer is that for low-lying $\omega$, we can solve for $\omega$ analytically by finding suitable approximations to this equation. In \eqref{phase-eq}, we can use the fact that for low values of $\omega$'s, ${\rm Arg}\left[\Gamma\left(i \frac{\omega}{a}\right)\right]$ can be approximated as a linear expression in $\omega/a$ as, 
\begin{equation} \label{Gamma-approx}
{\rm Arg}\left[\Gamma\left(i \frac{\omega}{a}\right)\right] =  - \frac{\pi}{2}-0.575 \left(\frac{\omega}{a}\right)+ ...
\end{equation}
The $0.575 ... $ here is a mathematical constant which is undoubtedly well-known to mathematicians. Using $\cos(x) = 0$ when $x = (2n-1)\pi/2$, we get an analytical expression valid for low $\omega$'s as follows 
\begin{equation} \label{Rindler-omega-a}
\frac{\omega}{a} = -\frac{n\pi}{a\tilde{\xi_o} + \log\left(\frac{J}{2aR}\right)}
\end{equation}
where, $ n \in \mathbb{Z}^+$ and $a\tilde{\xi_o} = a\xi_o + 0.575$ . We will refer to this result as the {\em Approximate Phase Equation} or as the {\em Analytic Low-Lying Spectrum}. This expression and variations of it for other black holes will be one of the main tools for us, in this paper.

A crucial fact about this solution is that it diverges when the denominator is zero for a given cutoff (negative $\xi_o$) which happens at large $J$. We will call this value of $J$,  $J_{max}$. The solution has completely broken down and diverges at this point, but we find that for values of $J$ that are about an order of magnitude smaller than $J_{max}$ or smaller (ie., $ J \lesssim J_{max}/10$), the analytic low-lying spectrum is a fairly good approximation to the exact spectrum\footnote{More concretely, it captures the entropy, temperature and the  ramp, as well as the exact spectrum does.}.  We can  easily determine $J_{max}$ to be
\begin{equation} \label{Jmax} 
J_{max} = 2aR \hspace{0.2cm}  e^{-0.575} e^{-a\xi_o}.
\end{equation}

An expression for very low $J$'s can also be written from \eqref{Rindler-omega-a}: 
\begin{equation} \label{ultra-approx}
\frac{\omega}{a} = -\frac{n\pi}{a\tilde{\xi_o}} + \frac{n\pi}{a^2 \tilde{\xi_o}^2} \log\left(\frac{J}{2aR}\right)
\end{equation}
This starts deviating from the spectrum (both low-lying and exact) pretty early -- but the reason why it is interesting is that this log form of the spectrum is able to see the linear slope $\sim 1$ ramp \cite{Sumit2}.
\begin{figure}
\centering
\begin{subfigure}{\textwidth}
  \centering
  \includegraphics[width=0.8\linewidth]{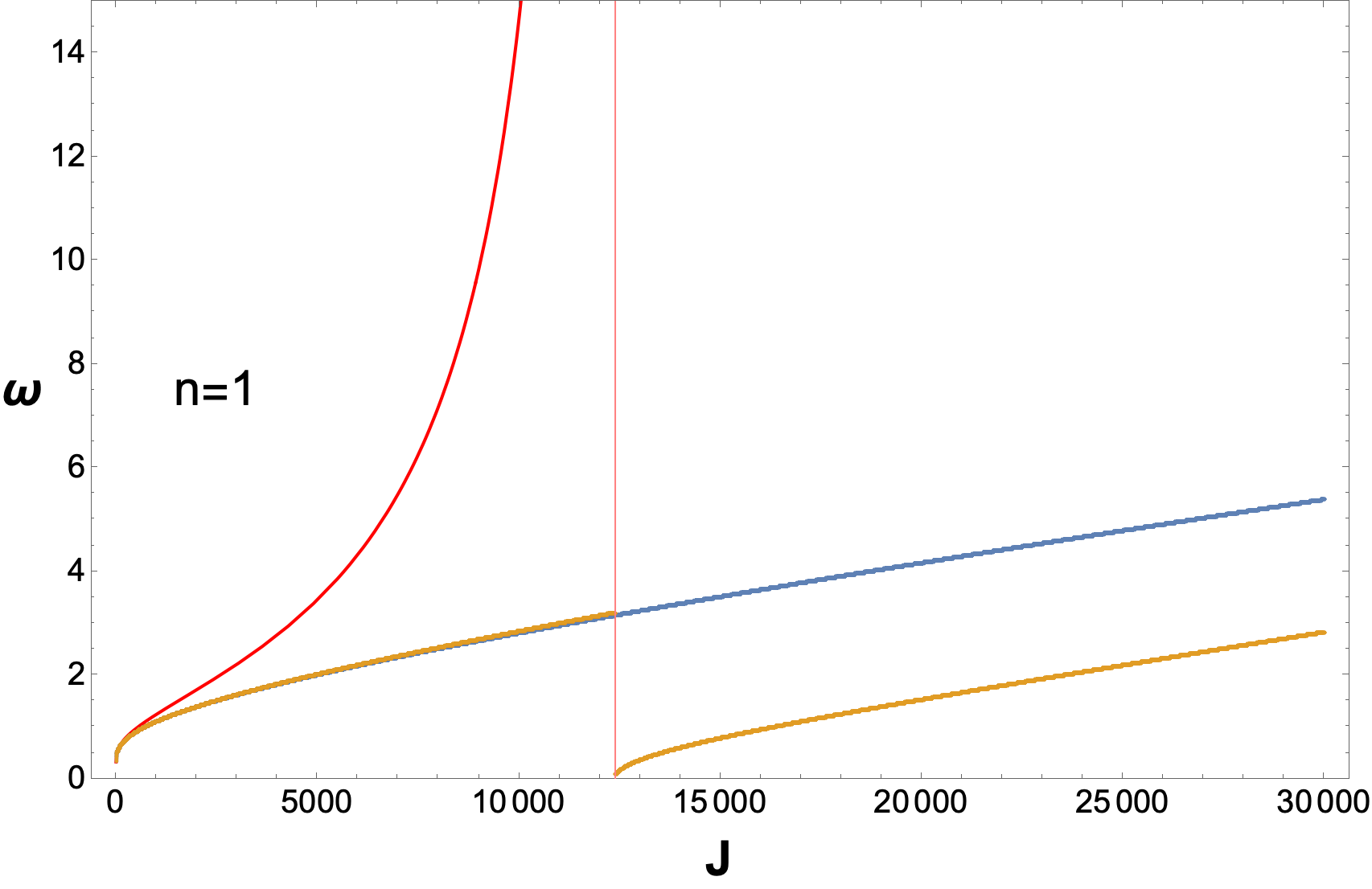}
  \caption{} % Label as Fig 4(a)
  \label{plot-large-J-n=1}
\end{subfigure}

\vspace{1cm} % Adjust the vertical spacing between the figures

\begin{subfigure}{\textwidth}
  \centering
  \includegraphics[width=0.8\linewidth]{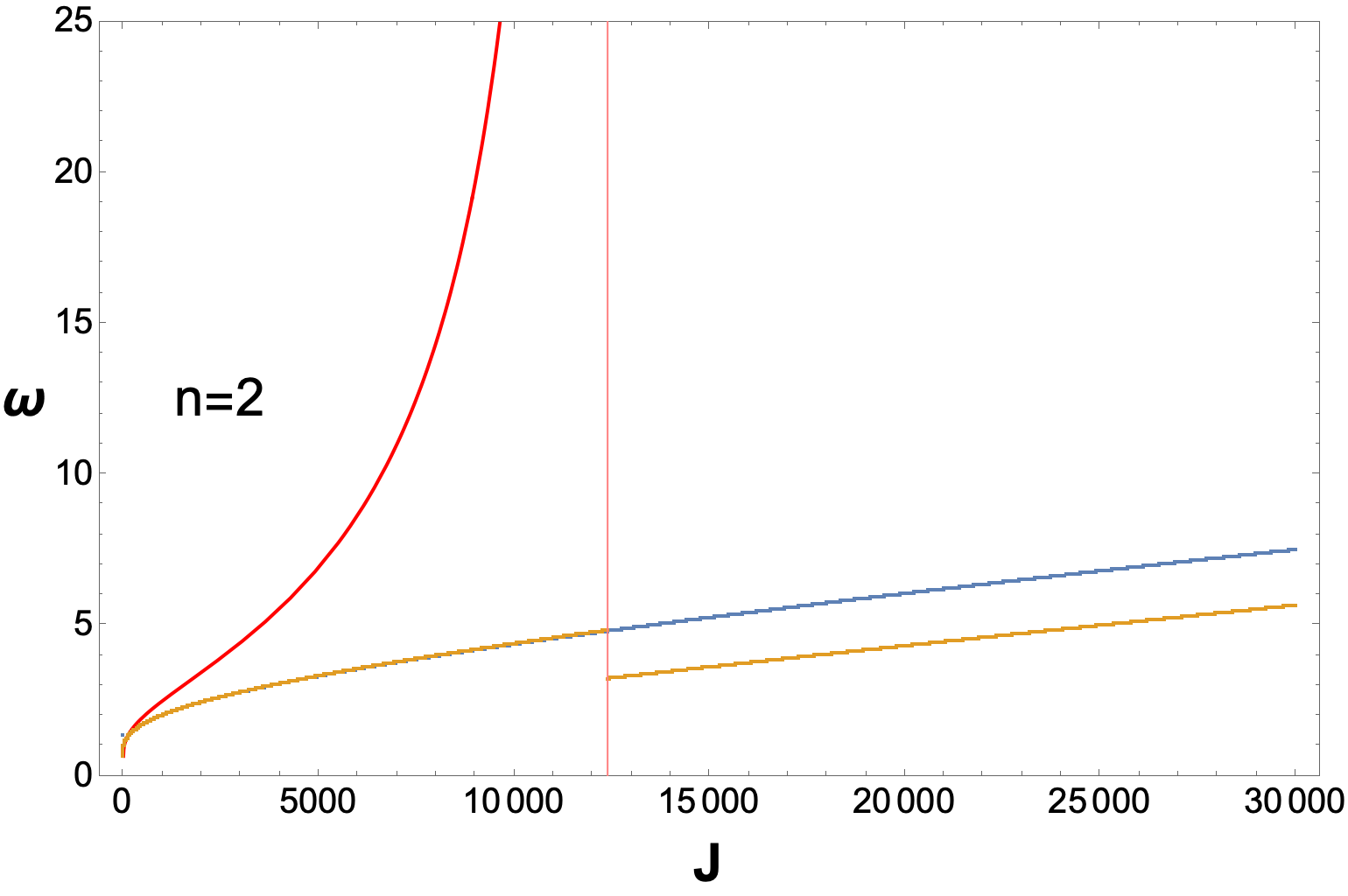}
  \caption{} % Label as Fig 4(b)
  \label{plot-large-J-n=2}
\end{subfigure}
\caption{Rindler Modes for $\xi_o = -10$ with $a = 1$ and $R = 1/2$: Blue stands for exact equation  \eqref{exact-equation}, yellow stands for phase equation \eqref{phase-eq}, red stands for the analytic low-lying spectrum \eqref{Rindler-omega-a} and the pink line denotes the breakdown point of the analytic low-lying spectrum, which is $J_{max} = 12400$ obtained from \eqref{Jmax}. The phase equation breaks down at $J = 12,394$ which is close to this.}. 
\label{plot-for-large-J}
\end{figure}

\begin{figure}
\centering
\begin{subfigure}{.5\textwidth}
  \centering
  \includegraphics[width=0.9\linewidth]{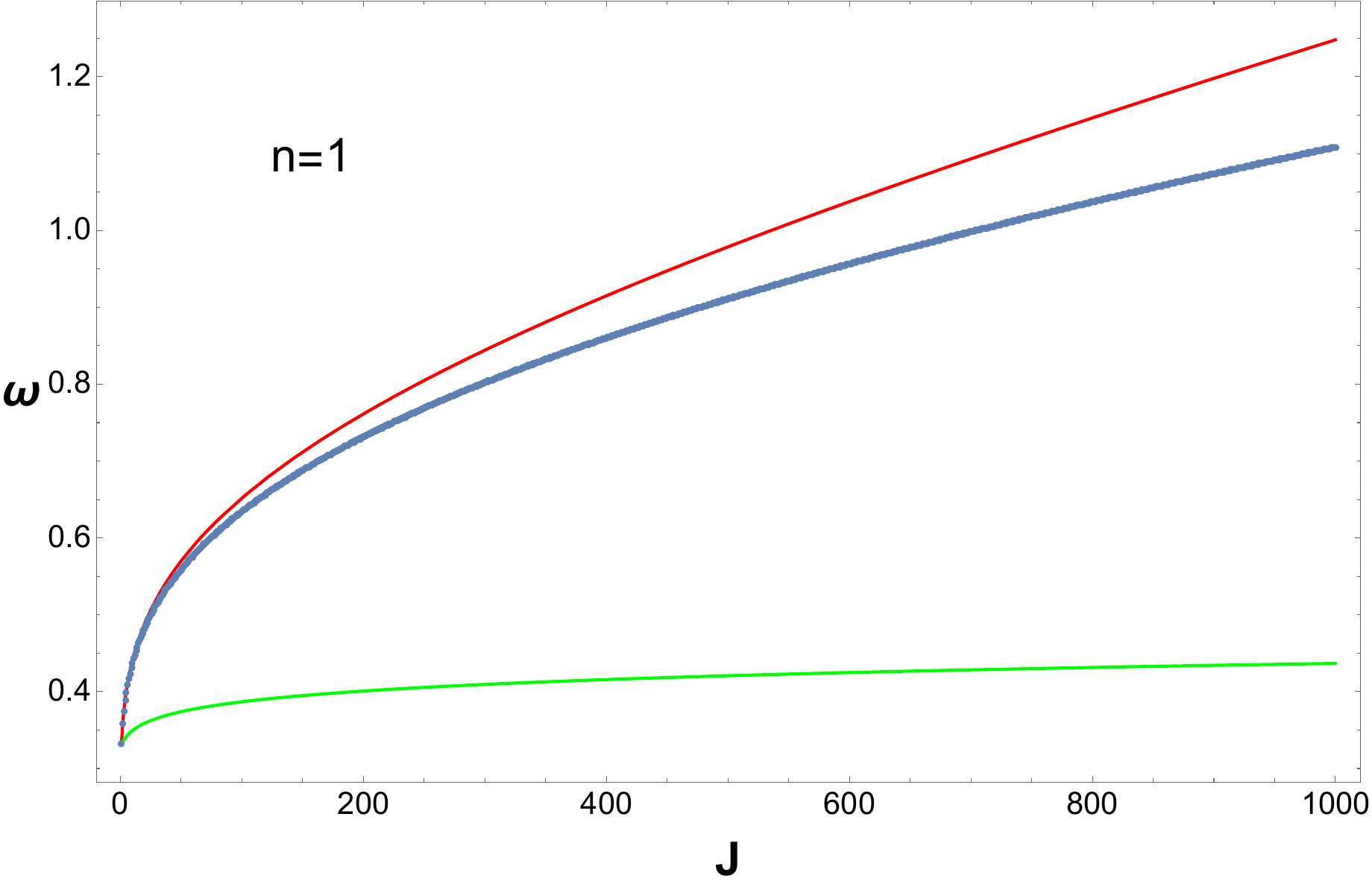}
  \caption{}
  \label{plot-for-small-J-n=1}
\end{subfigure}%
\begin{subfigure}{.5\textwidth}
  \centering
  \includegraphics[width=0.9\linewidth]{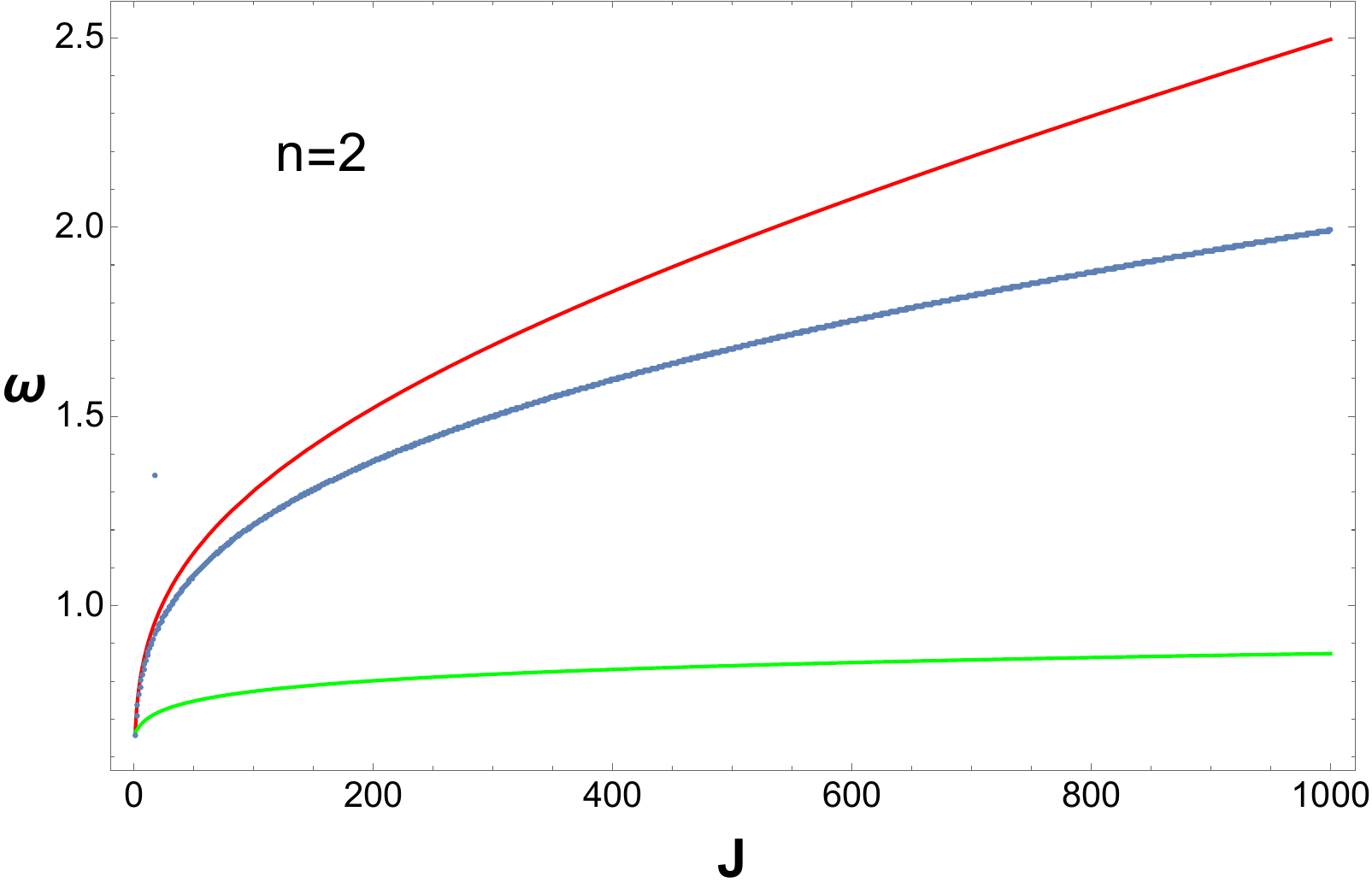}
  \caption{}
  \label{plot-for-small-J-n=1}
\end{subfigure}
\caption{Comparison between exact spectrum (blue), analytic low-lying spectrum (red) and the spectrum at low $J$'s (green) as in \eqref{ultra-approx} at $\xi_o = -10$ with $a = 1$ and $R = 1/2$.}. 
\label{plot-for-small-J}
\end{figure}

\begin{figure}[h]
       \centering
       \includegraphics[width=0.9\linewidth]{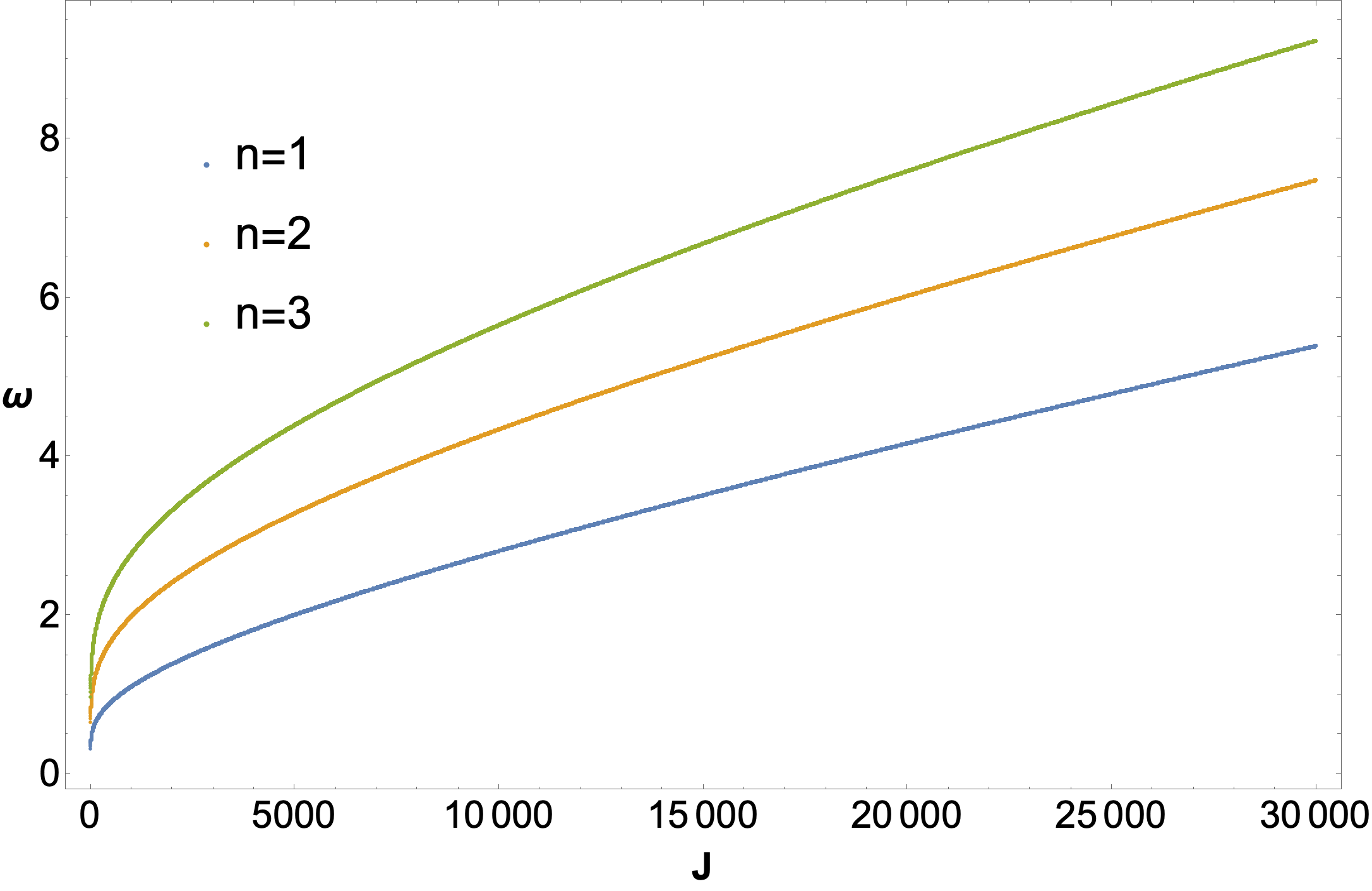}
       \caption{Plot of exact spectrum $\omega(n,J)$ for Rindler geometry $\xi_o = -10$ with $a = 1$ and $R = 1/2$ for $n = 1, 2$ and 3.}
       \label{many-n-one}
\end{figure} 

We would like to compare \eqref{exact-equation}, \eqref{phase-eq}, \eqref{Rindler-omega-a} and \eqref{ultra-approx} by actually solving for the $\omega$'s at a given cutoff. We collect our main observations below. 
\begin{itemize}
    \item From Fig \ref{plot-for-large-J}, we see that the analytic low-lying spectrum breaks down at $J_{max}$, while the solutions of the phase equation have a jump at a value close to $J_{max}$. Since we have access to the exact spectrum, the phase equation solutions are not of much interest in the Rindler case. But we will have more to say about perturbation theory and the jumps in the phase equation when we discuss the BTZ case\footnote{In the BTZ case, the exact equation involves hypergeometric equations and we have not been able to solve them directly for large values of $J$.}. Here, we will ignore them. The exact spectrum continues spectacularly without any breakdown.
    \item From Fig \ref{plot-for-small-J}, we notice that analytic low-lying spectrum is an approximate fit for the exact spectrum for range of $J \lesssim 0.1 J_{max}$ for low $n$ (with errors in the $10\%$ range at the higher end of $J$). Later in this paper, we will try to analytically model the full spectrum (more precisely, up to $J_{max}$). In the low-lying part we will use the analytic low-lying spectrum till $J = J_{inter} \sim 0.15 J_{max}$. Then we will match it with a linear fit from $J_{inter}$ to $J=J_{max}$. We will discuss the details of the linear fir, later. The point we will make here is that the low-lying part is the one which has the interesting physics. The larger $J$ parts of the spectrum are evenly spaced - such even spacing is what we will find in empty flat space with a hole instead of a horizon \cite{Sumit1} or a simple harmonic oscillator. But as long as we are interested in only the spectrum below $J_{max}$\footnote{We know from 't Hooft's calculation that the cut-off has to be lower than (or equal to) $J_{max}$. What we find in this paper is that the cut-off is in fact a bit lower, better captured by $J_{inter}$ than $J_{max}$.}, we find that the physics is controlled by modes below $\sim J_{inter}$. The slope of the ramp is unaffected by the higher modes (ie., modes in the range $J_{inter} < J < J_{max}$), and we are also able to capture the temperature and entropy. We will describe this in detail as we proceed.
    \item From the exact spectra in Fig \ref{many-n-one}, we see that the level spacing between $\omega$'s in the $n$ direction at a fixed $J$ is much larger than the level spacing between $J$'s at a given $n$. One can view this as an effective quasi-degeneracy in the $J$-direction. The plots appear continuous in the $J$-direction because of this, but they are actually plots of discrete points. This hierarchy in level spacings is important for the physics.
\end{itemize}

\subsection{Linear Fit of the Spectrum near $J = J_{max}$}

We would like to have some intuition for the spectrum near $J \sim J_{max}$. As we will see in discussions of 't Hooft's calculations and our generalizations, the region of the spectrum up to $J_{max}$ is what plays a role in the thermodynamics and chaos-aspects of black holes. In fact, we will eventually conclude that an even more convenient statement is true -- only the modes up to about an order of magnitude lower than $J_{max}$ are really significant for these questions. This is what we call $J_{inter}$ (or $J_{cut}$ in later sections, which is defined slightly differently, but again captures the same idea).

We will first solve the exact equation using a linear approximation near $J_{max}$. Between the analytic low-lying spectrum of the previous subsection and this linear high-lying approximation, we will have some intuition for the full spectrum up to $J_{max}$. We will refer to the two together, as the {\em analytic fitted spectrum}. We will compute the spectral form factor (SFF) for the analytic fitted spectrum and the exact spectrum (both for modes up to $J_{max}$) for various values of $J_{max}$ (or equivalently, the stretched horizon location) and see that the slopes of their ramps match with each other. As we slowly increase $J_{max}$ (or equivalently bring the stretched horizon closer to the horizon) the slope gets closer and closer to $\sim 1$. Together, we will take these observations as a strong hint that the analytic fitted spectrum is a good substitute for the exact spectrum, for some of the questions regarding chaos and thermodynamics of black hole horizons. 

Using the Bessel function formula $K[\nu,x] = \frac{\pi}{2} \frac{I[-\nu,x]-I[\nu,x]}{\sin(\pi\nu)}$ with $\nu = i\frac{\omega}{a}$, we can re-write the exact equation in terms of modified Bessel functions of $2^{nd}$ kind:
\begin{equation} \label{eq3.12}
    K\left[i\frac{\omega}{a},\frac{e^{a\xi_o} J}{aR}\right] = 0
\end{equation}
This is the form of the exact equation that we will often use. We can expand the LHS around $\omega = \omega_o$ and $J = J_{max}$, where $\omega_o$ is the energy value at $J = J_{max}$ and is the first root of the equation $K\left[i\frac{\omega}{a},\frac{e^{a\xi_o} J_{max}}{aR}\right] = K\left[i\frac{\omega}{a},2 e^{-0.575}\right] = 0$, where we have used \eqref{Jmax}. Note the striking fact that this fixes $\frac{\omega_o}{a}$ completely as a number. Its numerical value turns out to be about $\frac{\omega_o}{a} \approx 3.16... $

We can find a linear approximate solution around $J_{max}$ by doing a double Taylor expansion  
\begin{equation} \label{doubleTaylor}
\begin{split}
K\left[i\frac{\omega}{a},\frac{e^{a\xi_o} J}{aR}\right] &= K\left[i\frac{\omega_o}{a},\frac{e^{a\xi_o} J_{max}}{aR}\right] + \partial_{\omega} K\left[i\frac{\omega}{a},\frac{e^{a\xi_o} J}{aR}\right]_{\omega_o,J_{max}} \delta\omega\\
&+\partial_{J} K\left[i\frac{\omega}{a},\frac{e^{a\xi_o} J}{aR}\right]_{\omega_o,J_{max}} \delta J + O(\delta \omega^{2}, \delta J^2, \delta \omega \delta J) = 0\\
\end{split}
\end{equation}
Ignoring $O(\delta \omega^{2}, \delta J^2, \delta \omega \delta J)$ and higher order terms, we get a linear relation between $\omega$ and $J$. We will interested in solving it for $n=1$. We can re-write the above equation as:
\begin{equation} \label{eq3.14}
\begin{split}
\omega &= \omega_o - \frac{K\left[i\frac{\omega_o}{a},\frac{e^{a\xi_o} J_{max}}{aR}\right]+\partial_{J} K\left[i\frac{\omega}{a},\frac{e^{a\xi_o} J}{aR}\right]_{\omega_o,J_{max}} (J-J_{max})}{\partial_{\omega} K\left[i\frac{\omega}{a},\frac{e^{a\xi_o} J}{aR}\right]_{\omega_o,J_{max}}},\\
\omega &= \omega_o + c + m \frac{e^{a \xi_o}}{2R} J\\
\end{split}
\end{equation}
where the second equation is just a re-writing of the first, it simply defines $m$ (slope) and $c$ (intercept). Note also that the first term in the numerator on the RHS of the first equation is zero. Using the identity
\bea
\partial_{J} K\left[i\frac{\omega}{a},\frac{e^{a\xi_o} J}{aR}\right] =  -\frac{e^{a \xi_o} J}{2aR} \left(K\left[-1+i\frac{\omega}{a},\frac{e^{a \xi_o}J}{a R}\right]+K\left[1+i\frac{\omega}{a},\frac{e^{a \xi_o}J}{a R}\right]\right),
\eea
we can eventually write
\begin{equation} \label{eq3.15}
\begin{split}
c &= i\frac{a K\left[i\frac{\omega_o}{a},\frac{e^{a\xi_o} J_{max}}{aR}\right] + \frac{e^{a \xi_o} J_{max}}{2R} \left(K\left[-1+i\frac{\omega_o}{a},\frac{e^{a \xi_o}J_{max}}{a R}\right]+K\left[1+i\frac{\omega_o}{a},\frac{e^{a \xi_o}J_{max}}{a R}\right]\right)}{K^{(1,0)}\left[i\frac{\omega_o}{a},\frac{e^{a\xi_o} J_{max}}{aR} \right]}\\
m &= - i\frac{K\left[-1+i\frac{\omega_o}{a},\frac{e^{a \xi_o}J_{max}}{a R}\right]+K\left[1+i\frac{\omega_o}{a},\frac{e^{a \xi_o}J_{max}}{a R}\right]}{{K^{(1,0)}\left[i\frac{\omega_o}{a},\frac{e^{a\xi_o} J_{max}}{aR} \right]}}\\
\end{split}
\end{equation}
with $K^{(1,0)}[x,y]$ meaning partial derivative of BesselK function with respect to $x$. This fixes the linear fit completely.

We keep $a = 1, R = 1/2$ in all our numerics. A key observation is that $c$ and $m$ are independent of the cutoff (aka stretched horizon location). With these understandings, we can write a fully concrete expression for $\omega$ from the knowledge of Bessel functions: 
\begin{equation} \label{linear-approx-eqn}
\omega(1,J) = 1.4162 + 3.0973 e^{\xi_o}J 
\end{equation}
This equation we will call the {\em linear approximation equation} or the {\em linear high-lying approximation}.  The plot below in Fig \ref{ana-fit-spec} illustrates the kind of trade-off we are making when we replace the exact spectrum with the analytically fitted spectrum. A useful observation is that the shape (or ``curve") of the spectrum as a function of $J/J_{max}$ is independent of the stretched horizon cut-off (or equivalently $J_{max}$). The intersection between the analytic low-lying spectrum and the linear high-lying approximation is at a point we call $J_{inter}$: in the $J<J_{inter}$ region, we expect that we can get some sense of the spectrum using the analytic low-lying formula\footnote{One of the lessons of this paper is that the modes that are low-lying compared to $J_{max}$ by perhaps a few orders of magnitude, are enough to capture the chaos/thermodynamics aspects of black holes. This is the regime in which the analytic low-lying spectrum works best. $J_{inter}$ can be viewed as a sort-of worst case upper bound of this regime -- our discussions do not need us to go even as high as $J_{inter}$.} and in the $J\geq J_{inter}$ region, it is approximated by the linear approximation.  
\begin{figure}[h]
    \centering
    \includegraphics[width=0.6\linewidth]{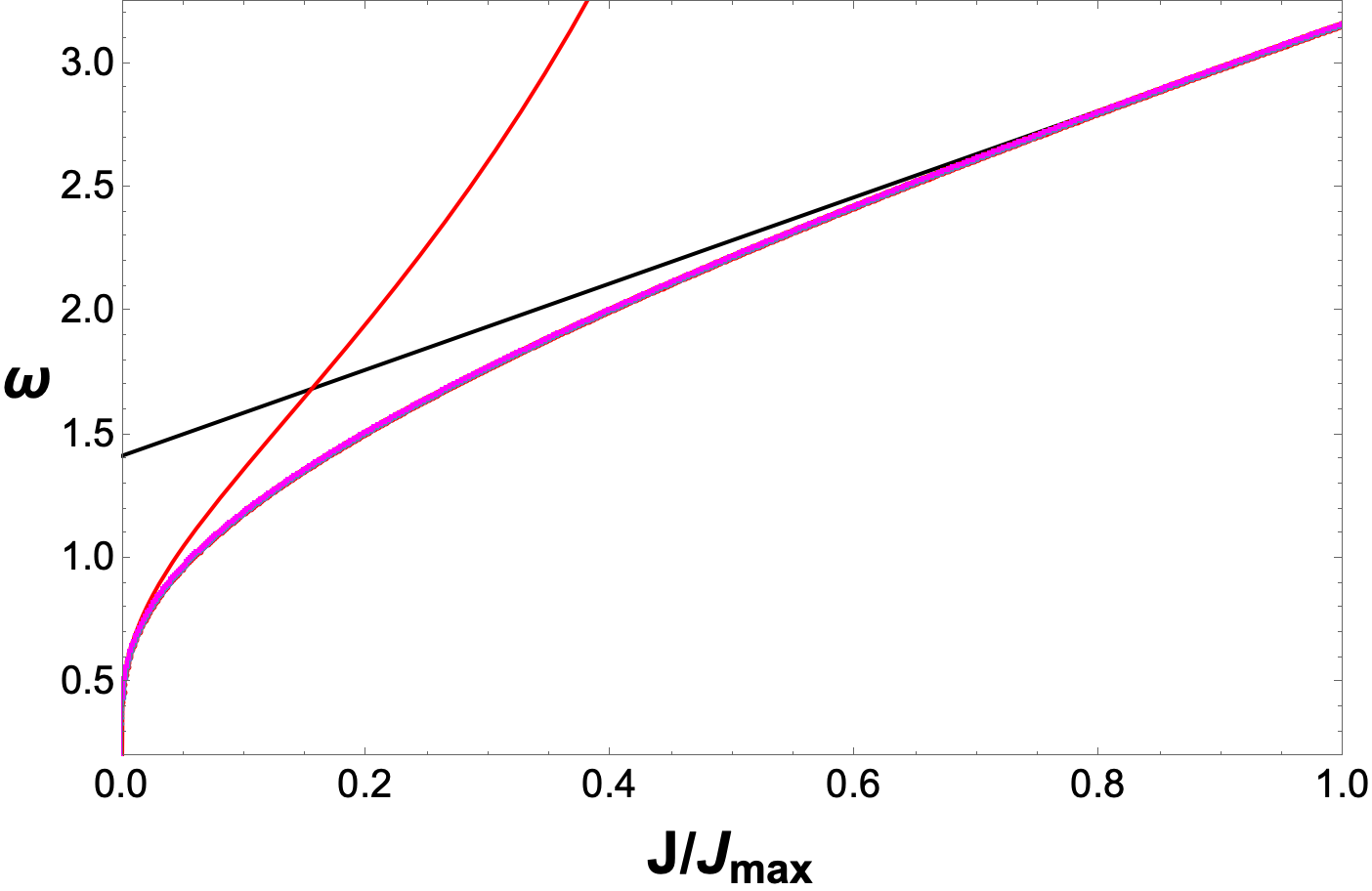}
    \caption{Plots of exact $n=1$ spectra for the cutoffs $\xi_o =-8, -10, -12, -14 \ {\rm and} -16$ together with the two constituents of the analytically fitted spectrum (the analytic low-lying spectrum, red, and the linear high-lying approximation, black). The exact spectra all hug the same pink curve, because the horizontal axis is scaled as $J/J_{max}$.}
    \label{ana-fit-spec}
\end{figure} 

\noindent
Let us make some comments about this plot:
\begin{itemize}
    \item The structure of the plots doesn't change along $\omega$-axis but we get access to more and more $J$'s along the horizontal axis as we increase (magnitude of) cutoff. This means that the consecutive neighbouring $\omega$ points along $J$-direction get closer and denser as we increase cutoff. This is what suggests that we plot $\omega$ vs $J/J_{max}$ for different cutoffs. The points get denser as we increase $J_{max}$ (or cut-off) but it should be kept in mind that the continuous looking pink curve in the figure actually stands for various cut-offs with discrete points, and the points do not overlap.  
    \item One can understand $J_{inter}$ from Fig \ref{ana-fit-spec}. We can explicitly calculate the intersection point from the equations for the red and black curves in the figure, and we find that $J_{inter}/J_{max} = 0.155292$. Note that this is not cut-off dependent, the cut-off dependence is implicit in the $J_{max}$. This is clear from the structure of \eqref{Rindler-omega-a} (note that we have made choices for $a$ and $R$) and \eqref{linear-approx-eqn}. Hence, the intersection point remains the same $\omega$ value and cannot be smoothed by going closer to the horizon. But we will see that the slope of the ramp does change with cut-off -- this is because the density and level-spacing of points on the pink curve changes with the cut-off. We find that the low-lying part {\em always} controls things like the SFF and the slope of the ramp -- and as we increase the cut-off the slope of the ramp tends closer to 1.   
\end{itemize}    
   
It is clear from the above discussion that the spectrum depends only upon \textbf{$j = J/J_{max}$} where $J_{max}$ encodes the information about the cutoff. So we can write the analytically fitted spectrum as a function of $j$ for $n=1$: 
    \begin{equation} \label{ana-fit-expression}
    \begin{split}
        \omega(1, j) &= -\frac{\pi}{\log\left(j\right)}, \hspace{2.3cm} \ \  0<j<j_{inter}\equiv 0.155292\\
        \omega(1, j) &= 1.4162 + 1.7428 \left(j\right), \hspace{0.4cm} \ \ j_{inter}\leq j \leq 1\\
    \end{split}
    \end{equation}
\noindent   
This has further implications, which we now discuss. 
    \begin{itemize}
        \item The difference between two consecutive omegas ($\Delta J = 1$) in the low-lying vs. the linear part of the spectrum can be computed from \eqref{ana-fit-expression}: 
        \begin{equation} \label{eq3.18}
        \begin{split}
            \Delta \omega &= \frac{\pi}{J \log\left(\frac{J}{J_{max}}\right)} \hspace{1cm}, 1\leq J<0.155292 J_{max}\\
            \Delta \omega &= \frac{1.7428}{J_{max}}\hspace{2.1cm}, 0.155292 J_{max}\leq J \leq J_{max}\\
        \end{split}
        \end{equation}
       These expressions show that the level-spacing is roughly evenly spaced around $J_{max}$ -- a fact that can be numerically checked from the exact spectrum. But at the low-lying part of the spectrum, the level-spacing has a more involved structure. This is presumably what is responsible for the slope $\sim 1$ linearity of the ramp as we increase the cut-off. Note that an evenly spaced (noisy) spectrum leads to a ramp with slope $\sim 2$, a fact noted in \cite{Sumit1, Sumit2}. 
              \item Even though it is hard to establish this conclusively, we suspect that the smaller level-spacing of the high-lying part of the spectrum is an indication that its influence arises only in the later part of the ramp. This is also plausible from our SFF plots in the next section, where find that the ramp has a slight upward tilt towards the top. Increase of the slope from $\sim 1$ to values of slope $ > 1$ is a feature noticed in \cite{Sumit2} as the spectrum changes from $\log n$ to positive power law $n^{\alpha >0}$ (including $\alpha=1$). So it stands to reason that this upward tilt is result of the part of the spectrum with constant (and small) level-spacing. This argument is not a proof, because it only pays attention to nearest-neighbor level-spacing.
        \item We have decided to demarcate the spectrum between its lower and upper parts via the $J_{inter}$ that we defined above.  This is a natural choice, but if our goal is simply to identify a location where the analytic low-lying spectrum is beginning to break down, we could also consider the $J$ at which the the latter has a point of inflection. We find in our calculations that these two choices are close to each other (we will illustrate this with an example in the next section). For high enough cut-off, we do not seem to see any substantial difference between the two choices (say in the linearity of the ramp).
\end{itemize}

\section{Using the Ramp to Locate Structure in the Spectrum}

It has been shown in \cite{Sumit1} that the normal modes of a free scalar on a black hole background with a stretched horizon gives rise to a dip-ramp-plateau (DRP) structure in the one-particle spectral form factor (SFF), which is defined as,
\begin{equation} \label{SFF Formula}
    g(\beta,t) = \frac{|Z(\beta,t)|^2}{|Z(\beta,0)|^2}
\end{equation}
where,
\begin{equation} \label{Partition Function for SFF}
    Z(\beta,t) = \sum_{\omega} e^{-(\beta-it)\omega} = \sum_{J=1}^{J=J_{cut}} \sum_{n=1}^{n_{cut}} e^{-(\beta-it)\omega(n,J)}
\end{equation}
It was noticed in \cite{Sumit1} that there is a quasi-degeneracy  in the $J$-direction (ie., dependence along $J$ direction is much weaker compared to that along the $n$-direction). It is the summation over $J$ modes that is correlated with the presence of the ramp. The DRP structure is largely unaffected by the choice of $\beta$. So we will set $\beta = 0$ and $n_{cut} = 1$ from the starting and use the expression
\begin{equation} \label{eq4.3}
    g(0,t) = \frac{\left|\sum_{J=1}^{J_{cut}} e^{it\omega(1,J)}\right|^2}{|J_{cut}|^2}
\end{equation}
We expect that the modes that are responsible for the black hole entropy are also loosely responsible for the linear ramp (i.e., slope  $= 1$). Later, we will see that the modes that are well-approximated by the analytic low-lying spectrum are the ones responsible for the thermodynamics of black holes. Here we will build evidence that the modes that contribute to the linearity of the ramp are also the same\footnote{In the papers \cite{Adepu,Sumit1} our goal was to use normal modes to identify features associated to chaos and random matrices in black holes. In this paper, our goal is morally the inverse. We wish to identify the {\em part} of the spectrum that is responsible for the signatures of quantum chaos in black holes.}.

We will first present evidence that the analytic fitted spectrum and the exact spectrum lead to the same slope of the ramp for a given cut-off. The utility of this observation is that this can be tested at relatively low cut-offs. Evaluation of the exact normal modes for very large $J$ at very large values of the stretched horizon cut-off is quite challenging. But once we build confidence that the analytic fitted spectrum is a good proxy for the exact spectrum, it is relatively easy to determine the latter at large cut-off and $J$. So we can compute the SFF using the analytic fitted spectrum, and find evidence that indeed the slope of the ramp tends to $\sim 1$ as the cut-off becomes large\footnote{Even here, we needed the help of friends to check numerically that the slope indeed tends to $\sim 1$ at large $J_{max}$. We thank Sumit Garg for testing some of our hypotheses in this section, numerically. We will only present the cases we have tested here, but we emphasize that the evidence for our claims is stronger than the plots we present.}. These experiments will allow us to distill the essence of these calculations into a toy model spectrum that captures the essential features of this spectrum, which will be discussed in the next subsection. All plots in this section are again done with $a=1, R=1/2$.

\begin{figure}
\centering
\begin{subfigure}{.5\textwidth}
  \centering
  \includegraphics[width=0.9\linewidth]{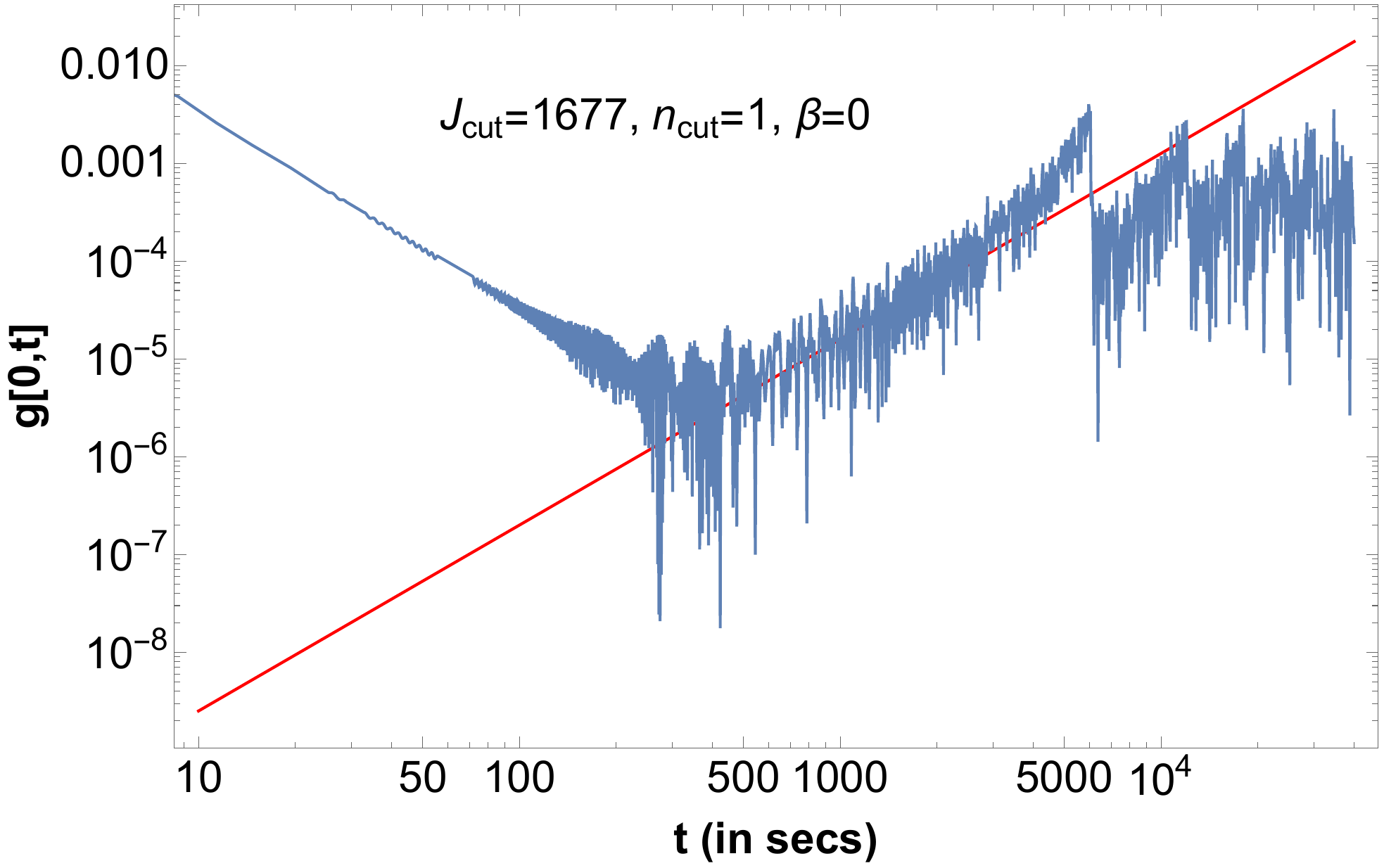}
  \caption{}
  \label{Exact SFF - (-8)}
\end{subfigure}%
\begin{subfigure}{.5\textwidth}
  \centering
  \includegraphics[width=0.9\linewidth]{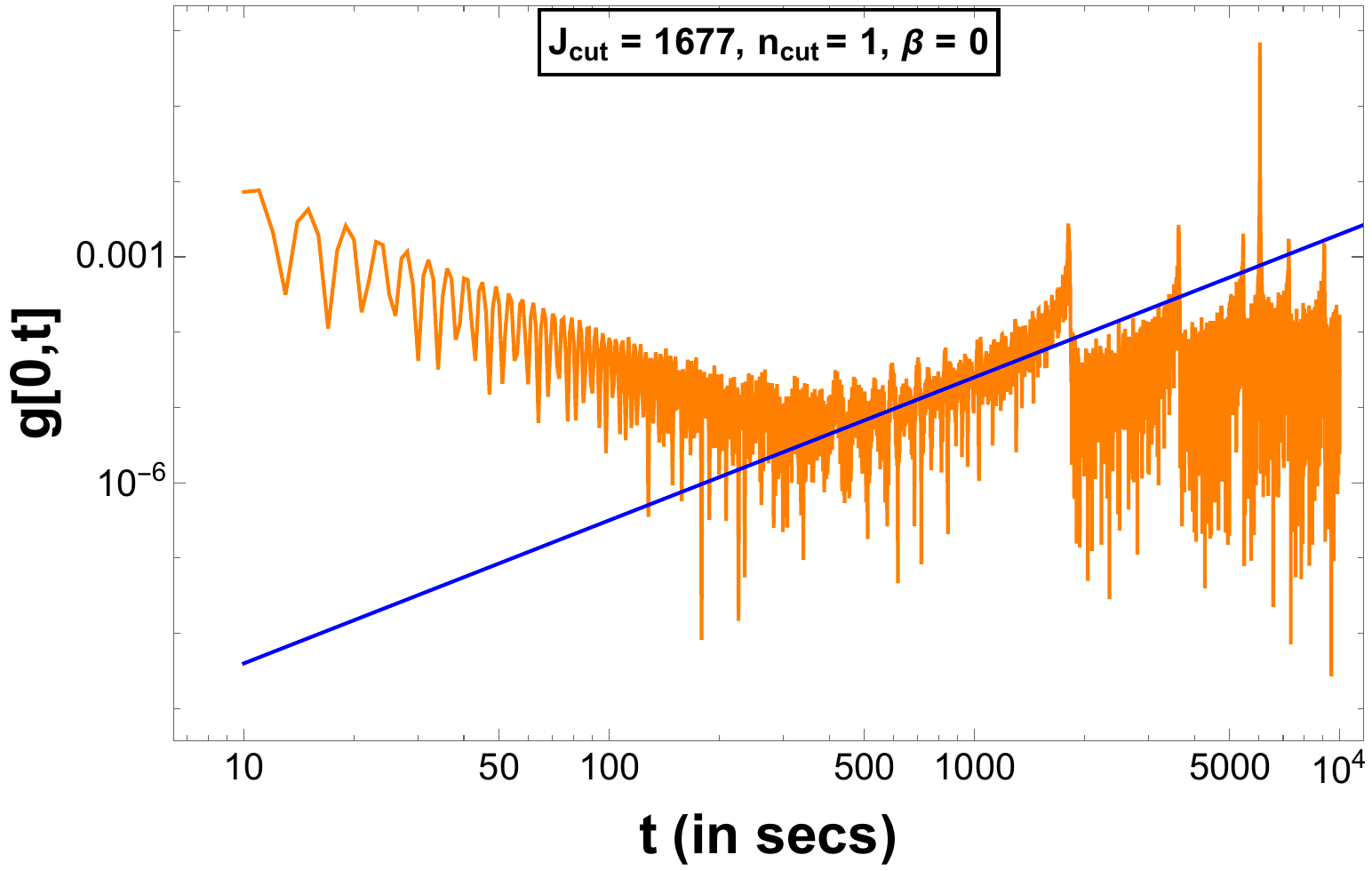}
  \caption{}
  \label{Analytical SFF - (-8)}
\end{subfigure}
\caption{Plot of exact (blue) and analytically fitted spectrum (orange) SFF for Rindler Geometry $\xi_o = -8$ with $a = 1$ and $R = 1/2$ for n = 1. Slope = 1.9 in both cases.}
\label{Exact and Analytical SFF (-8)}
\end{figure}

\begin{figure}
\centering
\begin{subfigure}{.5\textwidth}
  \centering
  \includegraphics[width=0.9\linewidth]{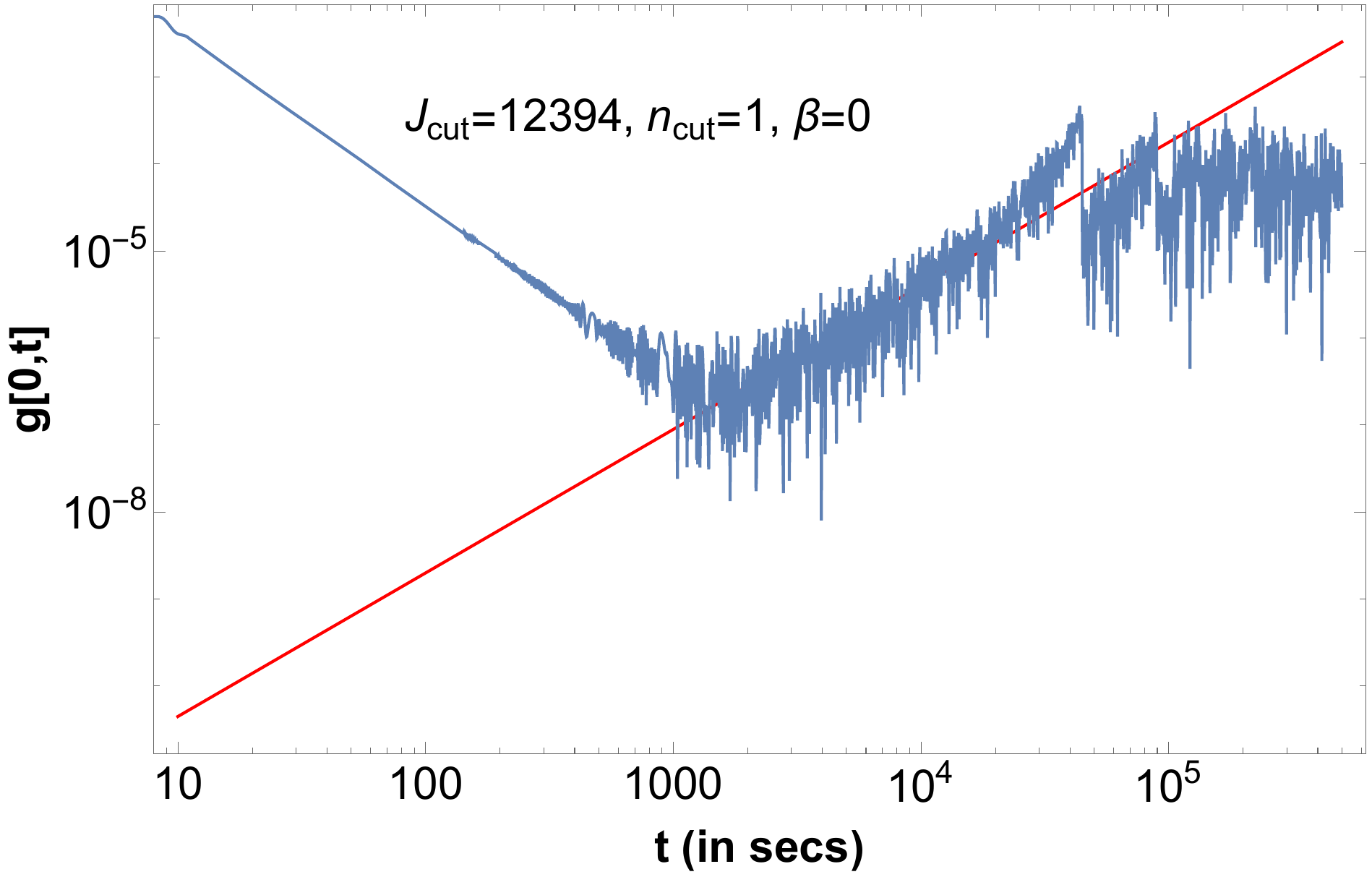}
  \caption{}
  \label{Exact SFF - (-10)}
\end{subfigure}%
\begin{subfigure}{.5\textwidth}
  \centering
  \includegraphics[width=0.9\linewidth]{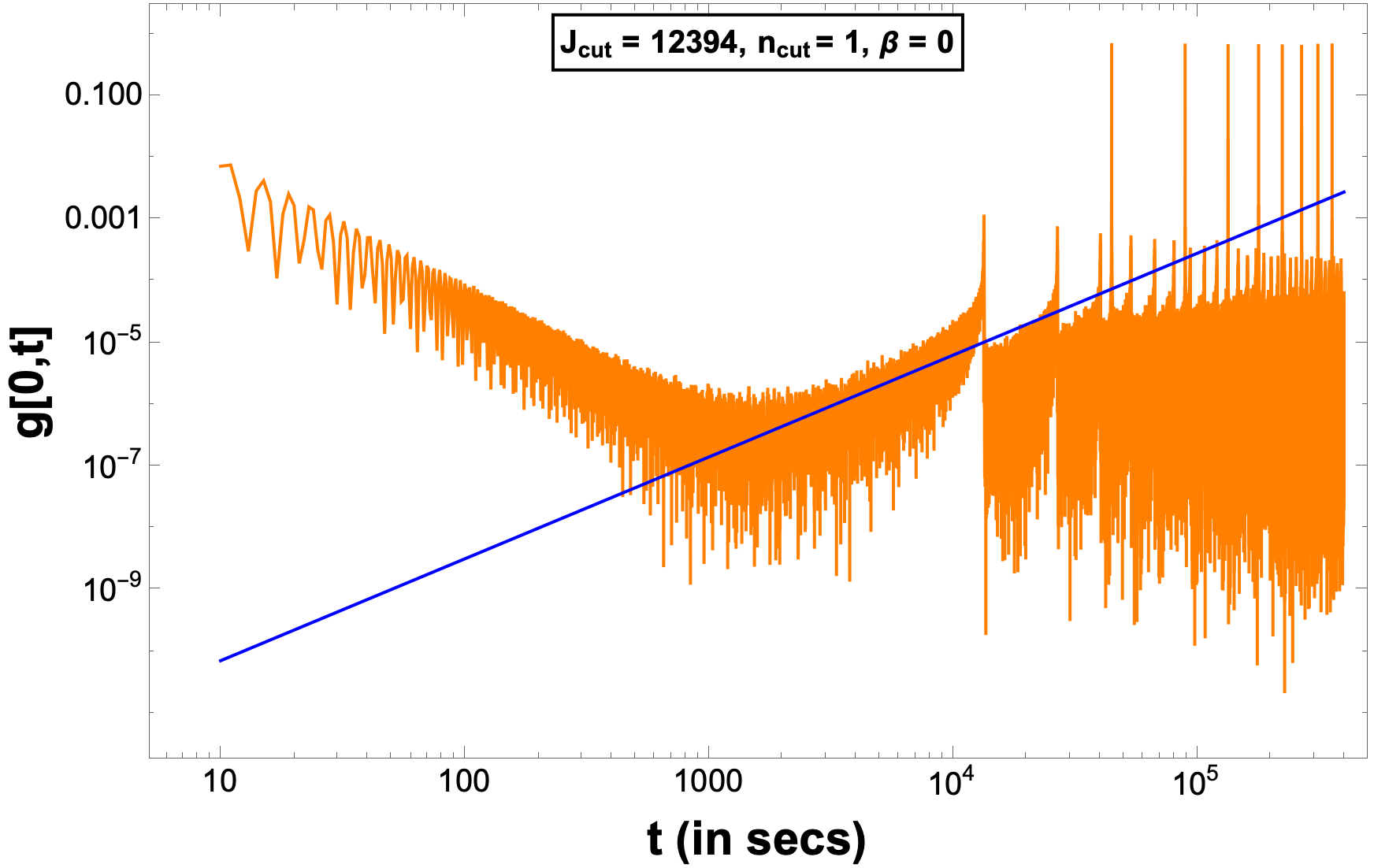}
  \caption{}
  \label{Analytical SFF - (-10)}
\end{subfigure}
\caption{Plot of exact (blue) and analytically fitted spectrum (orange) SFF for Rindler Geometry $\xi_o = -10$ with $a = 1$ and $R = 1/2$ for n = 1. Slope = 1.65 in both cases.}
\label{Exact and Analytical SFF (-10)}
\end{figure}

\begin{figure}
\centering
\begin{subfigure}{.5\textwidth}
  \centering
  \includegraphics[width=0.9\linewidth]{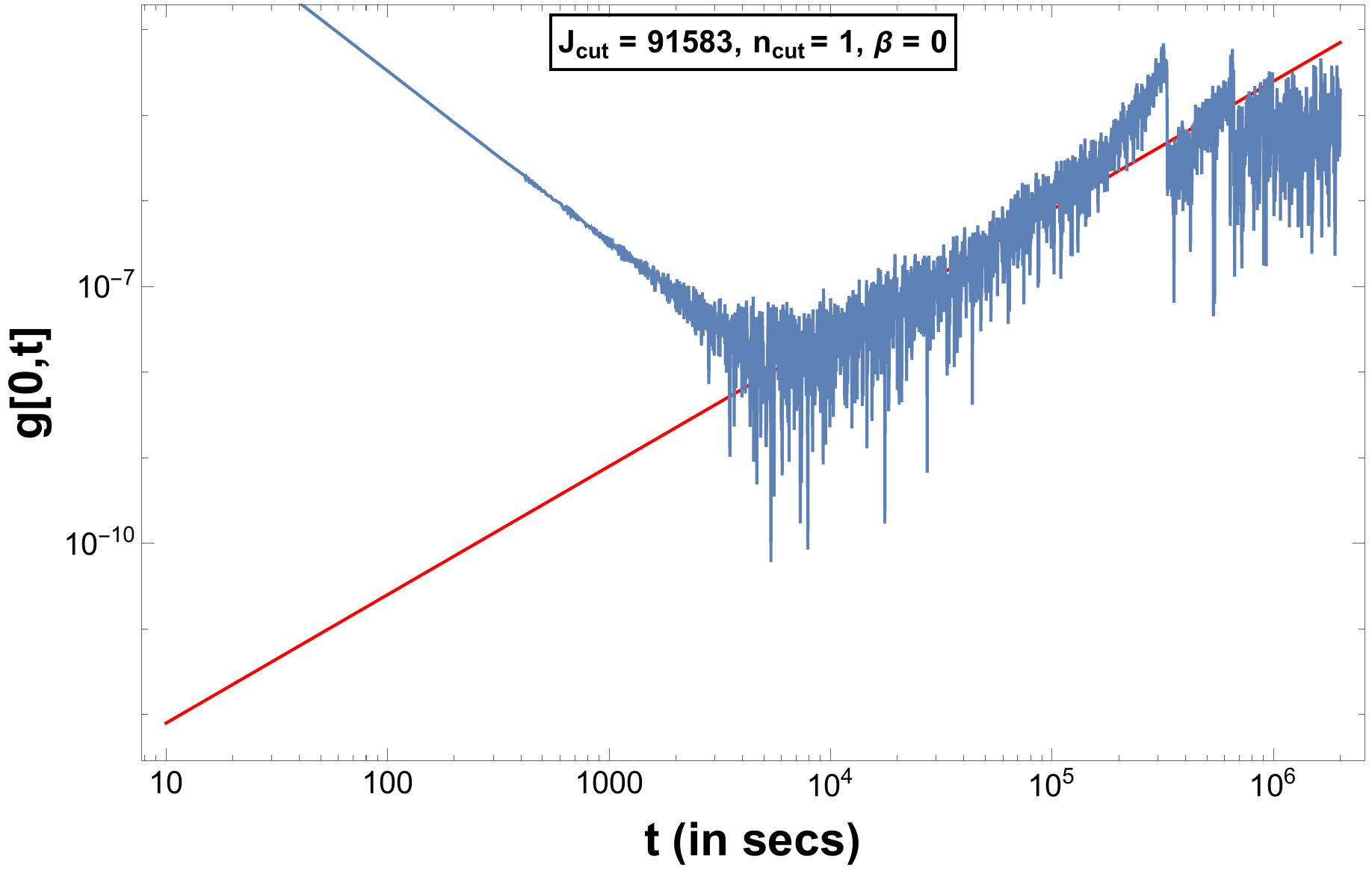}
  \caption{}
  \label{Exact SFF - (-12)}
\end{subfigure}%
\begin{subfigure}{.5\textwidth}
  \centering
  \includegraphics[width=0.9\linewidth]{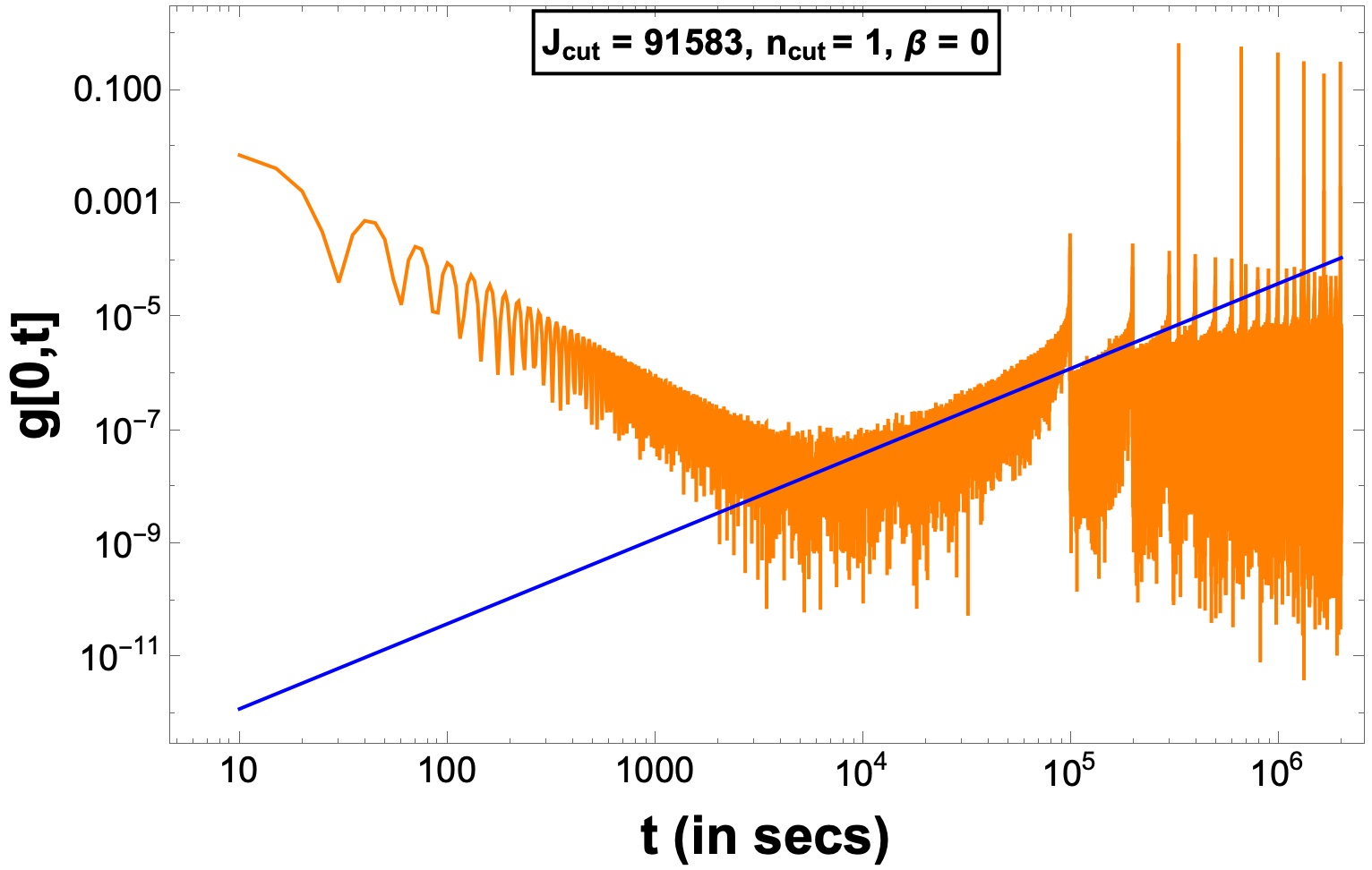}
  \caption{}
  \label{Analytical SFF - (-12)}
\end{subfigure}
\caption{Plot of exact (blue) and analytically fitted spectrum (orange) SFF for Rindler Geometry $\xi_o = -12$ with $a = 1$ and $R = 1/2$ for n = 1. Slope = 1.5 in both cases.}
\label{Exact and Analytical SFF (-12)}
\end{figure}

\begin{figure}
\centering
\begin{subfigure}{.5\textwidth}
  \centering
  \includegraphics[width=0.9\linewidth]{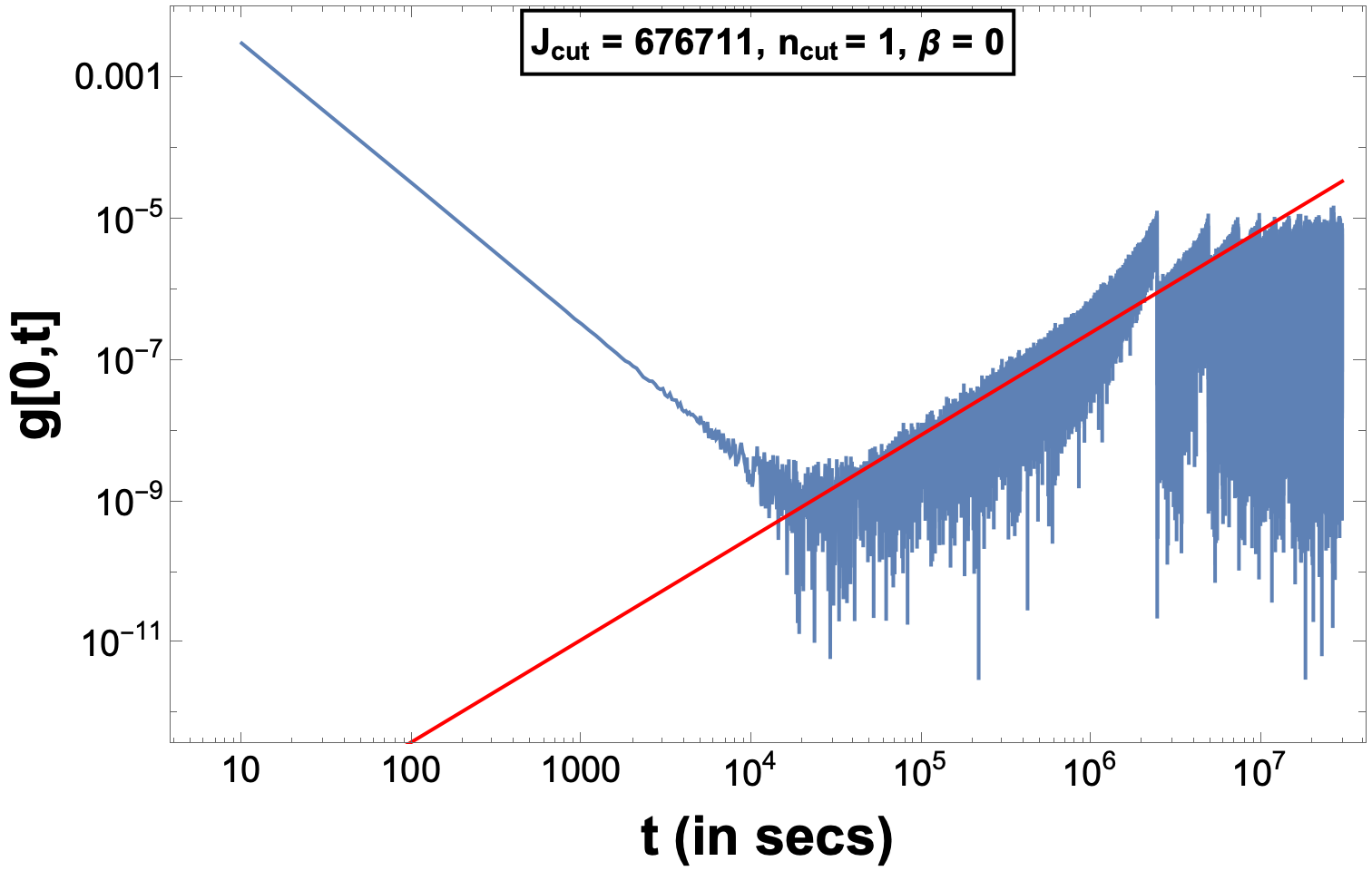}
  \caption{}
  \label{Exact SFF - (-14)}
\end{subfigure}%
\begin{subfigure}{.5\textwidth}
  \centering
  \includegraphics[width=0.9\linewidth]{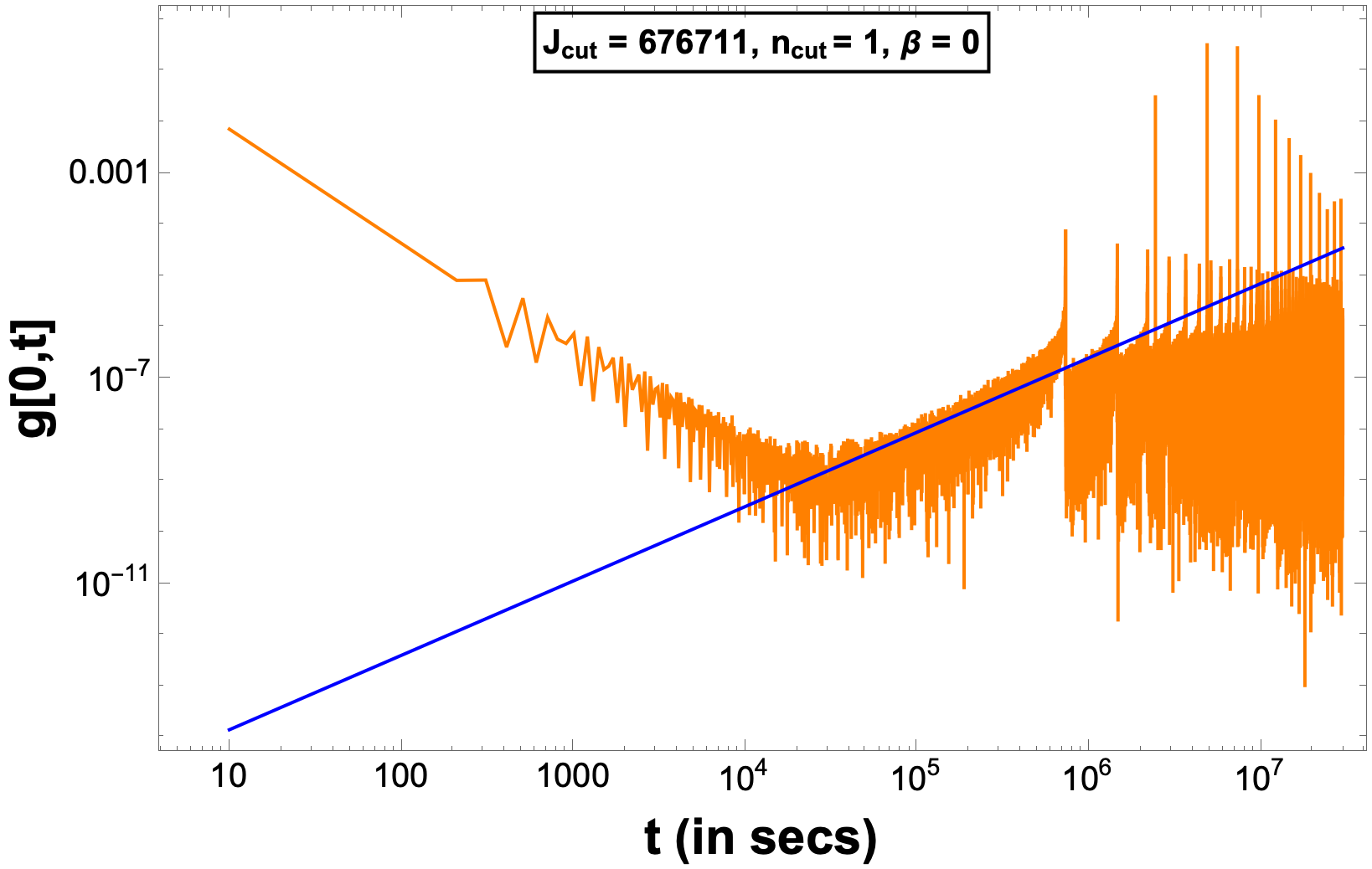}
  \caption{}
  \label{Analytical SFF - (-14)}
\end{subfigure}
\caption{Plot of exact (blue) and analytically fitted spectrum (orange) SFF for Rindler Geometry $\xi_o = -14$ with $a = 1$ and $R = 1/2$ for n = 1. Slope = 1.45 in both cases. }
\label{Exact and Analytical SFF (-14)}
\end{figure}

\begin{figure}
\centering
\begin{subfigure}{.5\textwidth}
  \centering
  \includegraphics[width=0.9\linewidth]{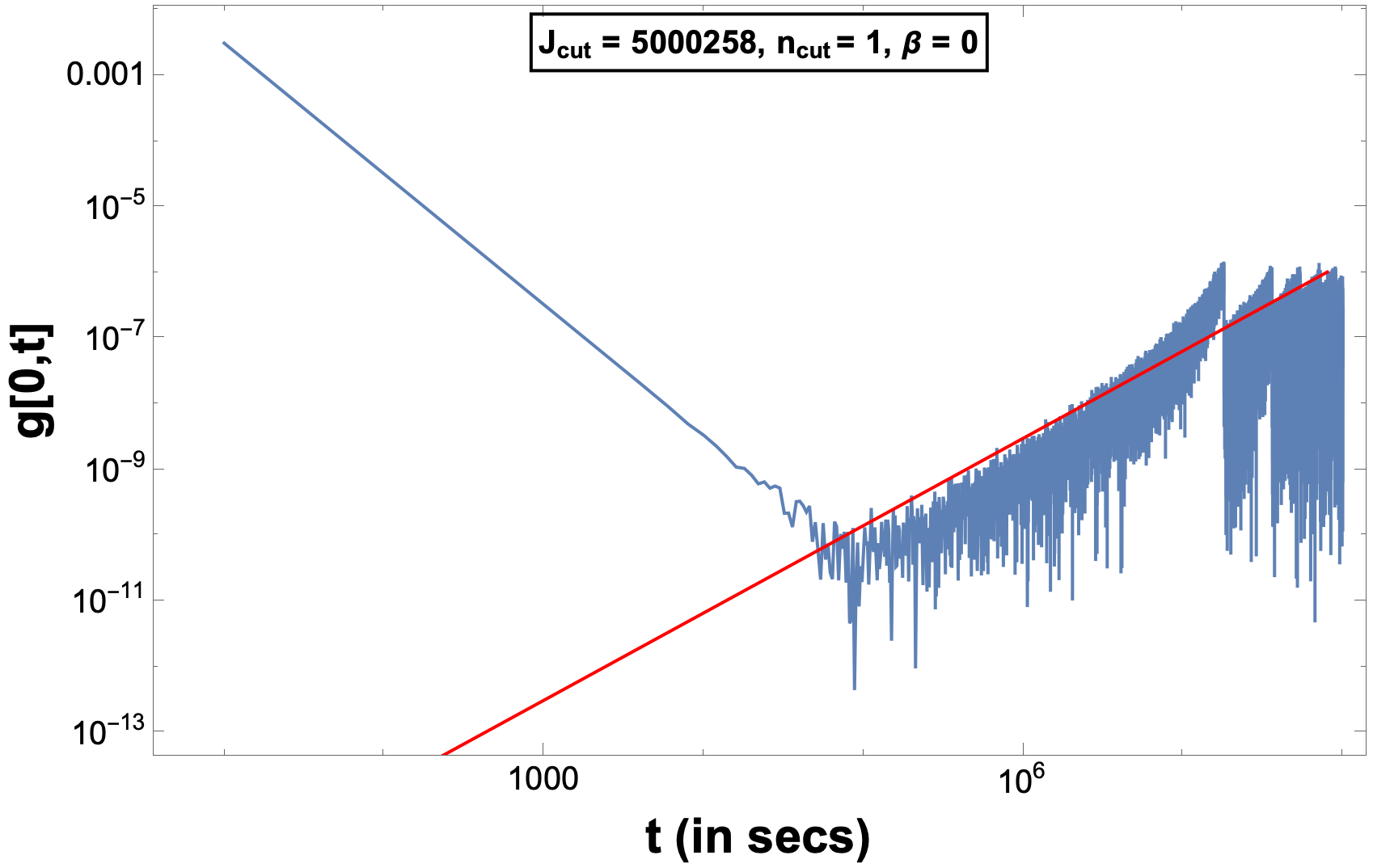}
  \caption{}
  \label{Exact SFF - (-16)}
\end{subfigure}%
\begin{subfigure}{.5\textwidth}
  \centering
  \includegraphics[width=0.9\linewidth]{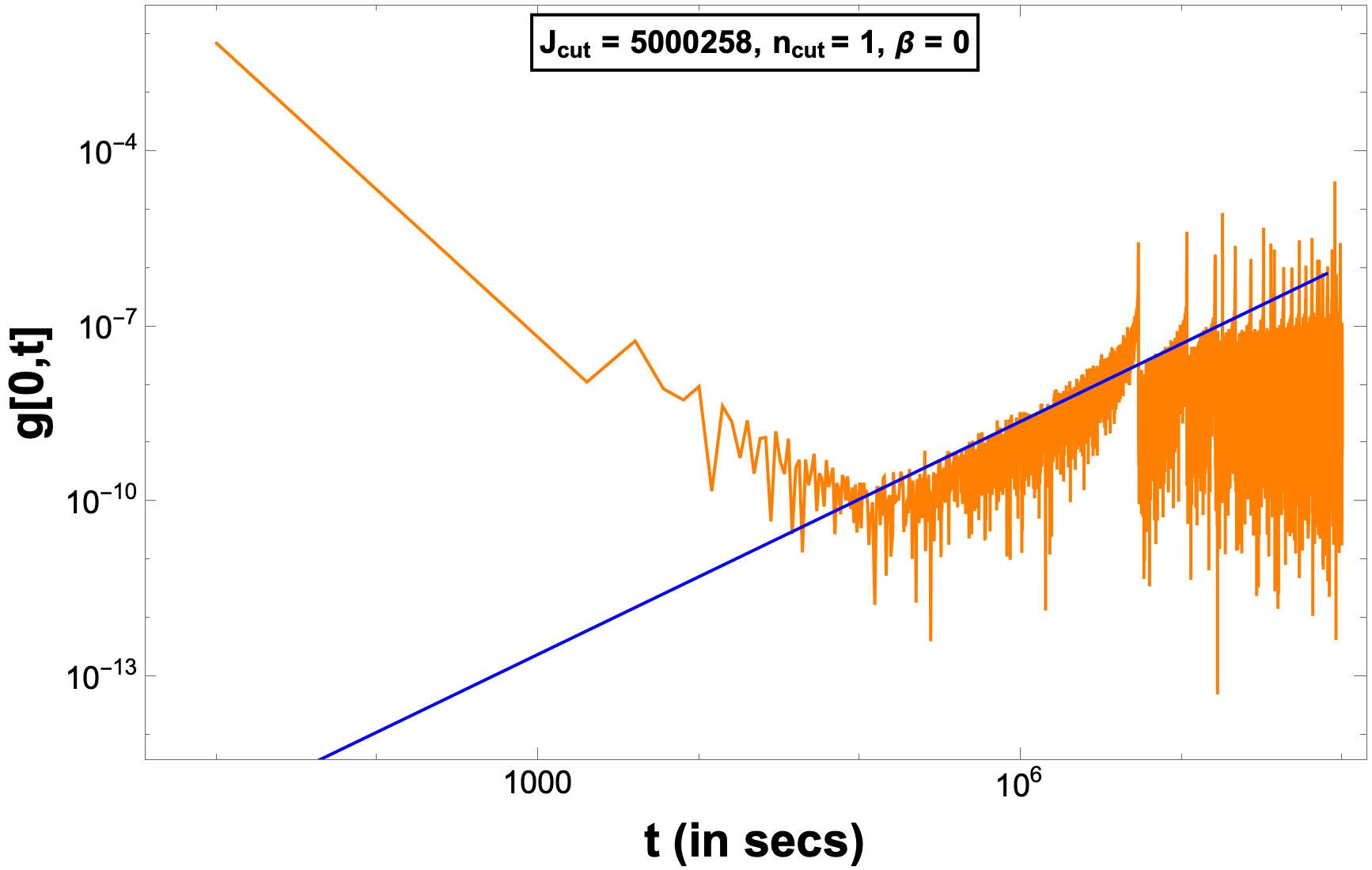}
  \caption{}
  \label{Analytical SFF - (-16)}
\end{subfigure}
\caption{Plot of exact (blue) and analytically fitted spectrum (orange) SFF for Rindler Geometry $\xi_o = -16$ with $a = 1$ and $R = 1/2$ for n = 1. Slope = 1.33 in both cases.}
\label{Exact and Analytical SFF (-16)}
\end{figure}

\begin{figure}
\centering
\begin{subfigure}{.5\textwidth}
  \centering
  \includegraphics[width=0.9\linewidth]{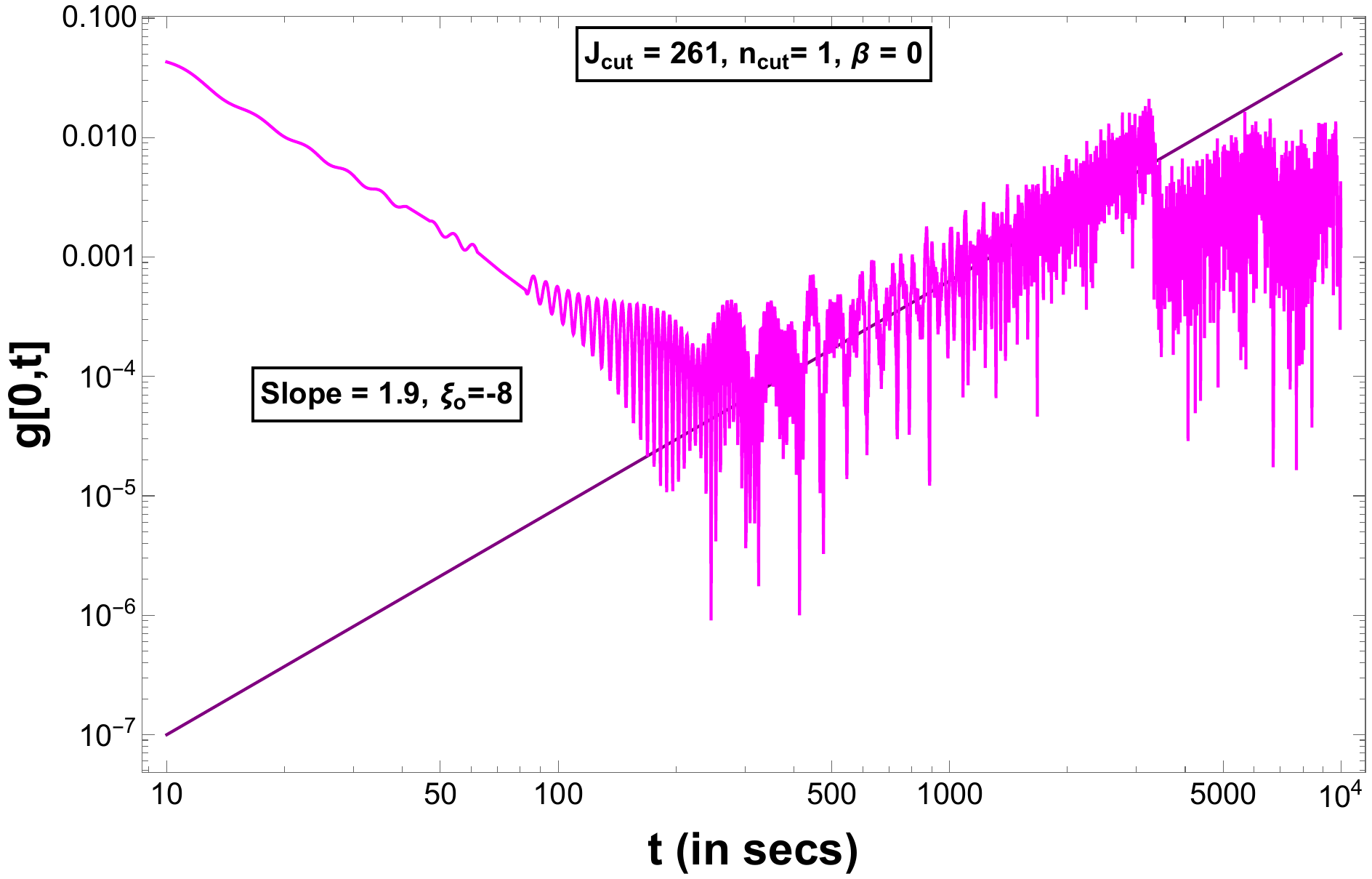}
  \caption{}
  \label{Exact Lower End SFF (-8)}
\end{subfigure}\\
\begin{subfigure}{.5\textwidth}
  \centering
  \includegraphics[width=0.9\linewidth]{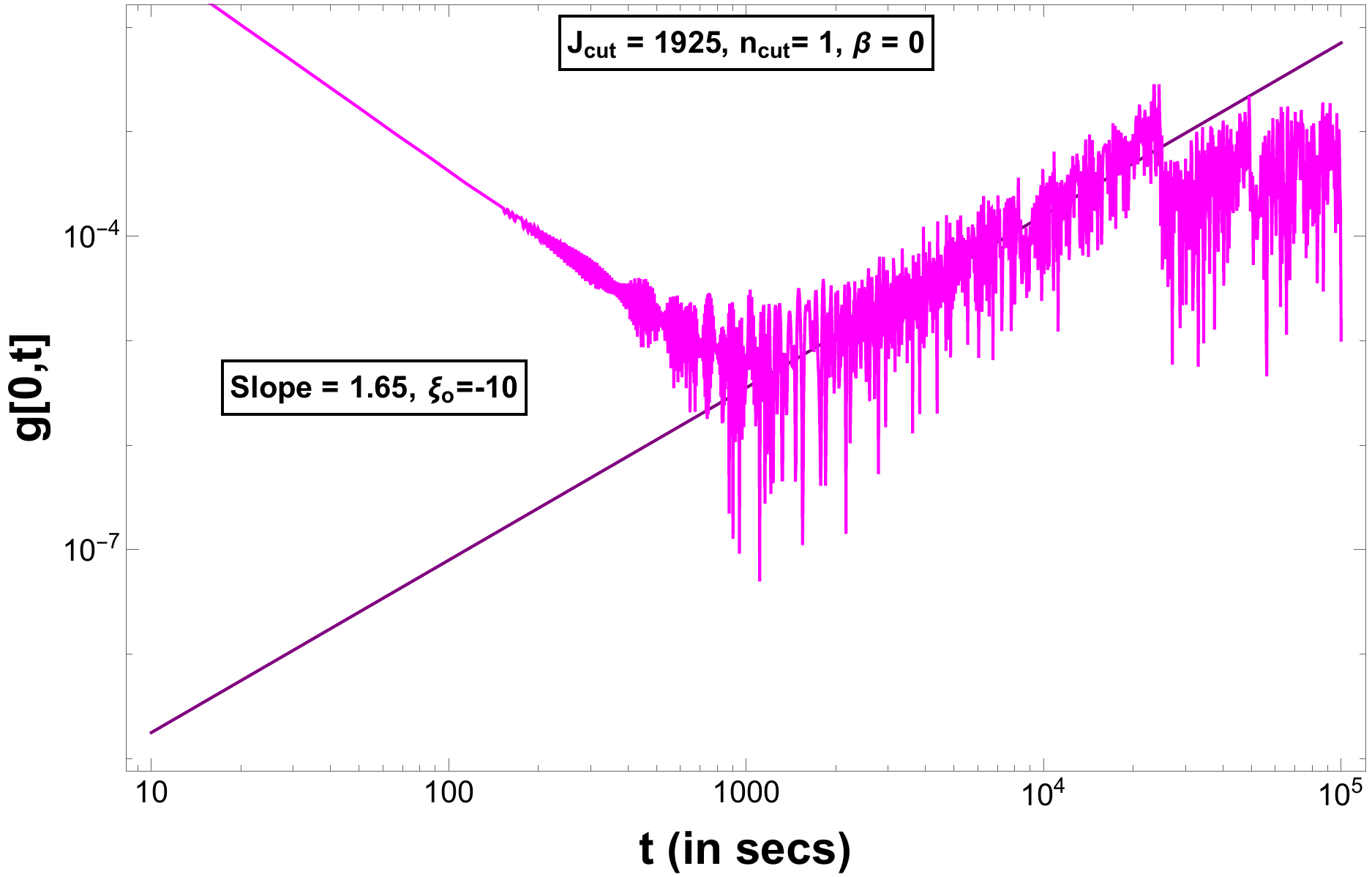}
  \caption{}
  \label{Exact Lower End SFF (-10)}
\end{subfigure}%
\begin{subfigure}{.5\textwidth}
  \centering
  \includegraphics[width=0.9\linewidth]{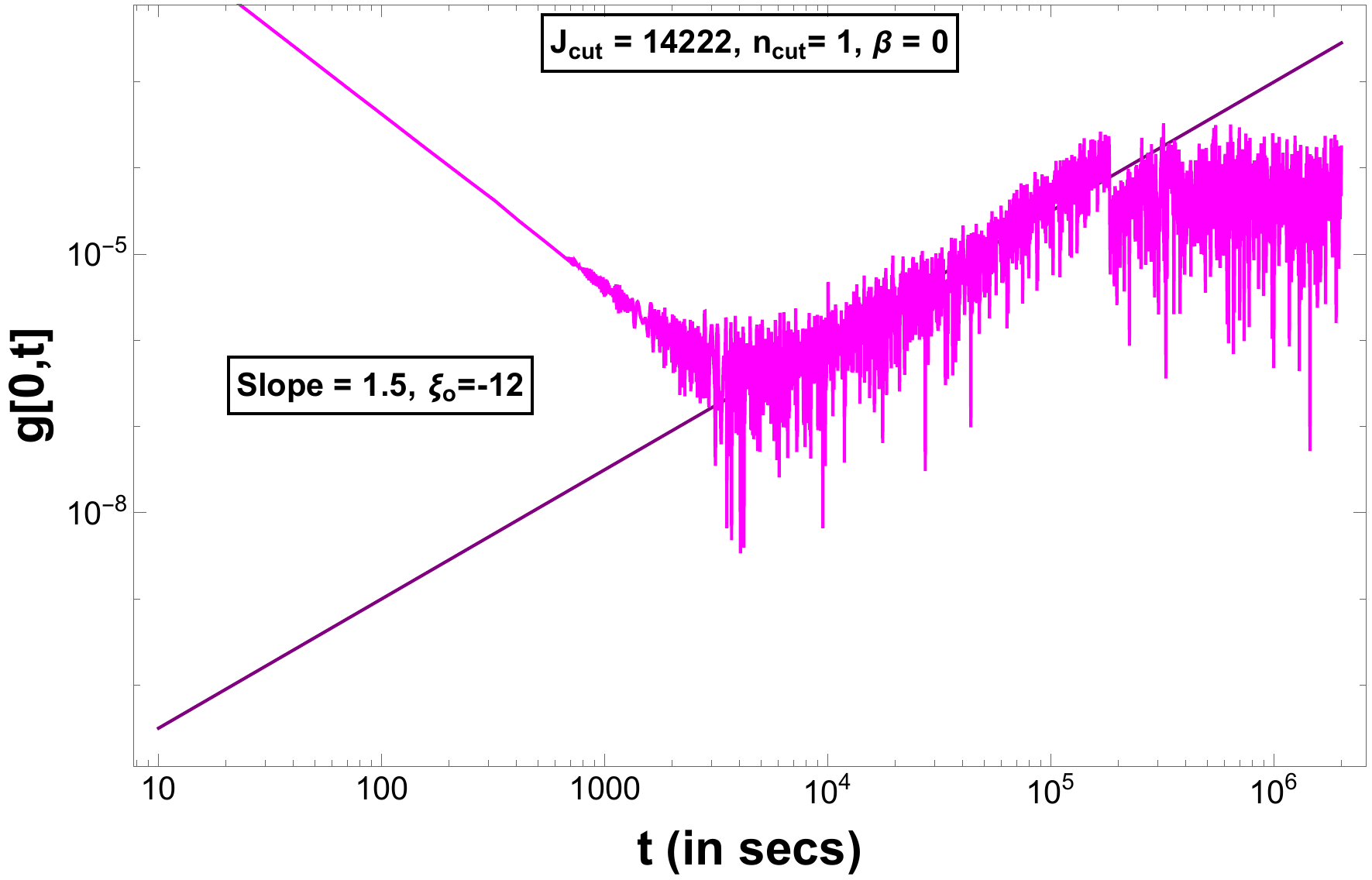}
  \caption{}
  \label{Exact Lower End SFF (-12)}
\end{subfigure}

\vspace{1cm} % Adjust the vertical spacing between the figures

\begin{subfigure}{.5\textwidth}
  \centering
  \includegraphics[width=0.88\linewidth]{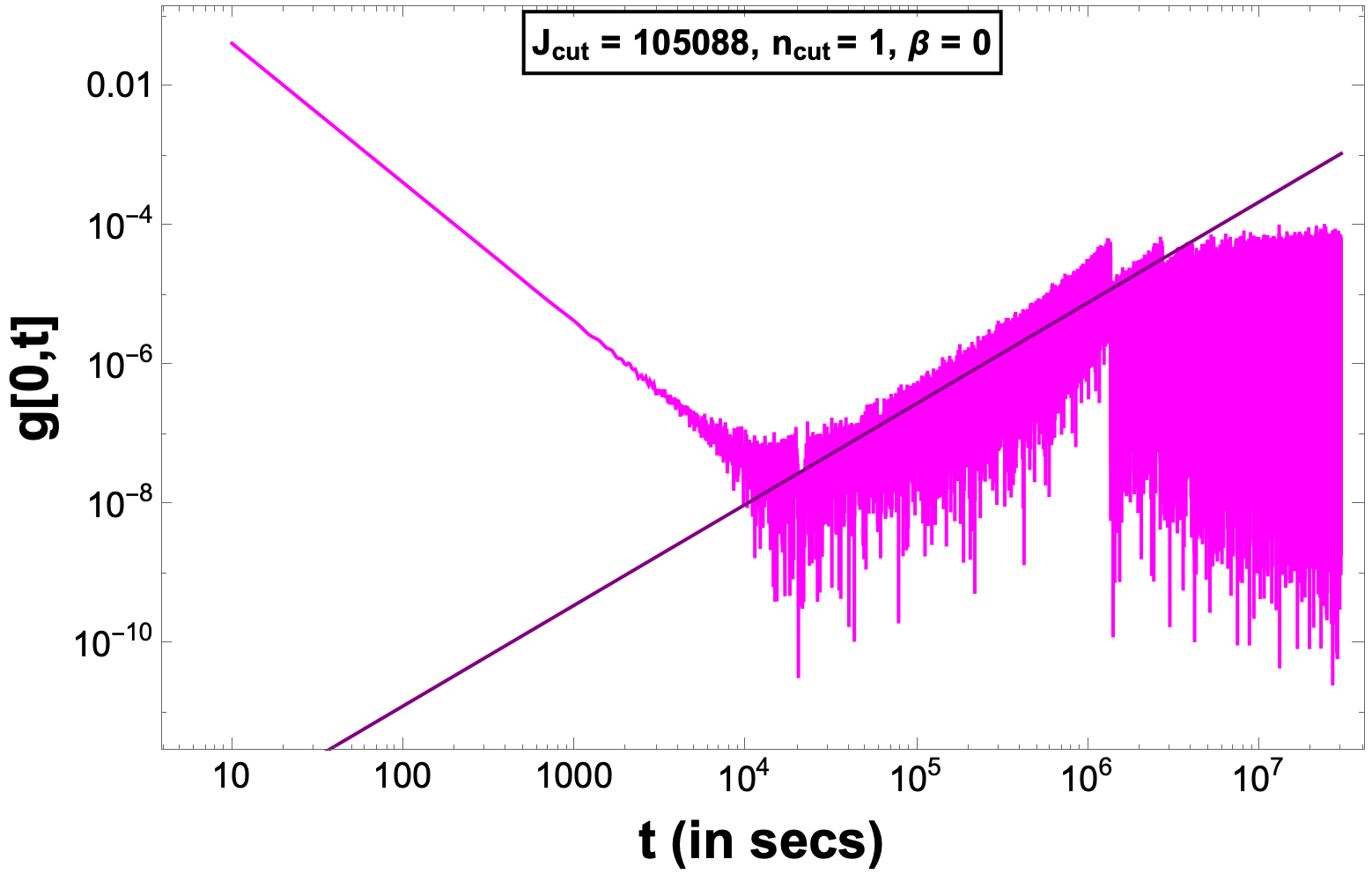}
  \caption{} % Label as Fig 7(c)
  \label{Exact Lower End SFF (-14)}
\end{subfigure}%
\begin{subfigure}{.5\textwidth}
  \centering
  \includegraphics[width=0.9\linewidth]{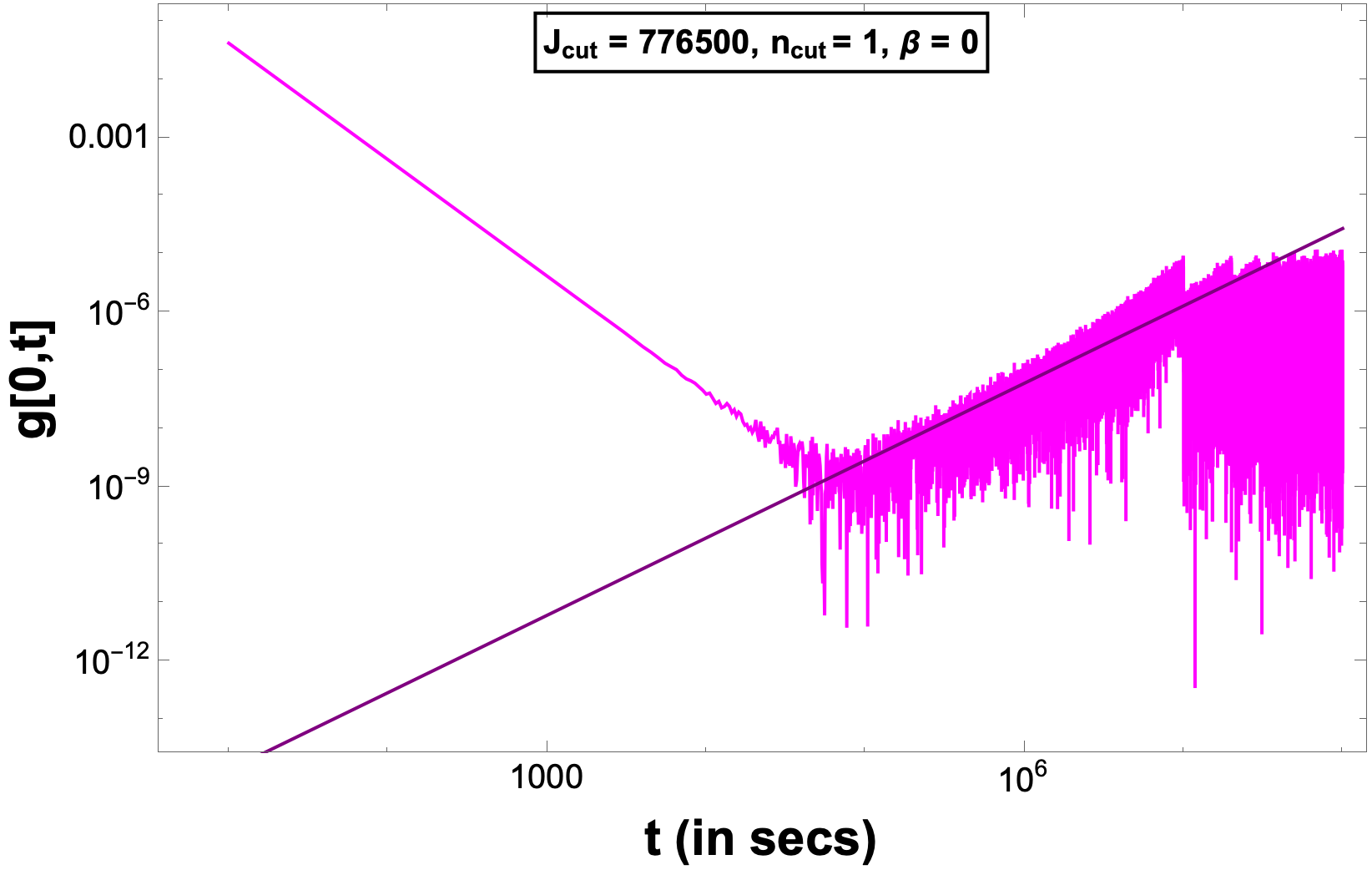}
  \caption{}
  \label{Exact Lower End SFF (-16)}
\end{subfigure}%
\caption{Plot of exact spectrum (magenta) for cutoffs $\xi_o = -8,-10,-12,-14, -16$ in increasing orders of $J_{max}$ for Rindler with $a = 1$ and $R = 1/2$ for $n = 1$ and $J$ going from $1$ to $J_{inter} = 0.155292 J_{max}$. Slopes are exactly the same as with full spectrum till $J=J_{max}$ for each of the given $J_{max}$'s.}
\label{Exact SFF with Lower Spectrum}
\end{figure}

\begin{itemize}
    \item Fig. \ref{Exact and Analytical SFF (-8)} to Fig.  \ref{Exact and Analytical SFF (-16)} shows the SFFs for the exact spectrum and the analytically fitted spectrum for $\xi_o = -8,\ -10, \ -12,\ -14 \ {\rm and} \ -16$ respectively. As seen from the graphs, the slopes remain exactly the same for both the spectra. Hence our analytically fitted spectrum is a good model of the exact spectrum, as far as slope of the ramp is concerned. The lowest we could go in slope without running into numerical difficulties is $1.33$ for $\xi_o = -16$. But S. Garg has been able to make significantly more progress in closely related examples by going to much higher values ($J_{cut} \sim10^{10-11}$) and get slopes as low as $\sim 1.07$. 
    \item Fig. \ref{Exact SFF with Lower Spectrum} is the plot of SFF with just the lower part of the exact spectrum (ie., below $J_{inter}$). We see that the slopes of the ramps remain the same as those for SFFs with the full spectrum up to $J_{max}$. This is another illustration that the ramps are controlled by the lower part of the spectrum.
    \item Another feature about these plots is that if one compares the blue (Fig. \ref{Exact and Analytical SFF (-8)} to Fig. \ref{Exact and Analytical SFF (-16)}) and magenta (Fig. \ref{Exact SFF with Lower Spectrum}) plots, we find that the dip time and the plateau time are about the same (for each of the cutoffs). Note that these are the exact spectrum plots with cut-offs up to $J_{max}$ vs those with cut-offs up to $J_{inter}$. The claims about dip and plateau times are harder to test precisely, so this observation is to be taken with a grain of salt -- nonetheless, it is worth pointing out that it is again consistent with the previous bullet point. 
\end{itemize}

\vspace{0.2cm}

\subsection{A Model Spectrum}

The above motivations have inspired us to look for a simplified spectrum\footnote{The $a$ here is not the same as in  the Rindler discussion. Its use is limited to this subsection and should not cause any confusion.}:
\begin{equation} \label{Toy Model Spectrum}
\omega_{toy}(n,J) = \frac{n}{a-\log(J)} = -\frac{n}{\log\left(j\right)}
\end{equation}
where $j=J/J_{max}$ and $J_{max} \equiv e^{a}$. This spectrum retains all the crucial features of the SFF, but keeps the parameters to the absolute minimum. Note for example, that a non-trivial extra parameter in the numerator or in the coefficient of the $\log J$ term can be absorbed into the unit of time when defining the SFF. So despite its simplicity, this spectrum is general enough when discussing the analytic low-lying spectrum.

\begin{figure}[h]
    \centering
    \includegraphics[width=0.7\linewidth]{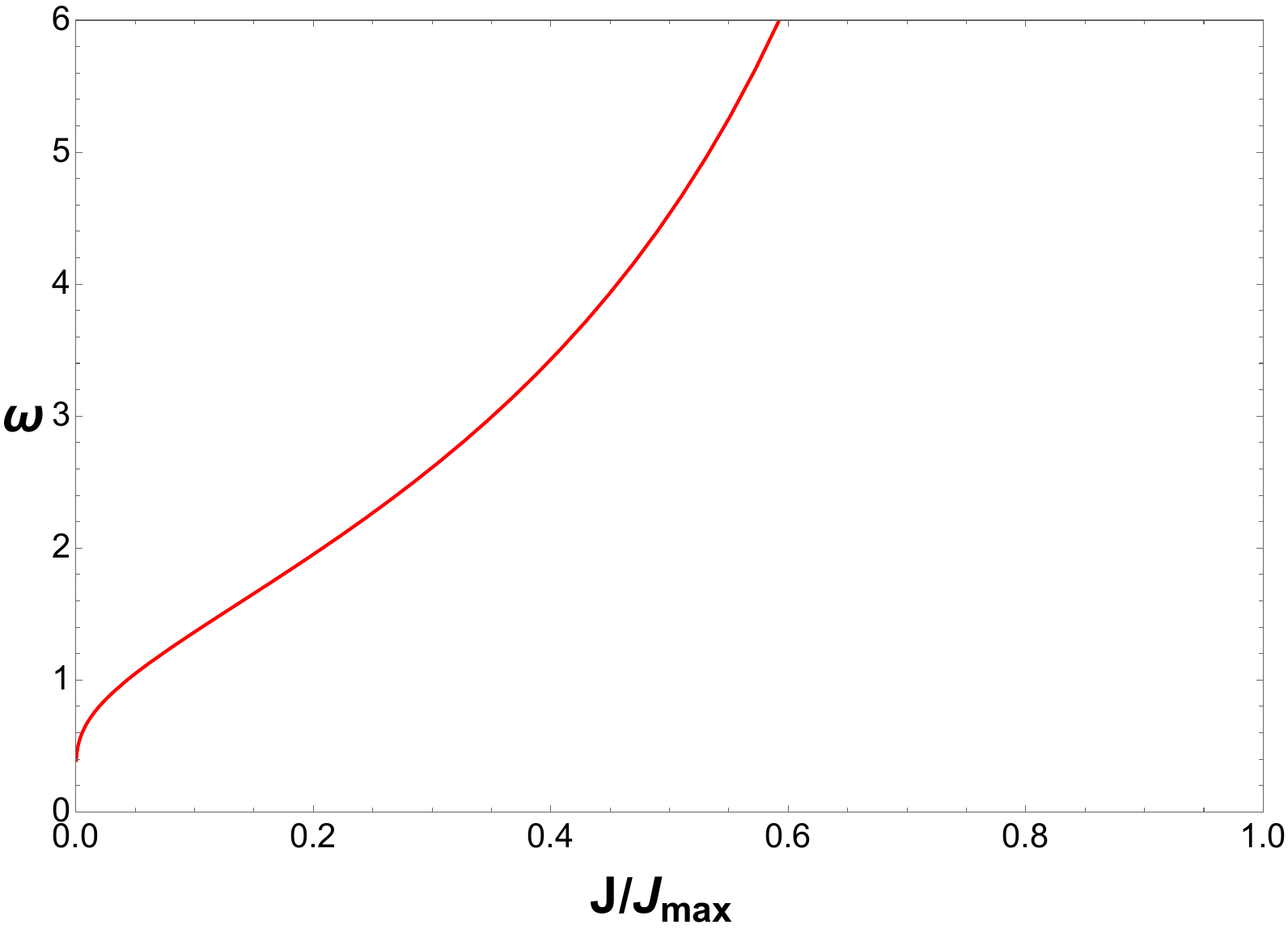}
    \caption{Toy model spectrum for arbitrary cut-off, plotted against $j=J/J_{max}$. }
    \label{figmodel}
\end{figure}

The spectrum has the shape we have seen already, see Fig \ref{figmodel}.
The question now is what is the analog of $J_{inter}$ which is a natural place to cut off the spectrum. We could work with $J_{inter}$ by considering an approximate high-lying spectrum, but instead we will explore the possibility of cutting it off at the inflection point of the above curve. This is also a natural candidate for where we can claim that the analytic low-lying spectrum is breaking down -- it is the place at which the nearest neighbor level-spacing of this approximation starts increasing. By setting the second derivative of \eqref{Toy Model Spectrum} to zero, we find $J_{crit} = e^{-2}J_{max}$ or $j_{crit}=e^{-2}=0.135335$, which is close to $j_{inter}$ we found in \eqref{ana-fit-expression}.

Fig. \ref{toy-many} below shows that the ramp structure of the SFF in the toy model follows the same phenomenology that we noted earlier. We see that the slope keeps decreasing as $J_{max}$ becomes higher and higher as earlier\footnote{Again, we remind the reader that results of S. Garg show that the slope gets much closer to $\sim 1$ as the cut-off is ramped up.}. We show a comparison plot here till $a=16$ which has slope of $1.39$. Beyond this $a$ (or $J_{max}$), the task is best done on systems that are more sophisticated than laptops.
\begin{figure}[h]
    \centering
    \includegraphics[width=1\linewidth]{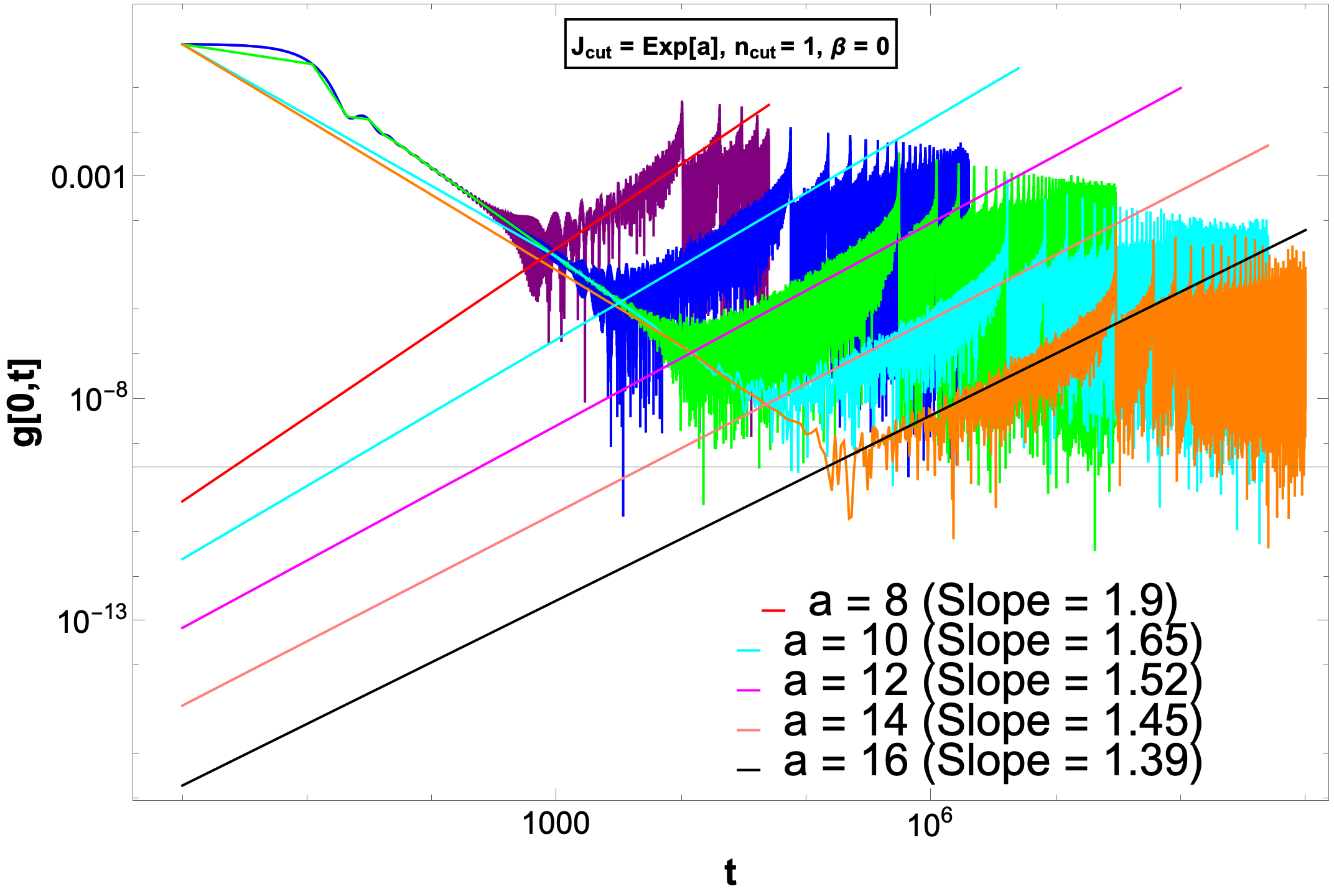}
    \caption{Plot of toy model spectrum for $a=-8, -10, -12, -14\ \ {\rm and}\ -16$, depicting gradual decrease of slope as we increase $a$, which controls the cutoff in the physical spectrum. }
    \label{toy-many}
\end{figure}

\section{BTZ Normal Modes}\label{BTZsec}

Now we turn to the BTZ black hole \cite{Adepu}. This is a harder case than Rindler, in the sense that we will not be able to solve the exact normal mode equation, directly (even numerically). But it is a good trade off between solvability and physics: unlike Rindler, we are dealing with an honest-to-God black hole and not just a near horizon geometry. Furthermore, unlike in higher dimensions, the wave equation is known in terms of hypergeometric functions and we will be able to make progress through suitable approximations. 

Even though we are not able to directly solve the BTZ exact equation, we will be able to solve it in an expansion around the horizon. The leading term in such an expansion is what we called the phase equation. The phase equation has the property that its solutions are quite a good approximation for the solutions of the exact equation, all the way to values of $J$ that are close to $J_{max}$\footnote{Recall that $J_{max}$ is defined as the value of $J$ where the analytic low-lying spectrum (which is the low-lying {\em approximate} solution of the phase equation), undergoes complete breakdown.}. Except for the difficulty with the exact equation, BTZ shares these features with Rindler. 

As we discussed, the phase equation is simply the leading term of the exact equation when expanded around the horizon, so the breakdown of its solutions is simply an indication that we need to incorporate the next order terms in the expansion of the exact equation. So in this section, we will incorporate the first corrections to the BTZ phase equation. We will see that once this is done, the qualitative behavior of the normal modes in BTZ become identical to those in Rindler all the way to $J_{max}$ and beyond\footnote{We expect that we can keep going to higher values of $J$ by taking further corrections into account. But as we alluded to earlier, the physics we are interested in is controlled by fairly low $J$'s (certainly not beyond $J_{max}$) so we will not need to.}. The main technical challenge we will overcome in this section is the solution of this corrected phase equation. 

We will work with non-rotating BTZ whose metric takes the form 
\begin{equation} \label{BTZmetric}
    ds^2 = -\frac{r^2-r^{2}_{h}}{L^2}dt^2 + \frac{L^2}{r^2-r^{2}_{h}}dr^2 + r^2 d\phi^2
\end{equation}
with $-\infty < t < \infty$, $0 < r < \infty$ and $0 \leq \phi < 2\pi$. We follow \cite{Sumit1} and focus on the massless scalar case ($m = 0$ or $\nu = 1$). After demanding normalizability at the boundary: 
\begin{equation} \label{BTZ-normalizable-soln}
    \phi_{\omega,J}(r) = C_{1} \hspace{0.1cm} r^{\frac{1}{2}-\frac{iJL}{r_h}} (r^2 - r^{2}_{h})^{-\frac{i\omega L^2}{2 r_h}}\left(-e^{-\frac{\pi JL}{r_h}} \left(\frac{r}{r_h}\right)^{\frac{2iJL}{r_h}} \frac{\gamma(J,1)}{\gamma(-J,1)} H(r) + G(r)\right)
\end{equation}
$J$ stands for angular quantum number, $\omega$ for energy, we suppress $\omega$ and $J$ indices on $C_1$, and $G(r)$, $H(r), \gamma(J,1)$ are given as
\begin{equation} \label{BTZ-sol-defs}
\begin{split}
    G(r) &= {}_{2}F_{1} \left(\frac{1}{2}\left(-\frac{i\omega L^2}{r_h}-\frac{iJL}{r_h}\right), \frac{1}{2}\left(2-\frac{i\omega L^2}{r_h}-\frac{iJL}{r_h}\right), 1-\frac{iJL}{r_h}, \frac{r^2}{r^{2}_{h}}\right)\\
    H(r) &= {}_{2}F_{1} \left(\frac{1}{2}\left(-\frac{i\omega L^2}{r_h}+\frac{iJL}{r_h}\right), \frac{1}{2}\left(2-\frac{i\omega L^2}{r_h}+\frac{iJL}{r_h}\right), 1+\frac{iJL}{r_h}, \frac{r^2}{r^{2}_{h}}\right)\\
    \gamma(J,1) &= \frac{\Gamma\left(1-\frac{iJL}{r_h}\right)}{\Gamma\left(\frac{1}{2}\left(2-\frac{i\omega L^2}{r_h}-\frac{iJL}{r_h}\right)\right) \Gamma\left(\frac{1}{2}\left(2+\frac{i\omega L^2}{r_h}-\frac{iJL}{r_h}\right)\right)}
\end{split}
\end{equation}
Tortoise coordinate $z$ is given by
\bea \label{BTZ-tortoise}
     z = \frac{L^2}{2 r_{h}} \log\left(\frac{r+r_h}{r-r_h}\right), \ \  
     r = r_{h} \coth\left(\frac{r_{h} z}{L^2}\right)
\eea
where, for $r$ lying between $r_h \leq r < \infty$, $z$ lies between 0 and $\infty$ and  $z = 0$ means AdS boundary and $z = \infty$ is the horizon.  Using the Kummer relation 
\begin{equation} \label{Kummer-BTZ1}
\begin{split}
    {}_{2}F_{1}(a,b,c,z) &= \frac{\Gamma(b-a)\Gamma(c)}{\Gamma(c-a)\Gamma(b)} (-z)^{-a} {}_{2}F_{1}\left(a,a-c+1,a-b+1,\frac{1}{z}\right)\\ 
    &+\frac{\Gamma(a-b)\Gamma(c)}{\Gamma(c-b)\Gamma(a)} (-z)^{-b} {}_{2}F_{1}\left(b-c+1,b,-a+b+1,\frac{1}{z}\right) ; z \notin (0,1)\\
\end{split}
\end{equation}
we can re-write the normalizable solution as  
\bea \label{BTZ-normalizable-single-term-form}
    \phi_{\omega,J}(r)= \Tilde{C} \left(\frac{r^2}{r_h^2}-1\right)^{\frac{-i\omega L^2}{2 r_{h}}} \left(\frac{r}{r_h}\right)^{-\frac{3}{2}+\frac{i\omega L^2}{r_h}} {}_{2}F_{1} \left(1+\frac{iL(J - \omega L)}{2 r_h} ,1-\frac{iL(J + \omega L)}{2 r_h}, 2, \frac{r^{2}_{h}}{r^2}\right) \nonumber \\
\eea
where, $\Tilde{C}$ is some $r$-independent constant. Imposing Dirichlet boundary condition at the stretched horizon $\phi_{hor} (z_o) = 0$ reads in tortoise coordinates as
\begin{equation} \label{BTZ-exact-tortoise}
    e^{i \omega \log (\cosh(z_o))} \tanh^{\frac{3}{2}}(z_o) \hspace{0.2cm} {}_{2}F_{1} \left(1+\frac{iL}{2 r_h} (J - \omega L),1-\frac{iL}{2 r_h} (J + \omega L), 2, \tanh^{2} (z_o)\right) = 0
\end{equation}
We call this the {\em BTZ Exact Equation}. Series expansion around $r = r_{h}$ in \eqref{BTZ-exact-tortoise}, and imposing Dirichlet boundary condition gives
\begin{equation} \label{BTZ-alpha-beta}
\begin{split}
    \cos(\alpha-\beta)+\cos(\theta) &= 0 \\
    \sin(\alpha-\beta)+\sin(\theta) &=0\\
\end{split}
\end{equation}
and $\alpha$, $\beta$ and $\theta$ are defined as, 
\begin{equation} \label{BTZ-alpha-beta-def}
\begin{split}
    \alpha &= \pi - \frac{\omega L^2}{2 r_h} \log(2) - {\rm Arg} \left[\Gamma\left(\frac{iJL}{r_{h}}\right)\right] + {\rm Arg} \left[\Gamma\left(1+\frac{i\omega L^2}{r_h}\right)\right] + {\rm Arg} \left[\Gamma\left(\frac{iL}{2r_{h}}(J-\omega L)\right)\right]\\ 
    &+ {\rm Arg} \left[\Gamma\left(1+\frac{iL}{2r_{h}}(J-\omega L)\right)\right] \\
    \beta &= \frac{\omega L^2}{2 r_h} \log(2) - {\rm Arg} \left[\Gamma\left(\frac{iJL}{r_{h}}\right)\right] - {\rm Arg} \left[\Gamma\left(1+\frac{i\omega L^2}{r_h}\right)\right] + {\rm Arg} \left[\Gamma\left(\frac{iL}{2r_{h}}(J+\omega L)\right)\right]\\
    &+ {\rm Arg} \left[\Gamma\left(1+\frac{iL}{2r_{h}}(J+\omega L)\right)\right] \\
    \theta &= {\rm Arg}\left[\left(\frac{r_o}{r_h}-1\right)^{\frac{i\omega L^2}{r_h}}\right]\\
\end{split}
\end{equation}
The two equations in \eqref{BTZ-alpha-beta} can be combined into the {\em BTZ phase equation} form 
\begin{align} \label{BTZ-phase-ew-1}
&\cos\left({\rm Arg} \left[\Gamma\left(\frac{i\omega L^2}{r_h}\right)\right]+{\rm Arg} \left[\Gamma\left(-\frac{iL}{2r_h} (J+\omega L)\right)\right] \right. \nonumber \\
&\qquad\left. + {\rm Arg} \left[\Gamma\left(\frac{iL}{2r_h} (J-\omega L)\right)\right] -\frac{\omega L^2}{2r_h} \log(2) - \frac{\omega L^2}{2 r_h} \log\left(\frac{r_o}{r_h}-1\right)\right) = 0.
\end{align}
Near the horizon, $z_o \approx \frac{L^2}{2 r_h} \log \left(\frac{2}{\frac{r_o}{r_h}-1}\right)$. In terms of the these coordinates, the BTZ phase equation can also be written as
\begin{align} \label{BTZ-phase-eq-2}
&\cos\left({\rm Arg} \left[\Gamma\left(\frac{i\omega L^2}{r_h}\right)\right]+{\rm Arg} \left[\Gamma\left(-\frac{iL}{2r_h} (J+\omega L)\right)\right] \right. \nonumber \\
&\qquad\left. + {\rm Arg} \left[\Gamma\left(\frac{iL}{2r_h} (J-\omega L)\right)\right] -\frac{\omega L^2}{r_h} \log(2) + \omega z_o \right) = 0.
\end{align}

\subsection{BTZ Analytic Low-Lying Spectrum}

As in Rindler, we can use the fact that for low values of $\omega$, Gamma functions can be approximated\footnote{The numerical factors in this case or in Rindler case are mathematical constants that we expect are known to the mathematicians. But we have not been able to find their values quoted anywhere, so we have determined them numerically. These numerical fits match with the exact results to $7^{th}$ or $8^{th}$ decimal places (we have checked this for $n=1,2,3$). We quote these values only to three decimals here, but they are trivial to determine more precisely.}:
\bea \label{BTZ-Gamma-approx}
   & {\rm Arg}\left[\Gamma\left(\frac{i \omega L^2}{r_h}\right)\right] = -0.575 \frac{\omega L^2}{r_h} -\frac{\pi}{2}\\
  &  {\rm Arg}\left[\Gamma\left(\frac{i L}{2 r_h}(J-\omega L)\right)\right] + {\rm Arg}\left[\Gamma\left(-\frac{i L}{2 r_h}(J+\omega L)\right)\right] = 0.636 - 0.989 \log\left(\frac{J L}{r_h}\right)
\eea
Then, using the fact that $\cos (x) = 0$ when $x = (2n-1)\pi/2$, we get an analytical expression valid for low $\omega$:
\begin{equation} \label{BTZ-analytic-low-lying-spectrum}
    \frac{\omega  L^2}{r_h} = \frac{n\pi}{\frac{r_h}{L^2}\tilde{z}_o - 0.989 \log\left(\frac{JL}{r_h}\right)}
\end{equation}
where, $n \in \mathbb{Z}^+$ and $\frac{r_h}{L^2}\tilde{z}_o = \frac{r_h}{L^2}z_o - 0.632$.  This is what we will call the {\em BTZ Approximate Phase Equation} or the {\em analytical low-lying spectrum for BTZ}. Like in Rindler, this solution also diverges at a value of $J$ that we call $J^{BTZ}_{max}$. It is given by 
\begin{equation} \label{Jmax-BTZ}
   J^{BTZ}_{max} = \frac{r_h}{L} e^{-0.639} e^{\frac{r_h z_o}{L^2}}= \frac{1.055 r_h L^{0.01}}{l_{p}^{1.01}}
\end{equation}
where in the last step we have used the geodesic distance to the stretched horizon as the Planck length:
\begin{equation} \label{Planck-BTZ-stretched-horizon}
    l_{p} = L \int_{r_h}^{r_h+x} dr \frac{1}{\sqrt{r^2 - r^{2}_{h}}} = 2 L\ e^{-\frac{r_h z_o}{L^2}}.
\end{equation}
An expression valid for very low $J$'s can also be written from \eqref{BTZ-analytic-low-lying-spectrum} as
\begin{equation} \label{btz-ultra-approx}
    \omega = \frac{n \pi}{z_o}+\frac{n \pi L^2}{r_h z^{2}_o} 0.989 \log\left(\frac{JL}{r_h}\right)
\end{equation}
This exhibits a quasi-degeneracy of the spectrum because of the logarithmic dependence on $J$.

Solving the BTZ exact equation is difficult since the function value grows very fast as we increase $J$. Already at $J = 1000$, it is around $10^{300}$ which is challenging for software like Mathematica. Fortunately, we can expand the exact equation in a suitable coordinate that vanishes at the horizon (like $1/r$ at the boundary of flat space) and set that coordinate to the stretched horizon value, and solve the equation order by order in the expansion. The BTZ phase equation that we looked at is the leading order result upon doing the above exercise, and going to the next order gives us its perturbative corrections\footnote{A hypergeometric function ${}_{2}F_{1} (a,b,c,w)$ is usually chosen to have a branch cut starting at $w=1$. Since in our BTZ exact equation this variable is controlled by hyperbolic tangent of tortoise coordinate $z_o$, it becomes indistinguishable from 1 even for $z_o \sim 10$. So, choosing higher cutoff can be problematic for numerical computations since the function will lie along the branch cut. This problem can be partially tackled by giving a small imaginary part to the tortoise coordinate. But this will still not solve the unbounded growth of the function after $J=1000$, so we will  resort to a perturbative analysis. Thankfully, the phase equation (and its perturbatively corrected version) yield very good approximations to the exact equation.}. The result of this exercise will be that the structure of the normal modes in the BTZ case is parallel to that in Rindler. By working out the Kerr case in an Appendix, we will later show that the structure has indeed some universal features. 

\subsection{Perturbative Analysis of the BTZ Exact Equation}

\begin{figure}
\centering
\begin{subfigure}{\textwidth}
  \centering
  \includegraphics[width=0.8\linewidth]{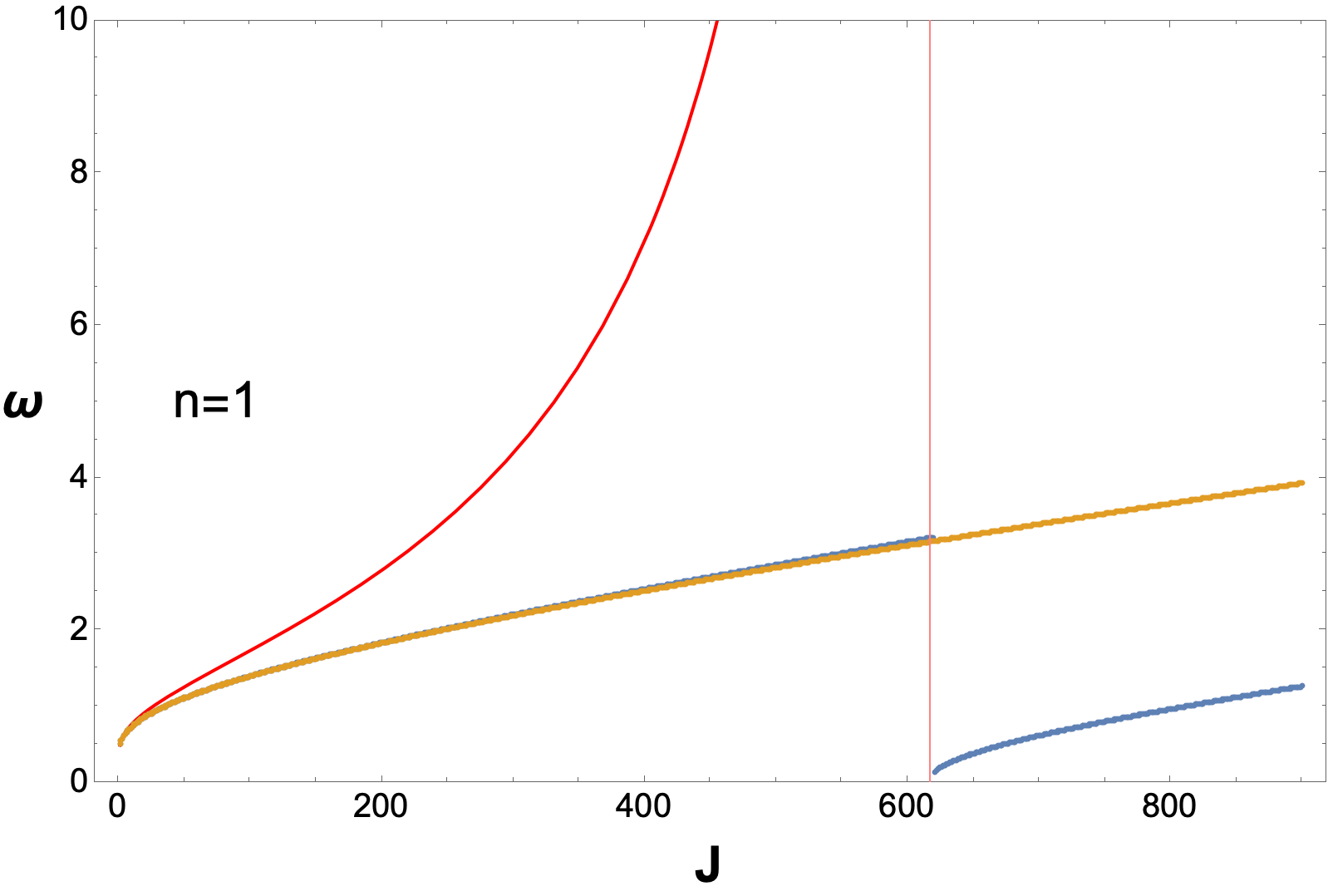}
  \caption{} % Label as Fig 4(a)
  \label{fig4a}
\end{subfigure}

\vspace{1cm} % Adjust the vertical spacing between the figures

\begin{subfigure}{\textwidth}
  \centering
  \includegraphics[width=0.8\linewidth]{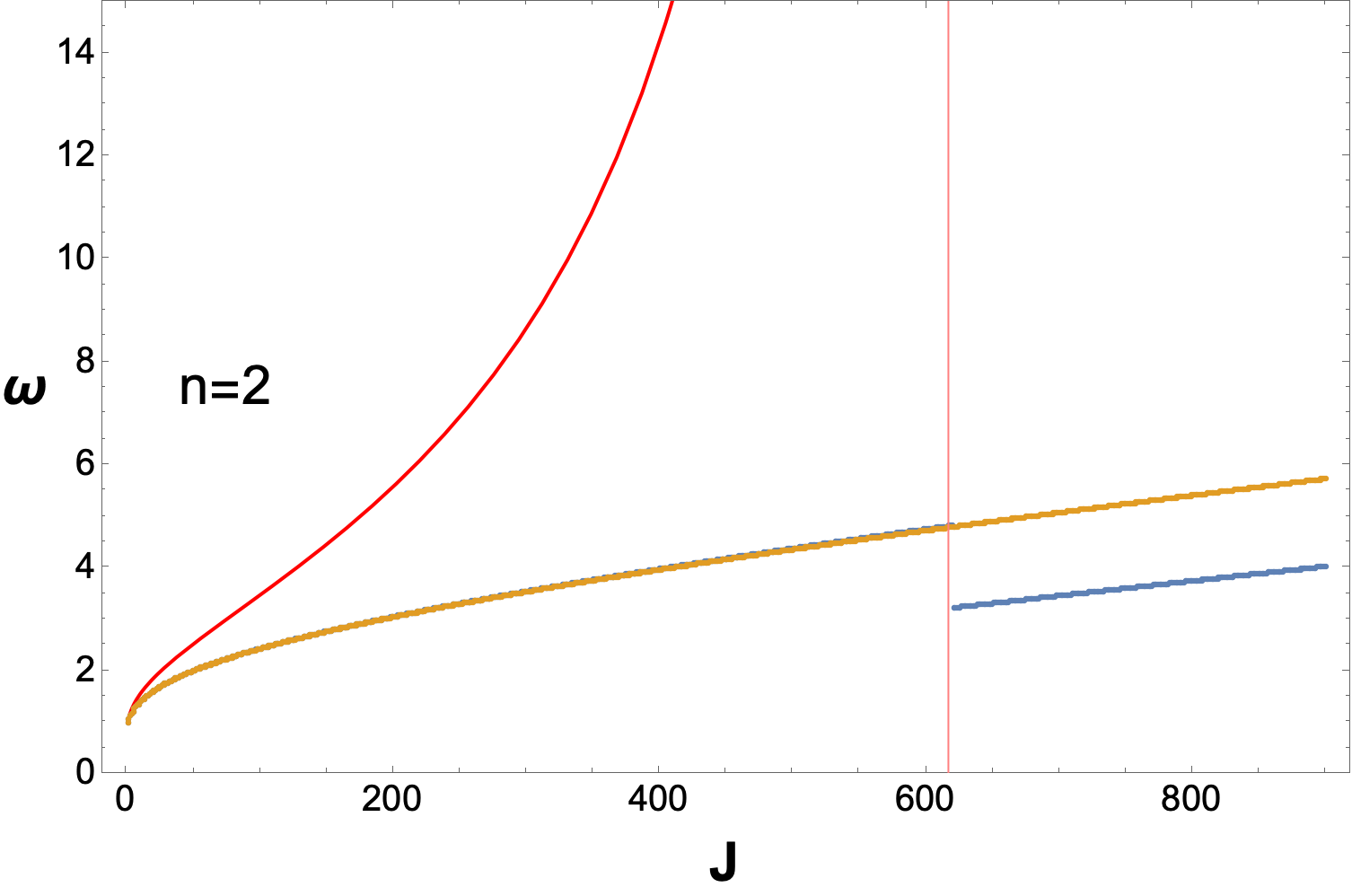}
  \caption{} % Label as Fig 4(b)
  \label{fig4b}
\end{subfigure}
\caption{BTZ Modes for $z_o = +7$ with $r_h = 1$ and $L = 1$: yellow
stands for {\em BTZ perturbed phase equation}  \eqref{BTZ-pert-phase}, blue stands for {\em BTZ phase equation} \eqref{BTZ-phase-ew-1}, red stands for the  {\em analytical low-lying spectrum for BTZ} \eqref{BTZ-analytic-low-lying-spectrum} and the pink line denotes the breakdown point of the low-lying analytic spectrum, which is $J_{max}= 626$ obtained from \eqref{Jmax-BTZ}. The phase equation breaks down at $J = 620$, which is close to this..} 
\label{fig13}
\end{figure}

\begin{figure}
\centering
\begin{subfigure}{.5\textwidth}
  \centering
  \includegraphics[width=0.9\linewidth]{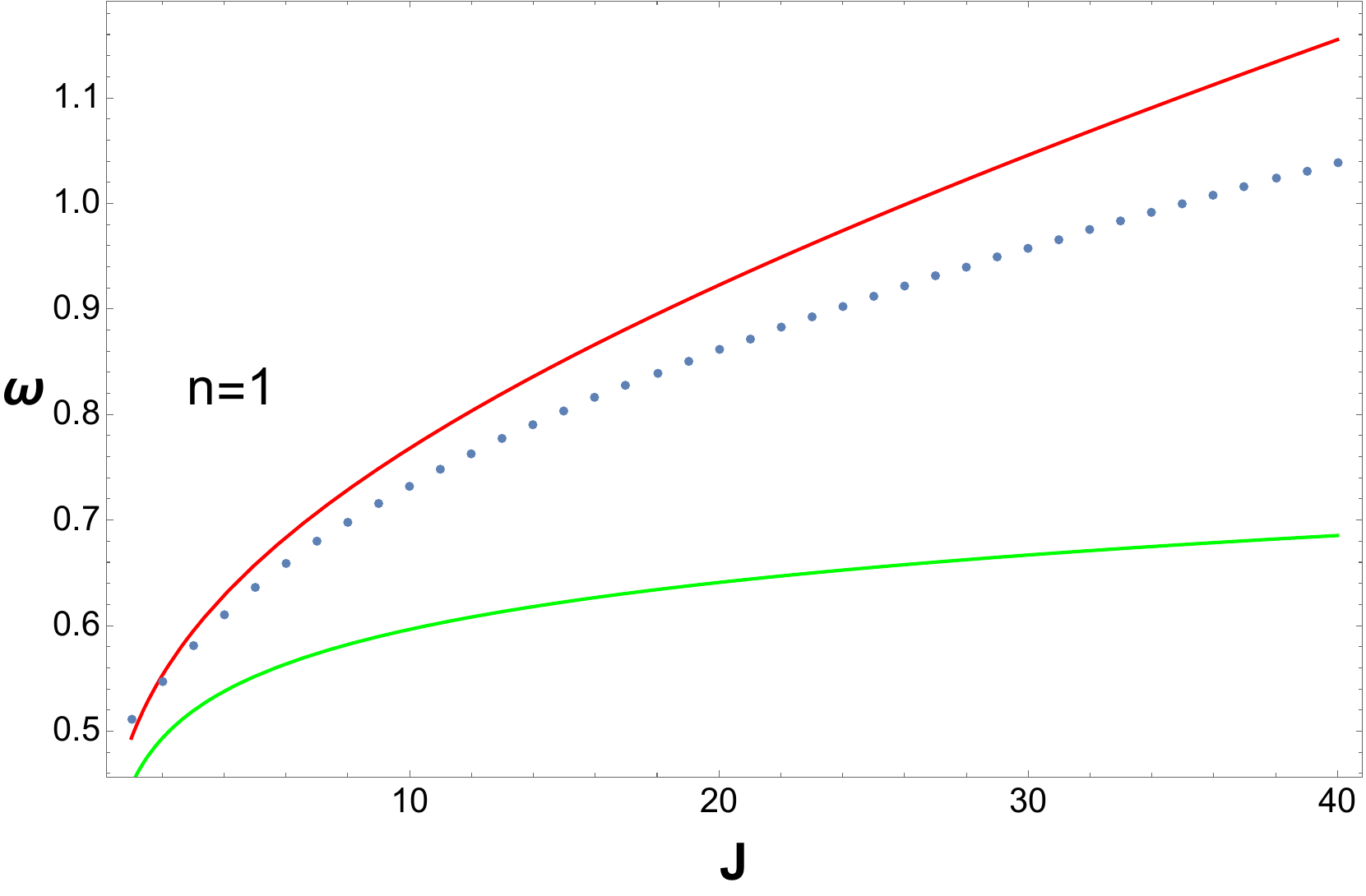}
  \caption{}
  \label{fig18(a)}
\end{subfigure}%
\begin{subfigure}{.5\textwidth}
  \centering
  \includegraphics[width=0.9\linewidth]{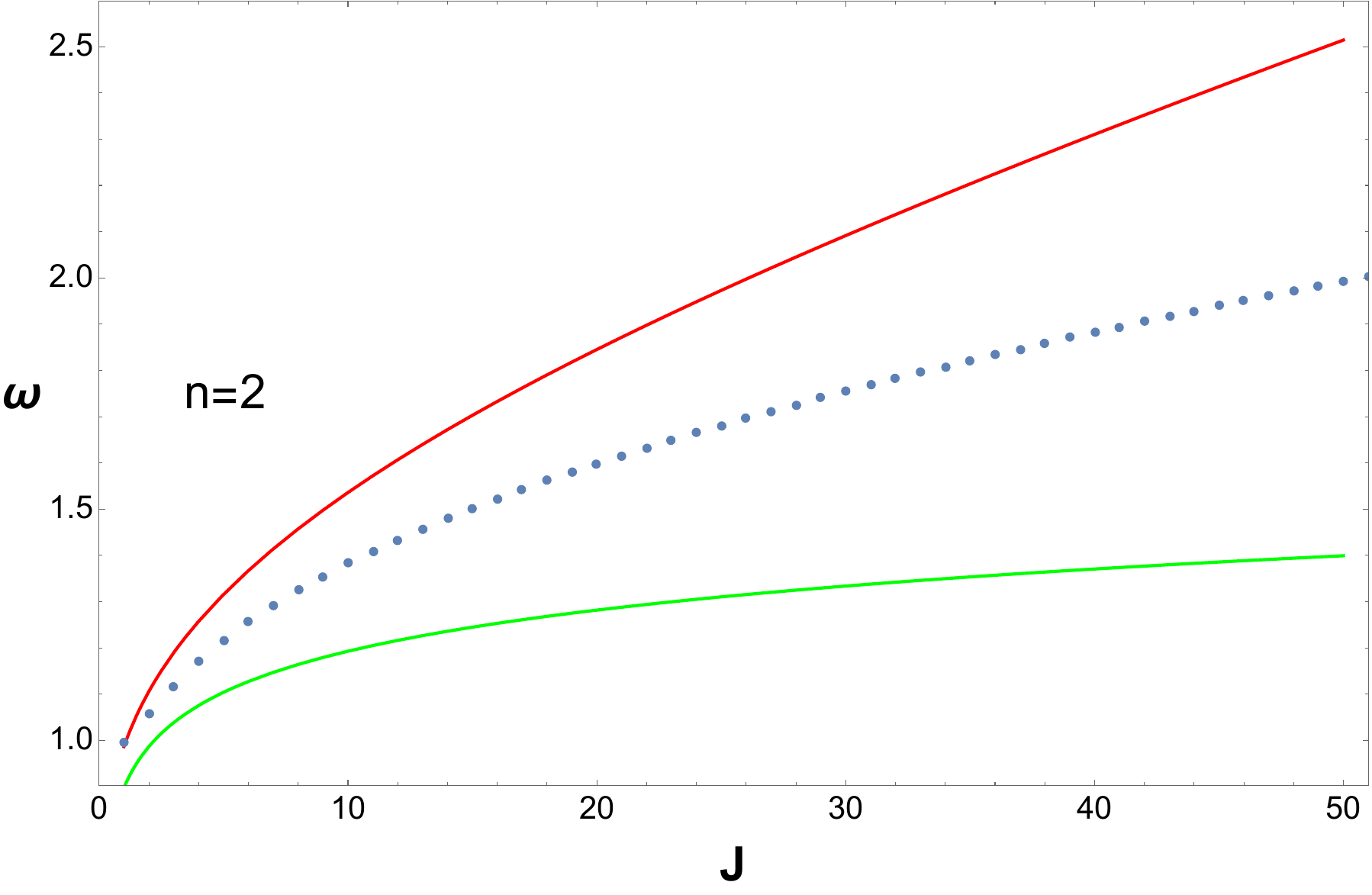}
  \caption{}
  \label{fig18(b)}
\end{subfigure}
\caption{Comparison between exact spectrum (blue), Approximate Phase Equation (red) and the spectrum at low $J$'s (green) as in \eqref{btz-ultra-approx} at $z_o = 7$ with $r_h = 1$ and $L = 1$}. 
\label{fig18}
\end{figure}

\begin{figure}[h]
       \centering
       \includegraphics[width=0.9\linewidth]{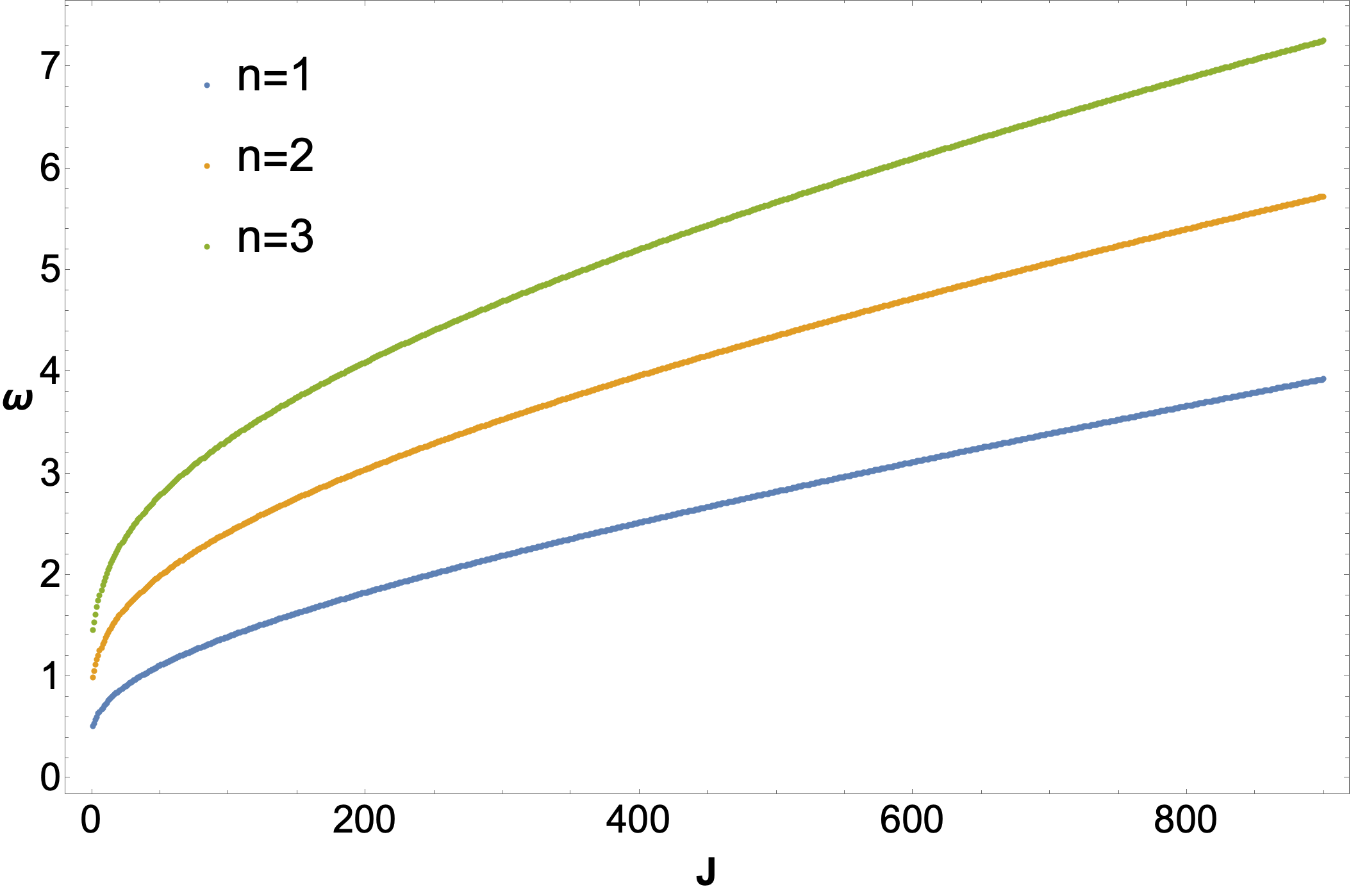}
       \caption{Plot of exact spectrum $\omega(n,J)$ for BTZ Geometry $z_o = 7$ with $r_h = 1$ and $L = 1$ for $n = 1, 2$ and 3.}
       \label{fig19}
\end{figure} 

We will use the following Kummer relation, the standard expansion of ${}_{2}F_{1} (a,b,c,z)$ about $|z|=0$ and two identities of Gamma functions in this subsection. 
\begin{equation} \label{Hyp-gamma-identities}
\begin{split}
   {}_{2}F_{1} (a,b,c,z) &= \frac{\Gamma(c) \Gamma(c-a-b)}{\Gamma(c-a) \Gamma(c-b)} {}_{2}F_{1} (a,b,a+b-c+1, 1-z) \\
   &+ \frac{\Gamma(c) \Gamma(a+b-c)}{\Gamma(a) \Gamma(b)} (1-z)^{c-a-b} {}_{2}F_{1} (c-a, c-b, 1+c-a-b,1-z)\\
   {}_{2}F_{1}(a,b,c,z) &= 1 + \frac{ab}{c} z + \frac{a(a+1)b(b+1)}{c(c+1)} \frac{z^2}{2!} + O(z^3), \ |z| = 0\\   
   \Gamma(ik) \Gamma(1-ik) &= -\frac{i \pi}{\sinh(\pi k)} ,\ \forall k \in \mathbb{R}\\
   \Gamma(ik) \Gamma(-ik) &= \frac{\pi}{k \sinh(\pi k)}\\
\end{split}
\end{equation}
Making a definition of $\epsilon$ via $\left(1-\frac{r_h}{r}\right) \approx -\tanh{r_h z_o/L^2}+1 = \coth{r_h z_o/L^2}-1 \equiv \epsilon$ and using the above identities, we find an equation that is strikingly similar in form to the zeroth order equation:
\begin{equation} \label{perturbed-phase-e} 
    \tilde{P}_{1} (\coth(z_o r_h/L^2)-1)^{-\frac{i\omega L^2}{2 r_h}} + \tilde{Q}_{1} (\coth(z_or_h/L^2)-1)^{\frac{i\omega L^2}{2 r_h}} =0
\end{equation}
where, 
\begin{equation} \label{eperturbed-btz-quantities}
\begin{split}
    D &= 2\frac{\left(1+\frac{iL}{2 r_h} (J-\omega L)\right) \left(1-\frac{iL}{2 r_h} (J+\omega L)\right)}{1-\frac{i\omega L^2}{ 2 r_h}}\\
    \tilde{P}_{1} &= P_{1}(1+D \epsilon)\\
    \tilde{Q}_{1} &= Q_{1}(1+D^{*} \epsilon)\\
\end{split}
\end{equation}
and $P_{1}$ and $Q_{1}$ are the same as defined in \cite{Sumit1} with $\nu=1$ for the massless case we are dealing with. From \cite{Sumit1}, we know that $|P| = |Q|$. From the above, we have
\begin{equation} \label{eq5.20}
\begin{split}
  |1+D \epsilon| &= \left|\frac{1+2\epsilon + \frac{\left(\frac{JL}{r_h}\right)^2 - \left(\frac{\omega L^2}{r_h}\right)^2} {2} \epsilon - (1+2 \epsilon) \frac{i\omega L^2}{r_h} }{1-\frac{i\omega L^2}{r_h}}\right|\\
  |1+D^{*} \epsilon| &= \left|\frac{1+2\epsilon + \frac{\left(\frac{JL}{r_h}\right)^2 - \left(\frac{\omega L^2}{r_h}\right)^2} {2} \epsilon + (1+2 \epsilon) \frac{i\omega L^2}{r_h} }{1+\frac{i\omega L^2} {r_h}}\right|\\
  |1+D \epsilon| = |1+D^{*} \epsilon| 
\end{split}
\end{equation}
One can show that even $|\tilde{P}_{1}| = |\tilde{Q}_{1}|$, which means that $\tilde{P}_{1}$ and $\tilde{Q}_{1}$ are pure phases. This leads us to conclude that the next order corrected equation is also a phase equation which takes a form similar to \eqref{BTZ-phase-eq-2}:
\begin{align} \label{BTZ-pert-phase}
&\cos\left({\rm Arg} \left[\Gamma\left(\frac{i\omega L^2}{r_h}\right)\right]+{\rm Arg} \left[\Gamma\left(-\frac{iL}{2r_h} (J+\omega L)\right)\right] \right. \nonumber \\
&\qquad\left. + {\rm Arg} \left[\Gamma\left(\frac{iL}{2r_h} (J-\omega L)\right)\right] -\frac{\omega L^2}{r_h} \log(2) \right. \nonumber \\
&\qquad\left. +{\rm Arg}\left[(1+2\epsilon)\left(1-\frac{i\omega L^2}{r_h}\right)+\frac{L^2}{2 r_h} (J^2 - \omega^2 L^2)\epsilon\right] - {\rm Arg} \left[1-\frac{i\omega L^2}{r_h}\right] +  \omega z_o\right) = 0.
\end{align}
This we will call the {\em BTZ perturbed phase equation}. 

This first order corrected equation is straightforwardly solvable numerically. And the normal modes that we get are qualitatively identical to the solutions of the exact equation in the Rindler case at the ranges of $J$ that we care about -- in particular, we can find solutions beyond $J_{max}$. The qualitative structure of the perturbed phase equation, phase equation, analytic low-lying spectrum etc. are analogous to our previous Rindler discussion. So we will not repeat the details here.

\section{Variations on 't Hooft's Calculation}

In this section we will review 't Hooft's calculation \cite{tHooft} of the entropy of the black hole\footnote{We will follow the notations of Solodukhin \cite{SolodukhinReview} till the end of the first subsection. The comparison between our notation in the rest of the paper and \cite{SolodukhinReview} is given in \eqref{SoloNotMatch}.}. We will also do the integrals in the calculation in a slightly different order, which will turn out to be instructive. This will also be useful in making sense of the more direct calculation of the thermodynamics of the black hole using our explicit normal modes.

The goal of 't Hooft's calculation was to compute the thermodynamics of the scalar field modes trapped between the angular momentum barrier and the stretched horizon (ie., brickwall, in the language of \cite{tHooft}). The fact that the modes are assumed to be trapped means that we are ignoring tunneling, and therefore the calculation is (semi-)classical. Note that normal modes of the system are the true eigenmodes of the scalar and therefore they are {\em not} semi-classical in this sense. 

Since we are only interested in these trapped modes, then we can get the entropy via a Bohr-Sommerfeld type approach which helps us to avoid the explicit computation of the modes. This was the clever path taken by 't Hooft. He solved the $l$-mode Schrodinger equation \eqref{TortoiseSchrodinger} using a WKB method, and noted that the Bohr-Sommerfeld quantization condition takes the form
\begin{equation} \label{BohrSomerfeld4BH}
n(\omega,l) = \frac{1}{\pi} \int_{r_{\epsilon}}^{r_{\omega,l}} \,dr \frac{1}{f(r)} \kappa(r,l,\omega)
\end{equation}
where, $r_\epsilon$ is the stretched horizon location, and is related to horizon radius via $r_\epsilon = r_+ + \frac{\pi \epsilon^2}{\beta_H}$ in the notation of \cite{SolodukhinReview}. The map between the notation in \cite{SolodukhinReview} and us is: 
\begin{equation} \label{SoloNotMatch}
    r_\epsilon \equiv r_s, \hspace{0.2cm} \epsilon \equiv l_s, \hspace{0.2cm} \beta_H \equiv 4 \pi r_+.
\end{equation}
Note that $\beta_H$ is the inverse Hawking temperature and therefore in our notation, stretched horizon distance is $r_s = r_+ + \frac{l_s^2}{4 r_+}$. We will follow Solodukhin's notation in the first two subsections. The near horizon metric is of the form, $f(r) = \frac{4 \pi}{\beta_H} (r-r_+)+O(r-r_+)^2$. The above equation comes from the phase in the WKB solution, where the radial mode is
\begin{equation} \label{WKBsol}
       R(r) = \rho(r) e^{\pm i \int \frac{dr}{f(r)}\kappa(r,l,\omega)}, \ \ {\rm with} \ \
       \kappa(r,l,\omega) = \sqrt{\omega^2 - \left(\frac{l(l+1)}{r^2}\right)f(r)}.\\
\end{equation}
Integration is real in the domain where $\kappa^{2}(r) \geq 0$. This allows one to fix the integration bound on $r$ from the brick wall radius ($r_\epsilon$) to the radius (for a given $l$) at which $\kappa(r_{\omega,l}) = 0$. This fixes $r_{\omega,l}$ to be\footnote{There are obviously two signs for this since it is a solution of a quadratic equation, but we take the one with negative sign, since for $r$ smaller than this $r_{\omega,l}$, $\kappa^{2}(r,l,\omega) > 0$ and this is only satisfied for negative root.}
\begin{equation} \label{romegal}
    r_{\omega,l} = \frac{2\pi l(l+1)}{\beta_{H} \omega^2} \left(1-\sqrt{1-\frac{\beta_{H}\omega^2 r_{+}}{\pi l(l+1)}}\right)
\end{equation}
An important observation is that $r_{\omega,l}$ becomes closer and closer to the horizon radius ($r_+$) as we make increments in $l$. 

For computation of the \eqref{BohrSomerfeld4BH} integral, the region close to the stretched horizon is the most important\footnote{A macroscopic distance away from the horizon, the thermodynamics of the modes is essentially that of a Planckian black body.}. Hence we shall often only be concerned with the lower limit $r_{\epsilon}$.  Explicitly, the number of modes with energy $\omega$ and angular quantum number $l$ is given by
\begin{equation} \label{explicit-Bohr-Sommerfeld}
n(\omega,l) = \frac{\beta_H}{4\pi^2}\int_{r_\epsilon}^{r_{\omega,l}} \,dr {\frac{1}{r-r_+} \sqrt{\omega^2 - \left(\frac{l(l+1)}{r^2}\right) \frac{4\pi}{\beta_H} (r-r_+)}}
\end{equation}
Each $l$-mode has a degeneracy of $2l+1$, so total number of modes at a fixed energy $\omega$ is 
\begin{equation} \label{eq}
    n(\omega) = \frac{\beta_H }{4\pi^2}\int_{r_\epsilon}^{r_{\omega,l}} \,dr {\frac{1}{r-r_+} \int \,dl (2l+1) \sqrt{\omega^2 - \left(\frac{l(l+1)}{r^2}\right) \frac{4\pi}{\beta_H} (r-r_+)}}
\end{equation}
Remarkably, it turns out that the $l$-integral here is explicitly doable, and the result is 
\begin{equation} \label{eq6.8}
    n(\omega) = \frac{\beta^{2}_H \omega^3}{24 \pi^3}\int_{r_\epsilon}^{r_{\omega,l}} \,dr {\frac{r^2}{(r-r_+)^2}}
\end{equation}
Since our focus is only on lower limit of the integral,  the leading order term of the expansion in $\epsilon$ (where $\epsilon \rightarrow 0$) is
\begin{equation} \label{n(omega)}
    n(\omega) = \frac{r^{2}_{+} \beta^{3}_{H} \omega^{3}}{24 \pi^4 \epsilon^2} + O(\log(\epsilon))
\end{equation}
With this expression, it is straightforward to obtain the area law for the entropy \cite{SolodukhinReview}, so \eqref{n(omega)} will be the target of our attention in the discussion below. The dependence on the physical quantities and the stretched horizon $\epsilon$ are the key things to note here -- 't Hooft's calculation is not precise enough to fix the overall numerical coefficient.

\subsection{Why did this Calculation Work?}

In the above calculation \cite{tHooft, SolodukhinReview}, we did the $l$-integral first, and then made an approximate evaluation of the lower end of the $r$-integral. It turns out that it is instructive to do the integrals in the reverse order. The result of the $r$-integral from the lower end turns out to be 
\begin{equation} \label{n(w,l)}
n(\omega,l) = \frac{\beta_H \omega}{2\pi^2} \log\left(\frac{r_+ \beta_H \omega}{\pi \epsilon (l^2 + l)^{\frac{1}{2}}}\right) + O(\epsilon^2)
\end{equation}
The implicit assumption in doing the integral, namely that the square root is real and positive leads to the condition
\begin{equation} \label{lbound}
l(l+1) \leq \frac{\omega^2 r^2}{f(r)}
\end{equation}
Since the integral is from near the horizon, $r$ can be replaced by $r_\epsilon$ to estimate the maximum value of $l$:
\begin{eqnarray} \label{lmax}
    l_{max}(l_{max}+1) &\approx & \frac{\beta_{H} \omega^2}{4 \pi} \frac{r^{2}_\epsilon}{r_\epsilon - r_+} 
    \approx  \frac{\beta^{2}_{H} \omega^2 r^{2}_{+}}{4 \pi^2 \epsilon^2} + \frac{\beta_{H} \omega^2}{2 \pi} r_{+} + O(\epsilon^2)\\
   \implies l_{max} &\approx & \frac{\beta_{H} \omega}{2 \pi} \frac{r_{+}}{\epsilon} \label{lmaxx}
\end{eqnarray}
Now, we can do the $l$-integral from 0 to $l_{max}$ (which again turns out to be exactly doable):
\begin{equation} \label{6.18}
\begin{split}
n(\omega) &= \frac{\beta_H \omega}{2\pi^2} \int_0 ^{l_{max}} \,dl (2l+1) {\log\left(\frac{r_+ \beta_H \omega}{\pi \epsilon l(1+1/l)^{\frac{1}{2}}}\right)}\\
 &= \frac{\beta_H \omega}{4\pi^2} l_{max}^2 \left[1- \log\left(\frac{\pi^2 \epsilon^2 l_{max}^2}{r_+^2 \beta_H^2 \omega^2}\right)\right]
\end{split}
\end{equation}
Using $l_{max}$ from \eqref{lmax}, we finally get $n(\omega)$ as
\begin{equation} \label{nomega-take2}
n(\omega) \approx \frac{r_+^2 \beta_H^3 \omega^3}{7 \pi^4 \epsilon^2} + ...
\end{equation}
which is the same as the expression in \eqref{n(omega)} except for the precise numerical factor. The approximations we did while reversing the order of integration are too crude, so this is reasonable.

The utility of this reversal of integration order is that \eqref{n(w,l)} allows a comparison with the explicit normal modes that we have determined in the previous section(s). In the Rindler case, using \eqref{Schwarzschild-Rindler-map} in \eqref{Rindler-omega-a}, we get
\begin{equation} \label{omega-Rindler}
    \omega(n,l) = \frac{n\pi}{r_{+}(2\log\left(\frac{2r_+}{l_s \sqrt{l^2+l+1}}\right)-1.15)}.
\end{equation}
This can be solved for $n$ trivially and we find
\begin{equation} \label{n-Rindler}
    n(\omega,l) = \frac{r_+\omega}{\pi}\left(2\log\left(\frac{2r_+}{l_s \sqrt{l^2+l+1}}\right)-1.15\right).
\end{equation}
Let us make a few comments.
\begin{itemize}
    \item A key difference between the two relations \eqref{n(w,l)} and \eqref{n-Rindler} is that the former has an extra dependence on $\log(\beta \omega)$, modulo which it matches with our explicit calculation of normal modes in Rindler. This difference is unsurprising. Our 't Hooftian calculation involved multiple approximations, not to mention the fact that we were working with semi-classical modes, not true normal modes.    
    \item But the crucial observation at this point is that even if one drops the $\log$ dependence in $n(\omega, l)$, we {\em still} get the same result for the total $n(\omega)$ modulo the unimportant numerical pre-factor. In other words, simply with $n(\omega, l) \sim  \frac{\beta_H \omega}{2 \pi^2}$, we find
\begin{eqnarray} 
    n(\omega) \approx \frac{\beta_H \omega}{2 \pi^2}\int_{0}^{l_{max}} dl (2l+1) \approx \frac{\beta_H \omega}{2 \pi^2} l^{2}_{max} \approx \frac{r^{2}_{+} \beta^{3}_H \omega^3}{8 \pi^4 \epsilon^2}
\end{eqnarray}
which already is the correct answer (up to the numerical factor), despite the absence of $\log (\beta_H \omega)$. We have used the fact from \eqref{lmaxx} that $l_{max}$ scales as $1/\epsilon$. Essentially all that is happening is that $\omega$ is just getting multiplied by $l^{2}_{max}$ in the $n(\omega,l)$ expression. This means that 't Hooft's answer is only reliant on the fact that the dependence of $\omega$ on $l$ is weak enough, and that the  dependence on $n$ is linear. As long as the $\omega$'s are  sufficiently degenerate along $l$ direction, we will reproduce the area scaling of the entropy.

    \item The above fact is evident in (say) our Fig. \ref{many-n-one}. We see that the difference between two consecutive $\omega$'s for a fixed $n$ is exponentially suppressed along $J$-direction as compared to the difference along the $n$-direction for a fixed $J$. This is what we mean by the statement that the $J$-direction is quasi-degenerate. So, all that 't Hooft needed was an $\omega \sim n/r_{+}$ loosely degenerate in $l$ so that when multiplied by $l^{2}_{max}$ we get the area scaling.

    \item Let us note from \eqref{omega-Rindler}, that $\omega$ is loosely $n \pi/r_{+}$ around $l_{max} \approx r_{+}/l_{s}$. To get estimates, we will often find it convenient to treat these modes as degenerate, with energy $n \pi/r_{+}$. 
    
  \item This is a good place to point out that it is precisely the absence of this quasi-degeneracy that gives rise to the volume law scaling of Planckian black body entropy. Typically, any spatial dimension leads to a harmonic oscillator-like (linear in some $n$) mode. What the black hole does is to remove that dependence for one of the angular dimensions (loosely, one can think of this direction as being captured by the Casimir of the horizon sphere).  This is the operational mechanism for the area scaling of entropy. It is clearly of interest to understand this more deeply.
\end{itemize}

\subsection{Temperature and Entropy from $\omega (n,l) \approx \frac{n \pi}{r_+}$}. 

Let us illustrate and strengthen the above discussion, by explicitly showing that a spectrum of the form
\bea
\omega(n,l) \sim \frac{n \pi}{r_+} \equiv n \omega_1
\eea
with a degeneracy of $l_{max}$ can reproduce both the temperature {\em and} the entropy up to numerical pre-factors, once the energy of the ensemble is specified to be the black hole mass. This should be viewed as an improvement on the calculation of 't Hooft, who did not have access to the actual spectrum and therefore needed to specify the temperature as well, to reproduce the entropy up to an $O(1)$ number. In the BTZ black hole where we have much more control on the details of the normal mode spectrum, we will in fact be able to fix both the temperature and entropy exactly while changing the cut-off in $l$ somewhat. This we will do in a later section, here our goal is only to illustrate the philosophy. So we will work with a Schwarzschild black hole, which was the original setting of 't Hooft's calculation and we will settle for fixing the quantities up to $O(1)$ numbers.

The energy integral in the canonical ensemble is 
\begin{equation} \label{EnergyIntegral}
E_{BH} =  \int \,dn \int \,dl {\frac{2l+1}{e^{\beta n \omega_1}-1} n \omega_1  \Theta[n^2 \omega_1^2 - V_l(r_{*s})}].
\end{equation}
The only point that needs explanation is the the $\Theta$-function. Its purpose is to impose the cut-off in $l_{max}$ that we discussed in the last subsection. In 't Hooft's approach it was justified via a semi-classicality demand, so we have chosen to do it the same way here. But we believe that this interpretation is an effective one, and that the bound is fundamentally on $l$ itself and not on the potential. This may be an important hint about the nature of the underlying UV degrees of freedom. We will adopt a slightly different approach to fix the range of the angular quantum number in later sections, which we believe is better motivated from the perspective of normal modes. 

\begin{figure}[h]
       \centering
       \includegraphics[width=0.5\linewidth]{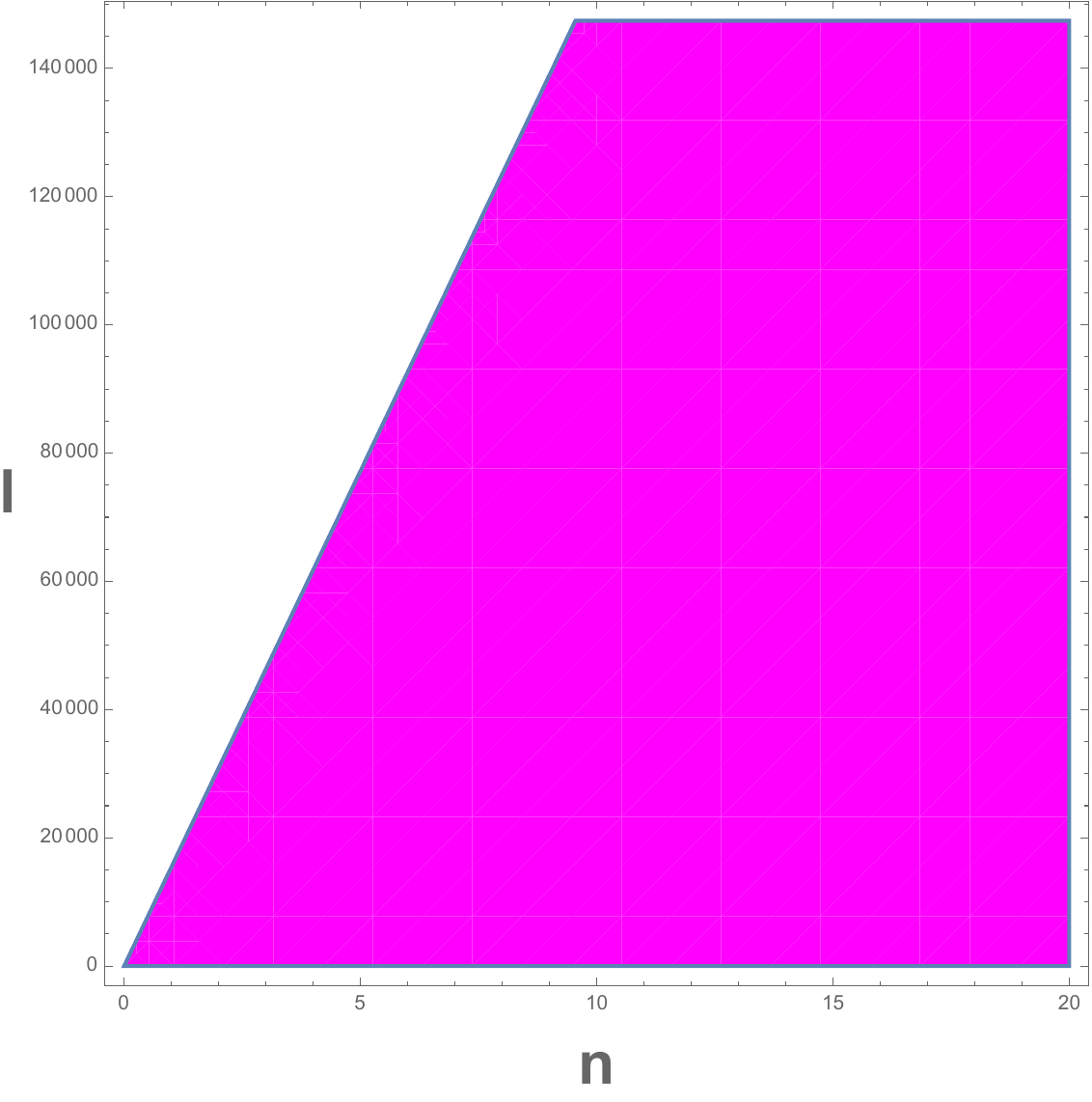}
       \caption{The shaded region is the one in which $\Theta$ function is non zero. The plot is for $\xi_{o} \equiv r_{*s} = -8$ and $r_{+} = 1/2$. }
       \label{lmax-vs-n}
\end{figure}

It is easy to see that the upper bound $l_{max} \approx \frac{2 r_+ \pi}{l_s} n$, see also Fig. \ref{lmax-vs-n}.
Doing the $l$-integral first from $0$ to $l_{max}$, we find
\bea \label{after-l-integral}
E_{BH}= \frac{4 r_+^2 \pi^2}{l_s^2} \int_{0}^{\infty} \,dn \frac{n^3 \omega_1}{e^{\beta n \omega_{1}}-1}
 = \frac{4 \pi^3 r_+^5}{15 \beta^4  l_s^2}
\eea
Now, using $E_{BH} = M$ and $r_+ = 2 M l_s^2$ we get\footnote{Note that $l_s=l_p$ for us.} 
\begin{equation} \label{temperature-thooft}
\begin{split}
   \beta &\approx M l_p^2 \approx  r_{+}\\
\end{split}
\end{equation}
This is the promised determination of temperature, now we turn to the determination of the entropy which was the famous result of 't Hooft.
 
To compute the entropy, we start with \cite{Pathria,Nomura}
\begin{equation} \label{Omega}
\begin{split}
\Omega(N_n^{{l}}) &= \prod_{l=0}^{\infty} \prod_{n=0}^{\infty} \frac{(2l+N_n^{(l)})!}{N_{n}^{(l)}! (2 l)!} \hspace{0.5cm}     (N_n^{(l)} \geq 0)\\
N_n^{l} &= \frac{2l+1}{e^{\beta \omega(n,l)}-1}
\end{split}
\end{equation} 
which gives the entropy as
\begin{equation} \label{Omega-entropy}
S = \ln \Omega 
   =  \int_{0}^{\infty} \,dn \int_{l=0}^{l_{max}} \,dl (2l+1) \left( {\frac{\beta n \omega_1}{e^{\beta n \omega_1}-1}}-\log (1-e^{-\beta n \omega_1})\right)\Theta[n^2 \omega_1^2 - V_l(r_{*s})] 
 \end{equation}
This leads to 
\begin{equation} \label{entropy-thooft}
S = \beta E_{BH} -  \frac{4 r^{2}_{+} \pi^2}{l_s^2} \int_0^ \infty \,dn n^2 \log(1-e^{-\beta n \omega_{1}}) = \frac{4}{3} \beta E_{BH} 
\end{equation}
where we used the standard result $\int_{0}^{\infty} \,dx x^2 \log(1-e^{-x}) = -\frac{\pi^4}{45}$. 

Thus, we see that $S \approx \beta E_{BH} \approx r_+^2/ l_p^2$. So we have reproduced both the temperature as well as the entropy, from a conventional statistical mechanics calculation. A key ingredient is the quasi-degeneracy in the $l$-direction, and an associated cut-off at $l_{max}$. Note also that we have used gravity in a couple of implicit places, one of which is in the Schwarzschild formula relating $r_+$ and $M$ -- this is the mechanism that connects geometry to the energy of the ensemble. Another technical point worth noting here is that the support for the $n$-integrals is coming from the  low $n$'s. So the physics is controlled by the low-lying part of the spectrum in $n$. 

We believe that the discussion in this section is the structural backbone of 't Hooft's calculation, strengthened to incorporate temperature as well, and simplified as much as possible (and hopefully not further) to retain (only) the key ingredients. As we already mentioned, an improved way to fix the range of the angular quantum number will be used in later sections. We will see that it is natural to have a somewhat lower cut-off for $l$.

There have been previous papers which have also attempted to reproduce both the temperature and entropy of the black hole, starting with the stretched horizon and using statistical mechanics. We make some comments about this in the next subsection. 

\subsection{Comments on Previous Work}

An interesting paper \cite{Nomura} also explores a philosophy similar to ours -- their goal is also to get the temperature and entropy by specifying the mass of the black hole as the energy of the stretched horizon ensemble. We discuss the similarities/differences between their results and ours below.  

\begin{itemize}
    \item The authors of \cite{Nomura} focus on the low-lying part of the normal mode spectrum. They do not fix it completely like we do, instead they argue that 
   \bea
   \omega(n+1,l)-\omega(n,l) \sim \frac{\pi}{2 r_+ \ln \left(\frac{r_+}{\sqrt{l(l+1)+1}l_s} \right)}
   \eea
by looking at an approximate version of the radial wave equation. Apart from minor differences, this is consistent with our results as well in the low-lying part of the spectrum. However, this approach means that $\omega(0,l)$ is not fully fixed. They assume based on heuristic arguments\footnote{The heuristic argument seems to be that this is the lowest value of the scalar effective potential, the value it takes on the stretched horizon. Note that ground state energies of quantum mechanical potentials do not have to be their lowest values. A good example is the half-harmonic oscillator  with a (brick-)wall on one side -- the ground state energy is fixed by the spring constant, not the lowest value of the potential.} that 
\bea
\omega(0,l)\equiv \omega^{(l)}_{0} \approx \frac{\sqrt{\lambda_{l}} l_s}{r_+} \label{nomura-w}
\eea
where, $\lambda_l = \frac{l(l+1)+1}{r^{2}_+}$. Our analytic low-lying normal mode expressions (say \eqref{omega-Rindler}) differ non-trivially from the resulting total $\omega(n,l)$ even at low values of $l$. At high values of $l$ note that one has to work with the exact spectrum, and these expressions badly break down (as clear from many of our plots). Note in particular that the (crude) degeneracy approximation and the resulting formula  $\omega \sim n \pi/r_+$ that we used in the last subsection goes {\em beyond} the analytic expression \eqref{omega-Rindler} which is only valid at low $l$. It used our knowledge of the exact spectrum. The analytic low-lying expression \eqref{omega-Rindler} is inescapably non-degenerate at large $l$ because the spectrum diverges.  
    \item The expression $\omega(0,l)$ above in \eqref{nomura-w} is crucial for the calculations in \cite{Nomura}. It is this term that gives rise to the saddle that they find when doing the integrals in the thermodynamic quantities. The result in their Section 3.1 is obtained via an integral of the form
\begin{equation}
    \begin{split}
        E_{BH} &=  \int_{l=0}^{\infty} dl\ 2l\left(\frac{l l_s}{r^{2}_{+}}\right) e^{-\beta \frac{l l_s}{r^{2}_{+}}} \int_{n=0}^{\infty} dn e^{-\beta \frac{n}{\left|r_{+} \log \left(\frac{r_{+}}{l l_s}\right)\right|}}\\
        &+\int_{l=0}^{\infty} dl\ \frac{2l}{\left|r_+ {\rm log}\left(\frac{r_{+}}{ll_s}\right)\right|} e^{-\beta \frac{l l_s}{r^{2}_{+}}} \int_{n=0}^{\infty} dn\ n\ e^{-\beta \frac{n}{\left|r_{+} \log \left(\frac{r_{+}}{ll_s}\right)\right|}}\\
        \implies E_{BH} &=  \int_{x=0}^{x=\infty} dx \frac{r^{5}_{+}}{\beta^4 l^{2}_{s}} (x^2 + x) e^{-x} \left|\log\left(\frac{\beta}{r_{+} x}\right)\right|\\
    \end{split}
    \end{equation}
We have defined $x = \frac{\beta l l_s}{r^{2}_+}$ in the last line. The crucial thing is that the saddle arises due to the $e^{-x}$ term without which integral would have been ever-growing. With the saddle the integral gets its support from $l \sim r_{+}/l_{s}$.  The end result is
    \begin{equation}
     E_{BH} = \frac{r^{5}_+}{\beta^4 l^2_{p}} \left|{\rm \log} \frac{\beta}{r_+}\right|(1+\mathcal{O}(1))
    \end{equation}
which can be compared to (3.11) of \cite{Nomura}.
    The crucial point is that this saddle is a spurious feature arising due to the presence of \eqref{nomura-w} in the spectrum. Furthermore, it happens near $l_{max}$ where the low-lying mode expressions have completely broken down. Also as a result, the expression relating $E_{BH}$ and $\beta$ contains an unpleasant log term, as opposed to our \eqref{after-l-integral}. If one works with it directly, it is not entirely convincing that we get the standard formulas of black hole thermodynamics.
\end{itemize}    

But despite all these intermediate differences, we believe their final claims have some moral similarities. Instead of a cut-off in $l$ that we find\footnote{Note that semi-classically, this cut-off is intuitive -- it arises because the angular momentum barrier is trapping the modes.}, they find a saddle at $l_{max}$. The fact remains that the angular quantum number is a crucial ingredient in the calculation both for us as well as them. We have decided to include this subsection because the cut-off interpretation is essential for some of our results.

\section{BTZ Thermodynamics from Normal Modes}

Our calculation in the previous section was suggestive and could be called an  improvement on 't Hooft's calculation.  But the philosophy has a few features which we find unsatisfactory. 
\begin{itemize}
\item 't Hooft uses the semi-classically trapped modes behind the angular momentum barrier to count the entropy. We find the normal modes of the stretched horizon to be a more fundamental characterization of the system than these semi-classical modes. Is there a way to formulate the calculation in terms of these normal modes, where the bounds on the angular quantum numbers can be motivated in some other way\footnote{In the previous section, our calculation used (crude approximations of) normal modes, but the cut-off in angular modes was motivated via (semi-)classical boundedness.}? Eventually we hope to understand the cut-off in $J$ in terms of properties of string modes.
\item Introducing the stretched horizon (at least superficially) is a direct violation of the principle of equivalence in bulk EFT. Such an act can only be justified, if the stretched horizon boundary condition is to be viewed as a proxy for UV completion and the fluctuations around these configurations can tell us about the UV complete microstates of the black hole. 
\item As we saw, what the semi-classical calculation accomplishes in practice is to introduce a cut-off in the angular momentum quantum numbers. This lead to the temperature and the entropy both getting fixed, but only up to $O(1)$ numbers. What if instead we fix the angular mode cut-off so that the entropy is precisely matched? What then happens to the temperature? In this section, we will show that in the case of the BTZ black hole, if we fix the angular cut-off so that the numerical coefficient of the BTZ entropy is correctly reproduced, the temperature is also matched {\em exactly}.
\item Interestingly, we also find that with some mild and plausible assumptions about the structure of the normal modes, the calculation generalizes readily to the rotating BTZ case as well, once we specify both the charges (mass and angular momentum of the black hole). In fact, we find that the calculation can be formulated naturally in a holomorphically factorized language. The angular momentum cut-off can be chosen to fix {\em both} the left and right moving entropies simultaneously, and this fixes the left and right moving (and via them the total) Hawking temperatures exactly. We will present the discussion in this section in this holomorphically factorized language, and specialize to the non-rotating case at the end of the discussion\footnote{We thank Vaibhav Burman for help in this calculation, and Suchetan Das for numerous related discussions.}. A calculation framed without the holomorphic factorization, for the non-rotating BTZ black hole, will be presented in \cite{Vaibhav} with somewhat different goals.
\item Interestingly, we find that the cut-off emerging in the angular modes in our approach in this section, is typically an order of magnitude or so {\em lower} than the $l_{max}$ we used in the previous one. This small hierarchy was invisible in the last section, presumably because of the numerical factors that one is not carefully with in the 't Hooftian calculation. Interestingly, this small hierarchy brings the calculation into the reliable part of the analytic low-lying spectrum of our previous sections\footnote{More discussion on this hierarchy can be found in the context of the Kerr black hole normal modes in Appendix \ref{KerrSec}.}. In light of our previous discussions, this means that the linear ramp is produced precisely by the modes responsible for the entropy. 
\end{itemize}

\noindent
The metric of rotating BTZ is given by \cite{CarlipBTZ} 
\begin{equation} \label{Rotating BTZ metric}
   ds^{2} = -(N^\perp)^2 dt^2 + f^{-2}dr^{2} + r^{2}(d\phi + N^\phi dt)^2
\end{equation}
where
\bea \label{shift and lapse rot.}
   N^{\perp} = f = \left(-8G {\cal M} +\frac{r^2}{L^2} + \frac{16 G^2 {\cal J}^2}{r^2}\right)^{1/2}, \ \ 
  N^\phi = - \frac{4G{\cal J}}{r^2}  
\eea
and $L$ is AdS length scale. The inner and outer horizon radii and their relation to ${\cal M}$ and ${\cal J}$ (the mass and angular momentum of the black hole) are
\bea \label{M-J-rplusminus}
   r^{2}_{\pm} = 4G{\cal M}L^2 \left(1 \pm \left[1-\left(\frac{{\cal J}}{{\cal M}L}\right)^2\right]^{1/2}\right), \ \  
   {\cal M} = \frac{r^{2}_+ + r^{2}_-}{8GL^2}, \ \ 
   {\cal J} &=& \frac{2 r_+ r_-}{8 G_N L}
\eea
The metric functions when expressed in terms of $r_+$ and $r_-$ using \eqref{M-J-rplusminus}, take a form that is a direct generalization of the expressions we considered in our previous normal mode discussion of the non-rotating BTZ. 

Similar to what was done in Section \ref{BTZsec}, Planck length $l_p$ in the rotating case can be calculated by demanding that it is the (radial) geodesic length from the horizon at $r_+$ to the stretched horizon at $r_s = r_+ + x$. The integral is exactly doable, but will only present its approximate form ($l_{p} = G$ in 2+1 dimensions):
\begin{equation}
  l_{p} = L\int^{r_{+}+x}_{r_{+}} dr \frac{r}{\sqrt{(r^2-r^{2}_+)(r^{2}-r^{2}_-)}} = L \sqrt{\frac{2 r_+ x}{r^{2}_+ - r^{2}_-}} + O(x^{3/2}) \label{BTZPlanck}
\end{equation}

We have learnt from our previous calculations, that an approximately degenerate (in the angular quantum number $J$) form of the low-lying normal modes\footnote{$J$ should not to be confused with the angular momentum of the black hole, which we denote by ${\cal J}$.} should be able to reproduce the thermodynamics of the black hole. In the rotating case, we will use the form\footnote{In the non-rotating black hole, $A$ was the horizon radius and the denominator had dimensions of $L^2$. For the rotating case the form can be read off from \eqref{analytic-BTZ Rot-low-lying}, but we will keep things general here. }
\begin{equation} \label{ansatz for rotating case}
   \Tilde{\omega}(n,J) \approx \frac{n\pi A}{L \ {\rm log}\left(\frac{B}{l_{p}}\right)}
\end{equation}
One minor subtlety is that when the black hole is rotating, it is more natural to work with the ``shifted" modes which have been co-rotated with the horizon, see eg. \cite{HawkHunterTaylor}. In the present case, this corresponds to working with 
\bea
\tilde \omega(n, J) \equiv \omega(n,J) - J \Omega_H \label{shifted}
\eea
where $\Omega_H$ is the angular velocity of the horizon. Indeed, these shifted modes $\tilde \omega$ were used to argue that the spectrum, SFF and ramp structure were qualitatively identical to the non-rotating case in \cite{Arnab}. Once the modes have been shifted, we expect to be able to do the calculation in terms of a factorized left and right moving CFTs at a left and right temperature. Note that the precise forms of $A$ and $B$ will turn out to be unimportant. This is unsurprising because, as we argued earlier, the robustness 't Hooft's calculation relies only on the linearity in $n$ and the degeneracy in $J$.

In terms of the shifted modes, we expect that the calculation will have a holomorphic factorization in terms of left and right movers\footnote{In fact in the BTZ case, the calculation can be done without this left-right factorization\cite{Vaibhav}. But we expect that the holomorphic factorization approach will suitably generalize to higher dimensions, so we will present it in that language here.}. This is something that arises in known string theory examples, {\em even} away from extremality and supersymmetry \cite{CveticLarsen}. Let us do the calculation first for the right movers ($J >0$).  

The partition function of the right moving modes is simply
\bea \label{Partition Function for Rotating Case (R)}
   \log Z_R &=& \frac{L \log\left(\frac{l_{p} }{B}\right)}{\pi A} \int_{0}^{\infty} d\Tilde{\omega} \sum_{J=0}^{J_{cut}} \log(1-e^{-\beta_R \Tilde{\omega}})\\
   \implies Z_R(\beta_R) & = & \exp\left[\log\left(\frac{B}{l_{p}}\right)\frac{\pi L J_{cut}}{6 \beta_R A}\right]
\eea
The first line is simply the sum over all the $n$'s and $J$'s with the cut-off at $J_{cut}$ imposed. The $n$-sum has been converted into an integral, and this integral is exactly doable. As we said earlier, we aim to fix $J_{cut}$ by matching it with the (right-moving) Bekenstein-Hawking entropy. Taking an inverse Laplace transform, we can get the density of states at a given\footnote{Note that this is the analogue of specifying the energy/mass in the 't Hooftian calculation of last section. By specifying $E_R$ and $E_L$ we specify ${\cal M}$ and ${\cal J}$. Both $\beta_R$ and $E_R$ are chosen to be dimensionless here for convenience.} energy $E_R$:
\begin{equation} \label{Density of States (R)}
   g_R(E_R) = \frac{1}{2\pi i} \int_{\beta-i \infty} ^{\beta +i \infty} d\beta_R \exp\left[\log\left(\frac{B}{l_{p}}\right)\frac{\pi L J_{cut}}{6 \beta_R A}+\beta_R E_R\right]
\end{equation}
where $E_R$ is the energy of the right modes. This can be evaluated by a saddle point approximation. The saddle is at\footnote{We will not distinguish the variable $\beta_R$ from its value at the saddle. This should not cause any confusion.} 
\bea
\beta_R = \sqrt{\log\left(\frac{B}{l_{p}}\right)\frac{\pi L J_{cut}}{6 A E_R}}.
\eea
The density of states and therefore the entropy at energy $E_R$ are given by
\bea \label{Saddle Pot. (R)}
   & g_{R}(E_R) \approx \exp\left[2 \sqrt{\log\left(\frac{B}{l_p }\right)\frac{\pi L E_R}{6 A}J_{cut}}\right]\\
   & S_R(E_R) = \log(g_R(E_R)) \approx 2 \sqrt{\log\left(\frac{B}{l_p}\right)\frac{\pi L E_R}{6 A}J_{cut}}
\end{eqnarray}
Now $E_R \equiv \frac{{\cal M}L+{\cal J}}{2}$ and the right mode entropy $S_R(E_R)=\pi \sqrt{\left(\frac{{\cal M}L+{\cal J}}{2G}\right)L}$  \cite{BTZStrominger}\footnote{$E_R$ and $S_R$ play conceptually different roles in this calculation. $E_R$ is simply fixing the charge, and this needs to be done in any statistical mechanics set up. But the scaling of $S_R \sim \sqrt{E_R}$ is a prediction of the stretched horizon paradigm, that in principle could have been falsified. That it does not, is non-trivial. Given that, we use the coefficient of $S_R$ in (say) \cite{BTZStrominger} to precisely fix $J_{cut}$.}. Note that $G \equiv l_p$ in 2+1 dimensions. Now $J_{cut}$ will be fixed in terms of the $l_{p}$, $A$ and $B$ as
\begin{equation} \label{m_cut right rotating (rotating BTZ)}
   J_{cut} = \frac{3\pi A}{2 l_p\ {\rm log}\left(\frac{B}{l_p}\right)}  
\end{equation}
For the left-movers, things work out precisely analogously -- except $E_L = \frac{{\cal M}L-J}{2}$ and $S_L(E_L)= \pi \sqrt{\left(\frac{{\cal M}L-J}{2G}\right)L}$ \cite{BTZStrominger}. An interesting fact is that the resulting $J_{cut}$ is the same as above.

Plugging back in, this fixes the temperatures precisely: 
\bea \label{Beta_L, S_L}
    \beta_{L} &=& \sqrt{\frac{\pi^2 L}{2 G ({\cal M}L-{\cal J})}} = \frac{2 \pi L}{r_+-r_-}, \\  \beta_{R} &=& \sqrt{\frac{\pi^2 L}{2 G ({\cal M}L+{\cal J})}} = \frac{2 \pi L}{r_++r_-}, \\ \beta_H &\equiv & \frac{1}{2} (\beta_{L}+\beta_{R}) = \frac{2\pi L r_+}{r^{2}_+ -
    r^{2}_-}
\eea
matching the standard expressions.

A couple of comments are in order. Firstly, note that the success of these calculations automatically means that the central charges of the CFTs will also work out correctly to yield the Brown-Henneaux formula (on both the left and the right). We have checked this. Secondly, the calculation in the non-rotating case can be done in two ways. Either by assuming that it is a rotating ensemble at zero angular momentum, or by working in the (non-rotating) canonical ensemble at fixed energy. The former calculation is a limiting version of the above calculation with $A=B=r_+$ and $r_-=0$, and so automatically works as a consequence of our above observations. The canonical ensemble version of the calculation also works for BTZ, and will be presented in \cite{Vaibhav}. 

\section{Concluding Remarks}

\begin{itemize}
\item{\bf Holomorphic Factorization and Higher Dimensions:} We have discussed the (rotating) BTZ case in detail in this paper and reproduced the thermodynamic quantities from the normal modes. The structure of the calculation suggested a holomorphic factorization between the left and right sectors. Such a factorization actually exists for very general black holes, including Kerr and even five dimensional black holes in string theory \cite{CveticLarsen}. The wave equations of these black holes in the near region can be solved using hypergeometric functions \cite{CveticLarsen, Strominger, CKHigherDimBH} and it seems very likely to us that the calculations of normal modes may allow a generalization of our BTZ calculations to these settings. It will be very interesting to revisit our Schwarschild and Kerr discussions from such a holomorphically factorized perspective. See some closely related comments in our Appendix \ref{KerrSec}.

\item{\bf Normal Modes of Ryu-Takayanagi Surfaces:} As illustrated by the Rindler example, our normal mode calculation was not restricted to black hole horizons. Indeed, we identified the normal modes of the Rindler wedge as well. The horizons of Rindler wedges are (hyper-)planes and they can be viewed as Ryu-Takayanagi (RT) surfaces in flat space, a perspective recently emphasized in  \cite{Abir}. This means that these RT areas can be understood as being computed by the thermodynamics of these Rindler normal modes. Similarly, it is clear that normal modes of AdS-Rindler wedges \cite{Casini} should be able to explain the RT entropies in AdS/CFT. This brings up an interesting puzzle. When we compute the horizon areas of black holes, we view those as thermal entropies. But the entropies associated to RT surfaces anchored to the boundary of AdS are von Neumann entropies. From the bulk point of view, if done using normal modes, both calculations are reminiscent of a thermal entropy calculation -- after-all this is the entropy of a gas of normal modes. How should one reconcile/understand this difference? Intuition about this question is likely to reveal some new insight about bulk locality in holography. 

\item{\bf Thermal vs Entanglement Entropy:} A closely related point is that the entropy computed by the brickwall, makes us think about entanglement entropy and area law. This is a very tantalizing perspective, see eg., \cite{Sorkin, SU}. Operationally however, the 
calculations that we discussed were simply thermal calculations and there was no entanglement entropy, anywhere. {\em We suspect that the correct picture here is that thermal entropy that one computes using normal modes should be viewed as an entanglement or von Neumann entropy in a holographic dual theory without gravity.} This perspective would resolve the question raised in our previous bullet point also, partially. It may be good to develop or rule out this speculation more concretely and give a mechanical understanding of holographic entanglement entropy from the {\em bulk}.

\item{\bf Normal Modes of de Sitter Space:} Even though we have not included it in this paper, we have worked out the normal modes of de Sitter space as well. This is particularly straightforward in dS$_3$. See an old related discussion in \cite{WonTaeKim}. We hope to present details of this calculation elsewhere after we gain some more confidence in the interpretational distinctions with black holes.

\item{\bf A New Hierarchy?} We noticed that $J_{cut} \sim J_{inter}$ was hierarchically lower than $J_{max}$. $J_{cut}$ was the more physically relevant quantity in our calculations, as opposed to 't Hooft's $J_{max}$. We were able to match both the entropy and temperature exactly using this, and we also noted that the linear ramp is controlled by the region of the spectrum below $J_{cut}$.  The hierarchy between $J_{cut}$ and $J_{max}$ is a relatively small one -- controlled approximately by the logarithm of the horizon radius to the Planck length (roughly, $\log S_{BH}$). This is a number of order 10-100 in the real world. But it was sufficient to make our calculations work --  we could use the analytic low-lying spectrum. It will be interesting to understand the origins of this hierarchy better. 
\end{itemize}

\section*{Acknowledgments}

We thank Vaibhav Burman and Suchetan Das for numerous discussions, which were crucial for this project. CK thanks Roberto Emparan for questions and encouraging comments at an early stage and Sumit Garg for numerically testing one of our claims about the linear ramp at a higher $J_{inter}$ than we managed in this paper. We thank Diptarka Das, Suman Das, Justin David, Sumit Garg, Arnab Kundu and Vyshnav Mohan for discussions.

\appendix

\section{Rotating BTZ}

We will discuss normal modes of some rotating black holes (rotating BTZ and Kerr) in the Appendices. We start with BTZ, which is simpler.

The metric for this background is given in \eqref{Rotating BTZ metric}. Solving massless scalar field in this geometry with the ansatz $\Phi(t,r,\phi) = \sum_{\omega,m} e^{-i\omega t} e^{i J \phi} \phi_{\omega,J}(r)$, we get the radial wave equation 
\begin{equation} \label{Rotating BTZ Radial Eqn}
   \frac{1}{r}\frac{d}{dr}\left(f^{2} \ r \frac{d \phi(r)}{dr}\right) + \left(\frac{1}{f^{2}}\left(\omega - \frac{\cal J}{2 r^2}J \right)^2- \frac{m^2}{r^2}\right) \phi(r) = 0
\end{equation}
where, $f$ and $\cal J$ are as defined in \eqref{shift and lapse rot.} and \eqref{M-J-rplusminus} respectively, and subscripts $\omega$,$J$ are suppressed in $\phi(r)$. This system has been analyzed in \cite{Arnab}, but we would like to obtain the low-lying spectrum analytically in a convenient form.  We have checked that our analytic results can be used to reproduce their numerical plots.

We will use $f = \frac{(r^2 - r^{2}_+)(r^{2}-r^{2}_-)}{r^2 L^2}$. Let us introduce a new dimensionless radial coordinate $z = \frac{r^2 - r^{2}_+}{r^2 - r^{2}_-}$ and a new function in this coordinate system $F(z) = z^{i\alpha}\phi(r)$, where $\alpha$ is defined as
\begin{equation} \label{alpha-BTZ Rot}
    \alpha = \frac{r_+L^2}{2(r^{2}_+ - r^{2}_-)}\Tilde{\omega}
\end{equation}
with $\Tilde{\omega}$ defined as in \eqref{shifted}. In this coordinate system, $z \rightarrow 0$ corresponds to the outer horizon, and $z \rightarrow 1$ corresponds to the AdS boundary.  The final form of the radial solution is
\begin{equation} \label{Rotating BTZ Radial Soln}
    F(z) = e^{-2 \pi \alpha} \ z^{2 i \alpha} \ C_{1} \ {}_{2}F_{1}(a,b,c,z) + C_{2} \ {}_{2}F_{1}(a^*, b^*, c^*, z)
\end{equation}
where $*$ denotes complex conjugate and $a, b, c$ are given by
\bea \label{abc-BTZ Rot}
  &a = -\frac{iL^2}{2 (r_+ + r_-)}\left(\omega + \frac{J}{L}\right) = -\frac{iL^2 \Tilde{\omega}}{2(r_{+}+r_{-})}+\frac{iLJ}{2r_+}\\
  &b = -\frac{iL^2}{2 (r_+ - r_-)}\left(\omega - \frac{J}{L}\right) = -\frac{iL^2 \Tilde{\omega}}{2(r_{+}-r_{-})}+\frac{iLJ}{2r_+}\\
  &c = 1 - 2 i \alpha
\eea
where, in $a$ and $b$ we have used \eqref{shifted}. Near $z \rightarrow 1$, radial wave equation in \eqref{Rotating BTZ Radial Soln} looks like
\bea \label{Near Boundary Soln BTZ Rot}
   &F_{z \rightarrow 1}(z) \approx (1-z) \left(C_2 \frac{\pi  {\rm Cosec} \left(\pi\left(a^* + b^* - c^*\right)\right) \Gamma \left(c^*\right)}{\Gamma \left(c^* - a^*\right) \Gamma \left(c^* - b^*\right) \Gamma
   \left(-a^*-b^*+c^*+1\right)} + C_1\frac{e^{-2 \pi \alpha}\pi
   {\rm Cosec} ((a+b-c) \pi) \Gamma (c)}{\Gamma (c-a) \Gamma (c-b) \Gamma (-a-b+c+1)}+ O\left(z-1\right)\right) \nonumber \\
   &+\pi\left(C_2 \frac{{\rm Cosec} \left(\pi\left(a^* + b^* - c^*\right)\right) \Gamma \left(c^*\right)}{\Gamma \left(c^* - a^*\right) \Gamma \left(c^* - b^*\right) \Gamma
   \left(-a^*-b^*+c^*+1\right)} - C_1\frac{e^{-2 \pi \alpha}\pi
   {\rm Cosec} ((a+b-c) \pi) \Gamma (c)}{\Gamma (c-a) \Gamma (c-b) \Gamma (-a-b+c+1)}+ O\left(z-1\right)\right)\hspace{-1cm}
\eea
Now demanding normalizability at the AdS boundary at $z \rightarrow 1$, one realises that the constant piece must go to zero thus giving a relationship between $C_1$ and $C_2$ as
\begin{equation} \label{C1 C2 BTZ Rot}
    C_2 = C_1 e^{-2 \pi \alpha} \frac{\Gamma (c) \Gamma \left(c^*-a^*\right) \Gamma \left(c^*-b^*\right)}{\Gamma \left(c^*\right)\Gamma (c-a) \Gamma (c-b)}
\end{equation}
Finally the radial wave equation looks like
\begin{equation} \label{Radial Soln Final BTZ Rot}
   F(z) = C_{1} \ e^{-2 \pi \alpha} \left(z^{2 i \alpha} \ {}_{2}F_{1}(a,b,c,z) + \frac{\Gamma (c) \Gamma \left(c^*-a^*\right) \Gamma \left(c^*-b^*\right)}{\Gamma \left(c^*\right)\Gamma (c-a) \Gamma (c-b)} \ {}_{2}F_{1}(a^*, b^*, c^*, z)\right)
\end{equation}
Solving the exact equation by setting \eqref{Radial Soln Final BTZ Rot} to zero at the stretched horizon ($z_o$) is again difficult. But by doing a near horizon expansion ($z \rightarrow 0$) of \eqref{Radial Soln Final BTZ Rot} we again get a structure useful for writing phase equation
\begin{equation} \label{Phase Eqn Rot BTZ}
    F_{hor}(z) \approx C_{1}(T_{1} + z^{2i\alpha}_o)
\end{equation}
where,
\bea
   &T_1 = \frac{\Gamma (c) \Gamma \left(c^*-a^*\right) \Gamma \left(c^*-b^*\right)}{\Gamma \left(c^*\right)\Gamma (c-a) \Gamma (c-b)}\\
   &z_o = \frac{2 x r_+}{r^{2}_+ - r^{2}_-} = \frac{l^{2}_p}{L^2}
\eea
In the last line of above equation we use the definition of $l_p$ from \eqref{BTZPlanck}. In \eqref{Phase Eqn Rot BTZ}, we demand a Dirichlet boundary condition at the stretched horizon radius, $z = z_o$  and we get the phase equation
\begin{equation}
    T_1 = -z^{2 i \alpha}_o
\end{equation}
One can easily check that $T_1$ is a pure phase given the definitions of $a, b, c$. So, the above equation contains only phases and thus can be written as
\bea
  & \cos(\alpha)+\cos(\theta) = 0\\
  & \sin(\alpha)+\sin(\theta) = 0
\eea
where $\alpha$ and $\theta$ are defined as,
\bea
  &\alpha = -\pi - 2 {\rm Arg}\left[\Gamma\left(\frac{iL^2 r_+ \Tilde{\omega}}{r^{2}_+ - r^{2}_-}\right)\right] - 2 {\rm Arg}\left[\Gamma\left(\frac{iL}{2r_+} \left(J - \frac{\Tilde{\omega} L r_+}{r_+ - r_-}\right)\right)\right] - 2 {\rm Arg}\left[\Gamma\left(-\frac{iL}{2r_+} \left(J + \frac{\Tilde{\omega} L r_+}{r_+ - r_-}\right)\right)\right] \nonumber\\
  &\theta = 2 \frac{L^2 r_+ \Tilde{\omega}}{r^{2}_+ - r^{2}_-} {\rm Log}\left(\frac{l_p}{L}\right)
\eea
Combining the two as in previous cases, we get
\begin{align} \label{BTZ Rot Phase Eqn Cosine}
&\cos\left({\rm Arg} \left[\Gamma\left(\frac{i L^2 r_{+} \Tilde{\omega}}{r^{2}_{+}-r^{2}_{-}}\right)\right]+{\rm Arg} \left[\Gamma\left(\frac{iL}{2r_+} \left(J - \frac{\Tilde{\omega} L r_+}{r_+ - r_-}\right)\right)\right] \right. \nonumber \\
&\qquad\left. + {\rm Arg} \left[\Gamma\left(\frac{-iL}{2r_+} \left(J + \frac{\Tilde{\omega} L r_+}{r_+ + r_-}\right)\right)\right] - \frac{\Tilde{\omega} L^2 r_+}{2(r^{2}_+ - r^{2}_-)} {\rm Log}\left(\frac{l_p}{L}\right)\right) = 0
\end{align}
In $r_- = 0$ limit this reduces to \eqref{BTZ-phase-eq-2}, as we would hope.

\subsection{Analytic Low Lying Spectrum}
We can approximate the phase equation in \eqref{BTZ Rot Phase Eqn Cosine} in the low $\Tilde{\omega}$ approximations of Gamma functions,
\bea \label{Gamma Function Appx BTZ Rot}
  &{\rm Arg} \left[\Gamma\left(\frac{i L^2 r_{+} \Tilde{\omega}}{r^{2}_{+}-r^{2}_{-}}\right)\right] = -\frac{\pi}{2} - 0.575\frac{L^2 r_{+} \Tilde{\omega}}{r^{2}_{+}-r^{2}_{-}} \nonumber \\
  &{\rm Arg} \left[\Gamma\left(\frac{iL}{2r_+} \left(J - \frac{\Tilde{\omega} L r_+}{r_+ - r_-}\right)\right)\right] + {\rm Arg} \left[\Gamma\left(\frac{-iL}{2r_+} \left(J + \frac{\Tilde{\omega} L r_+}{r_+ + r_-}\right)\right)\right] = -\frac{\Tilde{\omega}L^2 r_{+}}{r^{2}_+ - r^{2}_-}\left(0.009+0.998{\rm Log}\frac{JL}{2r_+}\right) \nonumber
\eea
to find
\begin{equation} \label{analytic-BTZ Rot-low-lying}
    \frac{\Tilde{\omega} L^2 r_{+} }{(r^{2}_{+} - r^{2}_{-})} = \frac{n \pi}{{\rm Log}\left(\frac{1.12 r_{+}}{J l_p}\right)}
\end{equation}
where, $n \in \mathbb{Z}^{+}$. This is the {\em Rotating BTZ approximate phase equation} or the {\em Rotating BTZ analytic low-lying spectrum}. This also exactly reduces to the non-rotating BTZ analytic low lying spectrum in \eqref{BTZ-analytic-low-lying-spectrum} as desired.

We have checked that these expressions can be used to reproduce the numerical results of \cite{Arnab}.

\section{Kerr}\label{KerrSec}

We will determine a class of normal modes of the Kerr black hole in this Appendix to further illustrate the universality of our results. The Kerr wave equation is too complicated to be directly solvable, so making progress requires some approximations. Our observation here is that the radial equation simplifies and become tractable in a certain limit. We will define this limit via
\bea
r \sim M \ll \frac{1}{\omega} \label{near-stretched}
\eea
The coordinate $r$ is the Boyer-Lindquist radial coordinate and $M$ is the mass of the Kerr black hole. A related limit has been previously suggested in the context of the ``hidden conformal symmetry" discussion \cite{Strominger, CKHigherDimBH}:
\bea
M \ll r \ll \frac{1}{\omega}.
\eea
This latter limit is not suitable for discussions of normal modes because the stretched horizon is close to the horizon and is about the same scale as $M$. So we want to avoid the hierarchy between $r$ and $M$. Fortunately, it turns out that \eqref{near-stretched} is sufficient to allow the simplifications we are seeking in the wave equation -- the goals of \cite{Strominger, CKHigherDimBH} were different and involved discussions of scattering and near-far regions which required the extra hierarchy. 

Of course, to make the ``horizon-skimming'' approximation \eqref{near-stretched} self-consistently, we will need the normal modes that we compute to satisfy something like
\bea
\omega \ll \frac{1}{r_+}.  \label{self-consistency}
\eea
Remarkably, this is precisely what is accomplished by the $\log (r_+/l_p)$ term in the low-lying normal modes that were crucial for the entropy/temperature discussion:
\bea
\omega \sim \frac{1}{r_+ \log (r_+/l_p)}  \ll \frac{1}{r_+}.
\eea
Note that the $\log$ term introduces a hierarchy which is precisely what we need for the approximation to be reliable -- in the real world, for a solar mass black hole, the log provides a factor of $\sim 90$. So we expect the low-lying normal modes of Kerr to satisfy \eqref{self-consistency}, and it is easy to check from our final results that this is indeed the case.

With this preamble lets us get to work. With parameters $M$ and $J = Ma$, where $a$ is the angular momentum per unit mass $M$, the Boyer-Lindquist form of the Kerr metric is
\begin{equation} \label{Kerr-Boyer-Lindquist}
    ds^2 = \frac{\rho^2}{\Delta} dr^2 - \frac{\Delta}{\rho^2} (dt - a \sin^{2}\theta d\phi)^2 + \rho^{2} d\theta^2 + \frac{\sin^{2}\theta}{\rho^2}((r^2+a^2) d\phi - a dt)^2
\end{equation}
Using the fact that $r_{+}$ and $r_{-}$ are outer and inner event horizons of the Kerr black hole in $G_N = 1$ units 
\begin{equation} \label{Kerr-variables}
\begin{split}
    &\Delta = r^2 + a^2 - 2 M r, \ \hspace{0.2cm}  \rho^{2} = r^2 + a^2 \cos^{2} \theta\\
    &r_{+} + r_{-} = 2M, \ \hspace{0.2cm} \sqrt{r_{+} r_{-}} = a\\
\end{split}
\end{equation}
For a massless scalar, after choosing the ansatz\footnote{Here $m$ is the azimuthal quantum number, the scalar is massless so there is no possibility of confusing it with the mass.} $\Phi(t,r,\theta,\phi) = e^{-i\omega t. + i m \phi} R(r) S(\theta)$, the angular and radial differential equations are:
\begin{equation} \label{Kerr-angular}
    \left(\frac{1}{\sin \theta} \partial_{\theta}(\sin \theta \partial_{\theta}) - \frac{m^2}{\sin^{2}\theta} + r_{+} r_{-} \omega^{2} \cos^{2} \theta\right) S(\theta) = \kappa_{l} S(\theta)\\
\end{equation}

\begin{align} \label{Kerr-radial}
&\left(\partial_{r}(\Delta \partial_{r}) + \frac{((r_{+} + r_{-})r_{+} \omega - m\sqrt{r_{+} r_{-}})^2}{(r-r_{+})(r_{+}-r_{-})} \right. \nonumber \\
&\qquad\left. - \frac{((r_{+} + r_{-})r_{-} \omega - m\sqrt{r_{+} r_{-}})^2}{(r-r_{-})(r_{+}-r_{-})} + (r^2 + (r_{+}+r_{-})(r+r_{+}+r_{-})) \omega^2\right) R(r) = \kappa_{l} R(r). 
\end{align}
Note that the separation constant is a Casimir-like quantity on the sphere, but is not analytically known, unlike in the spherically symmetric cases. The radial equation is again of Heun type. As it stands these are pretty intractable without numerical methods -- as we mentioned earlier, we would like to postpone the numerics to as late a stage as possible in the calculation. One of our successes in our previous examples was that we could write down an exact equation that defined the normal modes, and numerics only entered at that stage (or later). The horizon-skimming approximation \eqref{near-stretched} will accomplish precisely that.

\subsection{Skimming the Horizon}

We can drop the problematic terms of the radial and angular equations in the horizon-skimming approximation \eqref{near-stretched}. The effect of the approximation is simply to reproduce the same equations found in \cite{Strominger}, but the distinction is crucial to justify our calculations here. In any event, the radial equation becomes a hypergeometric equation and $\kappa_{l}$ can be identified as $l(l+1)$ with $S(\theta) e^{im\phi}$ the spherical harmonics $Y_{lm}(\theta,\phi)$.
\begin{equation} \label{horizon-skimming-equations}
\begin{split}
   \left(\frac{1}{\sin \theta} \partial_{\theta}(\sin \theta \partial_{\theta}) - \frac{m^2}{\sin^{2}\theta}\right) Y_{lm}(\theta,\phi) = -l(l+1) Y_{lm} (\theta, \phi), \ \hspace{0.5cm}  m = -l, -l+1,....&.., l+1, l\\
   \left(\partial_{r}(\Delta \partial_{r}) + \frac{((r_{+} + r_{-})r_{+} \omega - m\sqrt{r_{+} r_{-}})^2}{(r-r_{+})(r_{+}-r_{-})} - \frac{((r_{+} + r_{-})r_{-} \omega - m\sqrt{r_{+} r_{-}})^2}{(r-r_{-})(r_{+}-r_{-})}\right) R(r) &= l(l+1) R(r) \\
\end{split}
\end{equation}
Using, $\Delta = (r-r_{+})(r-r_{-})$ and $l \geq 0$ we get the radial modes in terms of hypergeometric functions
\begin{equation} \label{Kerr-radial}
   R(r) = (r-r_{+})^{\frac{iA}{r_{+}-r_{-}}} (r-r_{-})^{-\frac{iB}{r_{+}-r_{-}}}\left(e^{\frac{4\pi A}{r_+ - r_-}} \left(\frac{r-r_+}{r_+ - r_-}\right)^{-\frac{2iA}{r_+ - r_-}} C_{2} J(r) + C_{1} K(r) \right)
\end{equation}
where, $C_{1}$ and $C_{2}$ are arbitrary integration constants and $J(r)$, $K(r)$, $A$ and $B$ are given as, 
\begin{equation} \label{Kerr-radial-mode-defs}
\begin{split}
    K(r) &= {}_{2}F_{1} \left(-l+\frac{i(A-B)}{r_+ - r_-}, l+1+\frac{i(A-B)}{r_+ - r_-}, 1+\frac{2iA}{r_+ - r_-}, -\frac{r-r_+}{r_+ - r_-}\right)\\
    J(r) &= {}_{2}F_{1} \left(-l-\frac{i(A+B)}{r_+ - r_-}, l+1-\frac{i(A+B)}{r_+ - r_-}, 1-\frac{2iA}{r_+ - r_-}, -\frac{r-r_+}{r_+ - r_-}\right)\\
    A &= \sqrt{r_{+}(m\sqrt{r_-}-\sqrt{r_+}(r_+ + r_-) \omega)^2} = \sqrt{r_+} \left|m\sqrt{r_-}-\sqrt{r_+}(r_+ + r_-) \omega\right|\\
    B &= \sqrt{r_{-}(m\sqrt{r_+}-\sqrt{r_-}(r_+ + r_-) \omega)^2} = \sqrt{r_-} \left|m\sqrt{r_+}-\sqrt{r_-}(r_+ + r_-) \omega\right|\\
\end{split}
\end{equation}
Near $r \rightarrow \infty$, the radial wave solution in \eqref{Kerr-radial-mode-defs} looks like 
\begin{equation}
\begin{aligned}
\begin{split}
R_{r \rightarrow \infty}(r) \approx r^{-(l+1)} &\left[(r_{+}-r_{-})^{l+1+\frac{i(A-B)}{r_{+}-r_{-}}} \Gamma(-1-2l) \biggl(C_{1} \frac{\Gamma\left(1+\frac{2iA}{r_{+}-r_{-}}\right)}{\Gamma \left(-l+\frac{i(A-B)}{r_{+}-r_{-}}\right) \Gamma \left(-l+\frac{i(A+B)}{r_{+}-r_{-}}\right)} \right.\\
&\qquad\left. + C_{2} \frac{e^{\frac{4 A \pi}{r_{+}-r_{-}}} \Gamma\left(1-\frac{2iA}{r_{+}-r_{-}}\right)}{\Gamma \left(-l-\frac{i(A-B)}{r_{+}-r_{-}}\right) \Gamma \left(-l-\frac{i(A+B)}{r_{+}r_{-}}\right)}\biggr)+O(1/r^2)\right]+
\end{split}\\
\begin{split}
r^{l} &\left[(r_{+}-r_{-})^{-l+\frac{i(A-B)}{r_{+}-r_{-}}} \Gamma(1+2l) \biggl(C_{1} \frac{\Gamma\left(1+\frac{2iA}{r_{+}-r_{-}}\right)}{\Gamma \left(l+1+\frac{i(A-B)}{r_{+}-r_{-}}\right) \Gamma \left(l+1+\frac{i(A+B)}{r_{+}-r_{-}}\right)} \right.\\
&\qquad\left. + C_{2} \frac{e^{\frac{4 A \pi}{r_{+}-r_{-}}} \Gamma\left(1-\frac{2iA}{r_{+}-r_{-}}\right)}{\Gamma \left(l+1-\frac{i(A-B)}{r_{+}-r_{-}}\right) \Gamma \left(l+1-\frac{i(A+B)}{r_{+}r_{-}}\right)}\biggr)+O(1/r)\right]
\end{split}
\end{aligned}    
\end{equation}
Now, we demand that function should not blow up at $r \rightarrow \infty$. Since $l$'s are positive, the only divergences come from $r^{l}$ term. So, we can adjust $C_{1}$ and $C_{2}$ in a suitable way that the coefficient of $r^l$ goes to zero, thus killing off the divergences as $r \rightarrow \infty$\footnote{In BTZ, we demanded normalizability at the AdS boundary. What we are demanding here is essentially asymptotic flatness.}. The result is:
\begin{equation} \label{fixing-C-by-asymptotic-flatness}
    C_{2} = e^{-\frac{4 \pi A}{r_{+}-r_{-}}} \frac{\gamma(A,B,l)}{\gamma(-A,B,l)} C_{1}
\end{equation}
where $\gamma(A,B,l)$ is given by 
\begin{equation} \label{gamma-ABl}
    \gamma(A,B,l) = \frac{\Gamma\left(\frac{2iA}{r_{+}-r_{-}}\right)}{\Gamma\left(l+1+\frac{i(A-B)}{r_{+}-r_{-}}\right)\Gamma\left(l+1+\frac{i(A+B)}{r_{+}-r_{-}}\right)}.
\end{equation}
Finally the radial solution looks like
\begin{equation} \label{Radial-solution-Kerr}
   R(r) = C_{1} (r-r_{+})^{\frac{iA}{r_{+}-r_{-}}} (r-r_{-})^{-\frac{iB}{r_{+}-r_{-}}} \left(\left(\frac{r-r_+}{r_+ - r_-}\right)^{-\frac{2iA}{r_+ - r_-}} \frac{\gamma(A,B,l)}{\gamma(-A,B,l)} J(r) + K(r)\right)
\end{equation}

This is a good place to introduce tortoise coordinate ($y$) in this geometry. For this, we need radial null curves. So we set $ds^{2}=0$ and angular displacements to zero. Also, we choose $\theta = 0$ for simplicity. We get, 
\begin{equation} \label{Kerr-tortoise}
\begin{split}
    y &= \int \frac{r^2 + r_{+} r_{-}}{(r-r_{+})(r-r_{-})} dr\\
     &= r + \frac{r_{+}+r_{-}}{r_{+}-r_{-}}\left[r_{+} \log\left(\frac{r-r_{+}}{r_{+}-r_{-}}\right) - r_{-} \log\left(\frac{r-r_{-}}{r_{+}-r_{-}}\right) \right]
\end{split}
\end{equation}
Setting $r_{-} = 0$ in \eqref{Kerr-tortoise}, we reproduce the tortoise coordinate in Schwarschild geometry. Here again, for $r$ lying between $r_{+} \leq r < \infty$, $y$ lies between $-\infty$ to $\infty$, where negative infinity is near the outer horizon and positive infinity is the asymptotically flat region. Near the event horizon $y$ can be approximated as 
\begin{equation} \label{near-horizon-tort-kerr}
\begin{split}
    y_{o} &\approx r_{+} + r_{+} \left(\frac{r_{+}+r_{-}}{r_{+}-r_{-}}\right) \log\left(\frac{r_{o}-r_{+}}{r_{+}-r_{-}}\right) \\
    r_{o} - r_{+} &\approx (r_{+}-r_{-}) e^{\frac{(y_o-r_{+})(r_{+}-r_{-})}{r_{+}(r_{+}+r_{-})}}\\
\end{split}
\end{equation}\\
where, $y_o$ becomes large and negative as $r_o$ goes near $r_{+}$. 

Solving the exact equation obtained by setting the \eqref{Radial-solution-Kerr} to zero at the stretched horizon is difficult here, like in BTZ. But by doing a near-horizon expansion of \eqref{Radial-solution-Kerr} we can again get a structure useful for writing the phase equation:
\begin{equation} \label{near-horizon-wave-Kerr}
    R_{hor}(r) \approx C_{1} \left(R_{1} (r/r_{+}-1)^{-\frac{iA}{r_{+}-r_{-}}} + S_{1} (r/r_{+}-1)^{\frac{iA}{r_{+}-r_{-}}}\right)
\end{equation}
where, 
\begin{equation} \label{near-horizon-Kerr-def}
\begin{split}
   R_{1} &= (r_{+}-r_{-})^{\frac{i(2A-B)}{r_{+}-r_{-}}} r^{-\frac{iA}{r_{+}-r_{-}}}_{+} \frac{\Gamma\left(\frac{2iA}{r_{+}-r_{-}}\right) \Gamma\left(l+1-\frac{i(A+B)}{r_{+}-r_{-}}\right) \Gamma\left(l+1-\frac{i(A-B)}{r_{+}-r_{-}}\right)}{\Gamma\left(-\frac{2iA}{r_{+}-r_{-}}\right) \Gamma\left(l+1+\frac{i(A+B)}{r_{+}-r_{-}}\right)\Gamma\left(l+1+\frac{i(A-B)}{r_{+}-r_{-}}\right)} \\
   S_{1} &= (r_{+}-r_{-})^{-\frac{iB}{r_{+}-r_{-}}} r^{\frac{iA}{r_{+}-r_{-}}}_{+}\\
\end{split}
\end{equation}
In \eqref{near-horizon-wave-Kerr} we demand a Dirichlet condition at the stretched horizon radius, $r = r_o$, to get the phase equation 
\begin{equation} \label{Kerr-phase-eq}
    \frac{R_1}{S_1} = -\left(\frac{r}{r_+} - 1\right)^{\frac{2 i A}{r_+ - r_-}} .
\end{equation}
Since $l$, $A$ and $B$ are real quantities, one can easily check from \eqref{near-horizon-Kerr-def} that $|R_{1}| = |S_{1}|$ so \eqref{Kerr-phase-eq} contains only phases and thus can be written as
\begin{equation} \label{eq8.18}
\begin{split}
   \cos(\alpha-\beta) + \cos(\theta) &=0 \\ 
   \sin(\alpha-\beta) + \sin(\theta) &=0 \\
\end{split}
\end{equation}
where, $\alpha$, $\beta$ and $\theta$ are defined as, 
\begin{equation} \label{eq8.19}
\begin{split}
    \alpha = {\rm Arg}[R_{1}] &= {\rm Arg}[(r_{+}-r_{-})^{\frac{i(2A-B)}{r_{+}-r_{-}}}] + {\rm Arg}[r^{-\frac{iA}{r_{+}-r_{-}}}_{+}]+2 {\rm Arg}\left[\Gamma\left(\frac{2iA}{r_{+}-r_{-}}\right)\right]\\
    &+ 2 {\rm Arg}\left[\Gamma\left(l+1-\frac{i(A+B)}{r_{+}-r_{-}}\right)\right] + 2 {\rm Arg}\left[\Gamma\left(l+1-\frac{i(A-B)}{r_{+}-r_{-}}\right)\right]\\
    \beta = {\rm Arg}[S_{1}] &= {\rm Arg}[(r_{+}-r_{-})^{-\frac{iB}{r_{+}-r_{-}}}] + {\rm Arg}[r^{\frac{iA}{r_{+}-r_{-}}}_{+}]\\
    \theta &= {\rm Arg}\left[\left(\frac{r}{r_{+}}-1\right)^{\frac{2iA}{r_{+}-r_{-}}}\right]\\
\end{split}
\end{equation}
Combining the two as in previous cases, we get
\begin{align} \label{Kerr-phase-eq-1}
&\cos\left({\rm Arg} \left[\Gamma\left(\frac{2iA}{r_{+}-r_{-}}\right)\right]+{\rm Arg} \left[\Gamma\left(l+1-\frac{i(A-B)}{r_{+}-r_{-}}\right)\right] \right. \nonumber \\
&\qquad\left. + {\rm Arg} \left[\Gamma\left(l+1-\frac{i(A+B)}{r_{+}-r_{-}}\right)\right] -\frac{A}{r_{+}-r_{-}} \log \left(\frac{r_{o}-r_{+}}{r_{+}-r_{-}}\right) \right) = 0
\end{align}
Using the near horizon tortoise coordinate from, \eqref{Kerr-phase-eq-1} can be written in the form which we call the official {\em Kerr phase equation}:
\begin{align} \label{Kerr-phase-eq-2}
&\cos\left({\rm Arg} \left[\Gamma\left(\frac{2iA}{r_{+}-r_{-}}\right)\right]+{\rm Arg} \left[\Gamma\left(l+1-\frac{i(A-B)}{r_{+}-r_{-}}\right)\right] \right. \nonumber \\
&\qquad\left. + {\rm Arg} \left[\Gamma\left(l+1-\frac{i(A+B)}{r_{+}-r_{-}}\right)\right] -\frac{A}{r_{+}(r_{+}+r_{-})} y_{o} + \frac{A}{r_{+}+r_{-}} \right) = 0
\end{align}
This equation contains the azimuthal quantum number, which is not an insurmountable complication \cite{Future}, but we will settle here for working out the case $m=0$ which makes the structure quite parallel to the non-rotating BTZ case. This will enable us to get to (a restricted class of) the normal modes quite quickly. 

\subsection{$m=0$}

Once we set $m=0$, the variables $A$, $B$ take the form
\begin{equation} \label{eq8.22}
\begin{split}
   A &= r_{+} (r_{+}+r_{-})\omega \\
   B &= r_{-} (r_{+}+r_{-})\omega \\
\end{split}
\end{equation}
Using a Kummer formula we can write the radial mode in this case as
\begin{equation} \label{Kummer-Kerr-mode}
\begin{split}
    R(r) &= \tilde{C} (r-r_{+})^{\frac{i r_{+}(r_{+}+r_{-})\omega}{r_{+}-r_{-}}} (r-r_{-})^{-\frac{i r_{-}(r_{+}+r_{-}) \omega}{r_{+}-r_{-}}} \left(\frac{r-r_{+}}{r_{+}-r_{-}}\right)^{-l-1-i (r_{+}+r_{-}) \omega}\\
    &{}_{2}F_{1} \left(l+1-\frac{i (r_{+}+r_{-})^2\omega}{r_{+}-r_{-}}, l+1+i (r_{+}+r_{-})\omega, 2l+2, -\frac{r_{+}-r_{-}}{r-r_{+}}\right)\\
\end{split}
\end{equation}
Near the outer event horizon $r=r_o$, RHS of \eqref{Kummer-Kerr-mode} can be put to zero in tortoise coordinates ($y_o$) to get the {\em Kerr exact equation, with $m=0$}: 
\begin{equation} \label{eq8.24}
\begin{split}
   & e^{\frac{(y_o-r_{+})(r_{+}-r_{-})}{r_{+}(r_{+}+r_{-})} \left(-l-1 + \frac{ir_{-} (r_{+}+r_{-}) \omega}{r_{+}-r_{-}}\right)} (r_{+}-r_{-})^{i(r_{+}+r_{-})\omega} \times\\
   &\times {}_{2}F_{1} \left(l+1-\frac{i (r_{+}+r_{-})^2\ \omega}{r_{+}-r_{-}}, l+1+i (r_{+}+r_{-})\omega, 2l+2, -e^{-\frac{(y_o-r_{+})(r_{+}-r_{-})}{r_{+}(r_{+}+r_{-})}}\right)=0\\
\end{split}
\end{equation}
Because of the hypergeometric, this is again hard to solve directly. So we take the leading near horizon expansion of it, which is the {\em Kerr phase equation with $m=0$}:
\begin{align} \label{Kerr-phase-m=0}
&\cos\left({\rm Arg} \left[\Gamma\left(\frac{2i r_{+} (r_{+}+r_{-}) \omega}{r_{+}-r_{-}}\right)\right]+{\rm Arg} \left[\Gamma\left(l+1-i(r_{+}+r_{-}) \omega\right)\right] \right. \nonumber \\
&\qquad\left. + {\rm Arg} \left[\Gamma\left(l+1-\frac{i (r_{+}+r_{-})^2 \omega}{r_{+}-r_{-}}\right)\right] - \omega y_{o} + \omega r_{+} \right) = 0
\end{align}

\subsubsection{Analytic Low-Lying Spectrum}

\begin{figure}
\centering
\begin{subfigure}{\textwidth}
  \centering
  \includegraphics[width=0.8\linewidth]{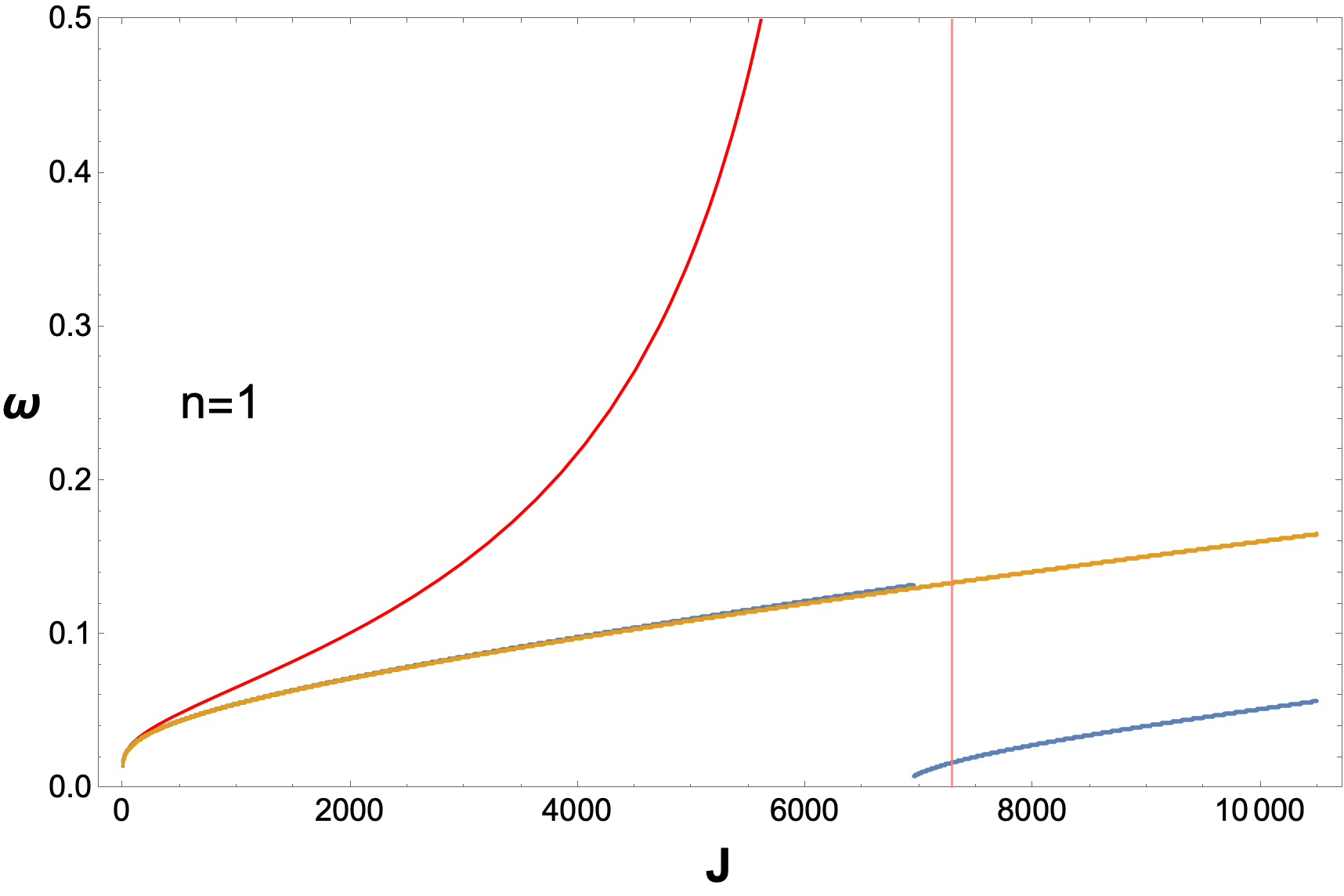}
  \caption{} % Label as Fig 4(a)
  \label{fig4a}
\end{subfigure}

\vspace{1cm} % Adjust the vertical spacing between the figures

\begin{subfigure}{\textwidth}
  \centering
  \includegraphics[width=0.8\linewidth]{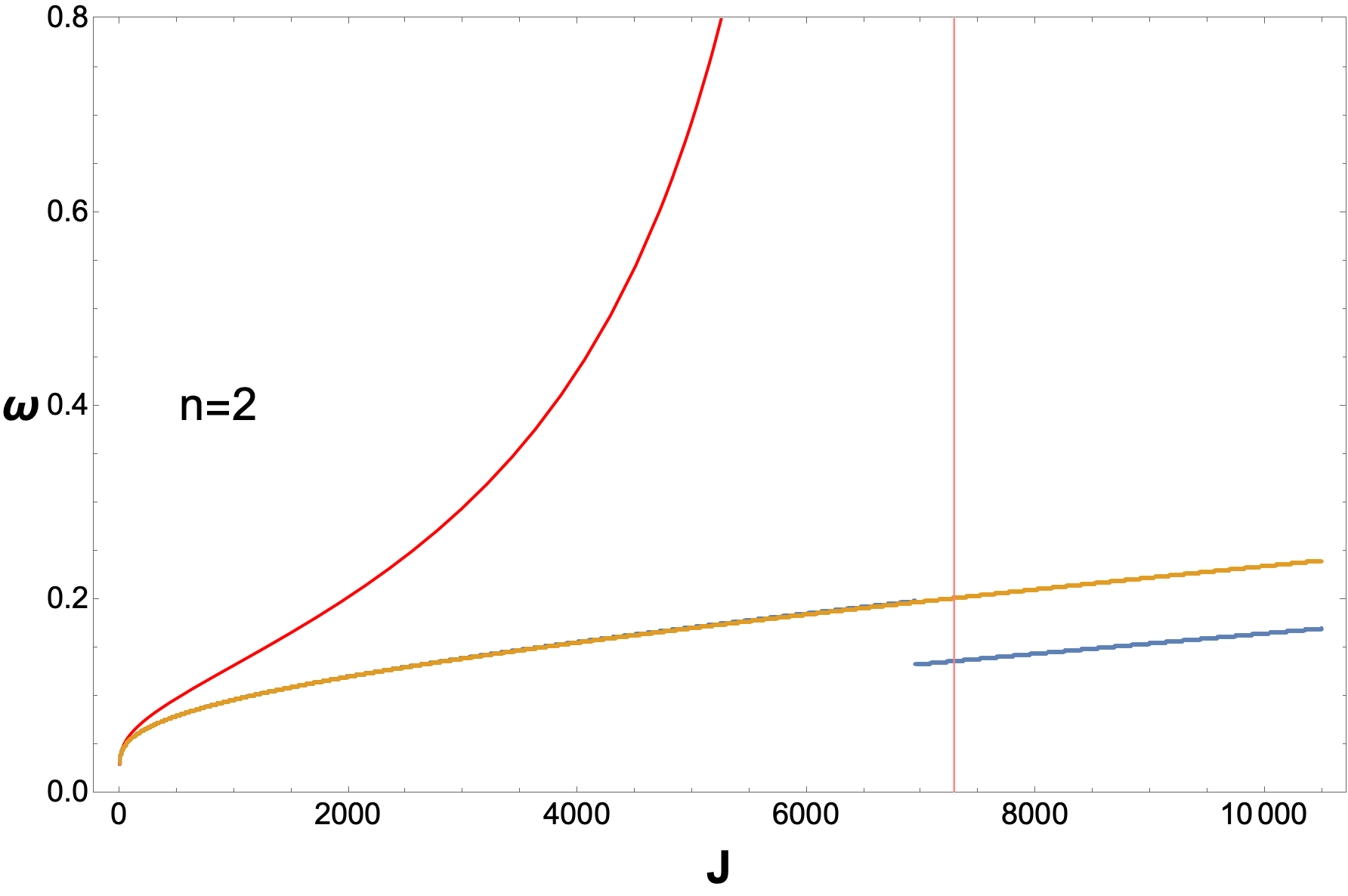}
  \caption{} % Label as Fig 4(b)
  \label{fig4b}
\end{subfigure}
\caption{Kerr Modes for $y_o = -220$ with $r_+ = 10$ and $r_- = 1$: yellow
stands for {\em perturbed Kerr phase equation}  \eqref{perturbed-Kerr-phase-equation}, blue stands for {\em Kerr phase equation} \eqref{Kerr-phase-m=0}, red stands for the analytically fitted spectrum {\em Kerr analytic low-lying spectrum} \eqref{analytic-Kerr-low-lying} and the pink line denotes the breakdown point of the low-lying analytic spectrum, which is $l_{max}= 7290$ obtained from \eqref{lKerrmax}. The phase equation breaks down at $J = 6950$, which is close to this.
.} 
\label{fig17}
\end{figure}

\begin{figure}
\centering
\begin{subfigure}{.5\textwidth}
  \centering
  \includegraphics[width=0.9\linewidth]{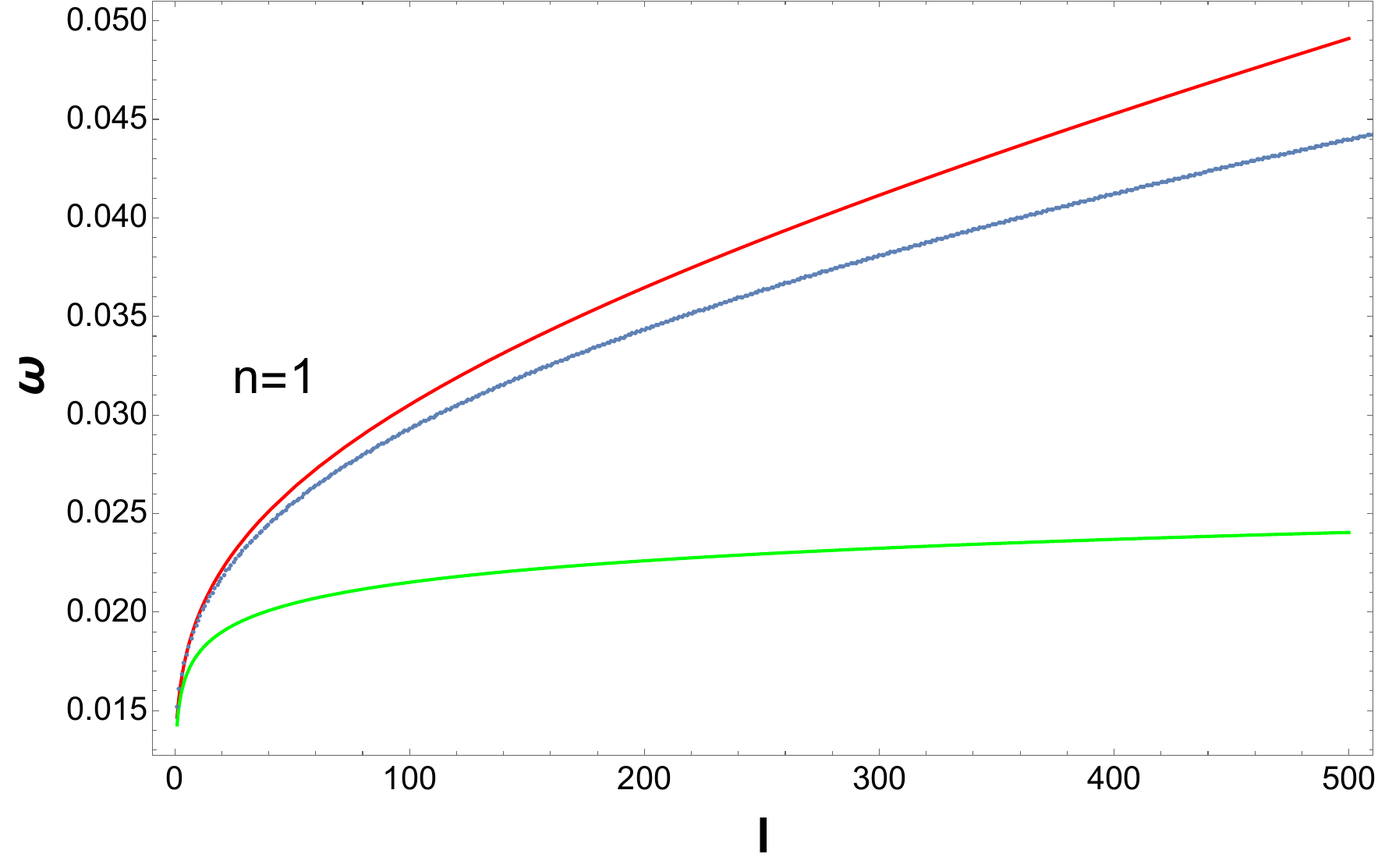}
  \caption{}
  \label{fig5(a)}
\end{subfigure}%
\begin{subfigure}{.5\textwidth}
  \centering
  \includegraphics[width=0.9\linewidth]{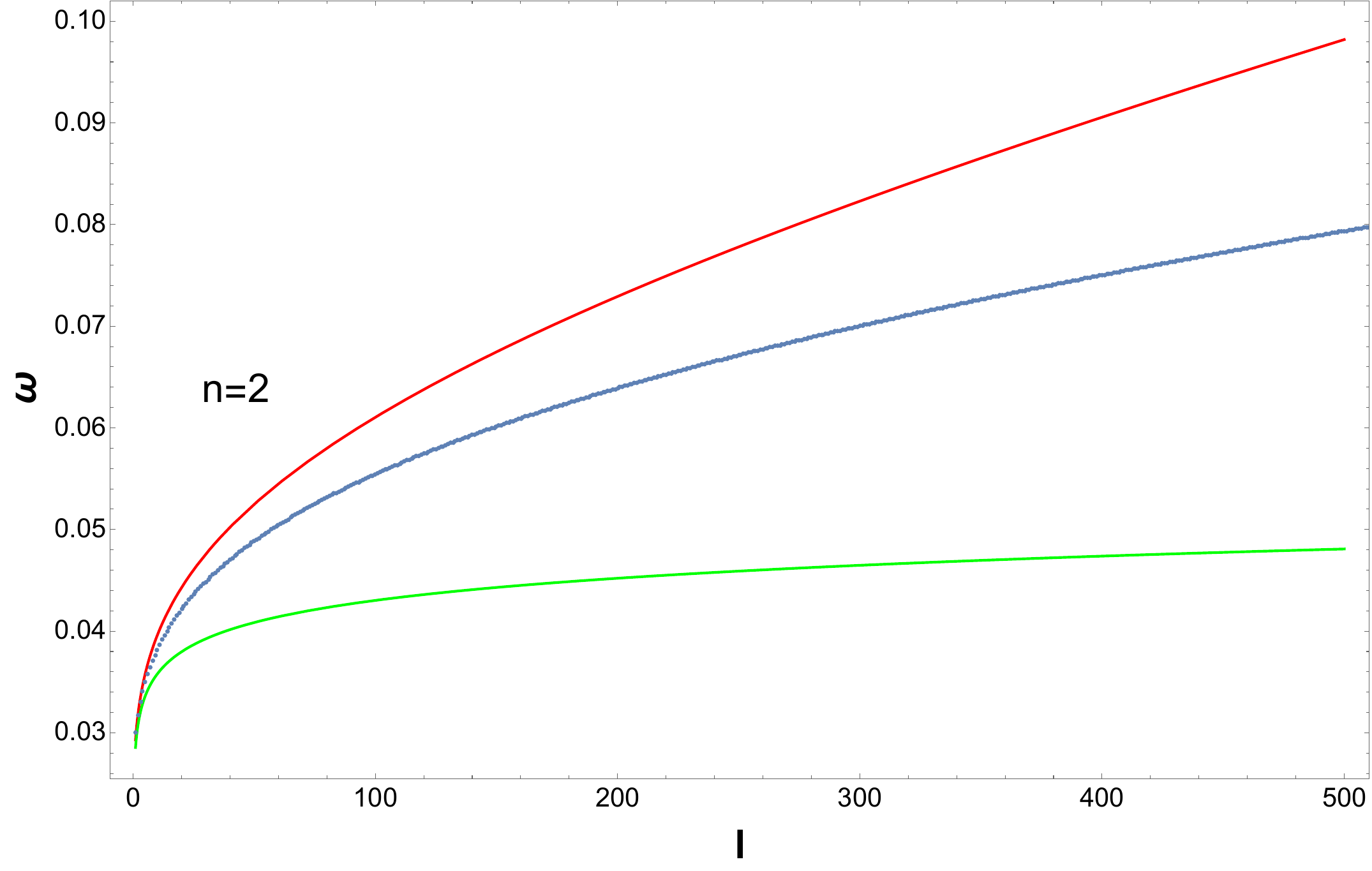}
  \caption{}
  \label{fig5(b)}
\end{subfigure}
\caption{Comparison between exact spectrum (blue), Approximate Phase Equation (red) and the analytical spectrum at low $l$'s (green) as in \eqref{ultra-low-Kerr} at $y_o = -220$ with $r_+ = 10$ and $r_- = 1$.}. 
\label{fig18}
\end{figure}

\begin{figure}[h]
       \centering
       \includegraphics[width=0.9\linewidth]{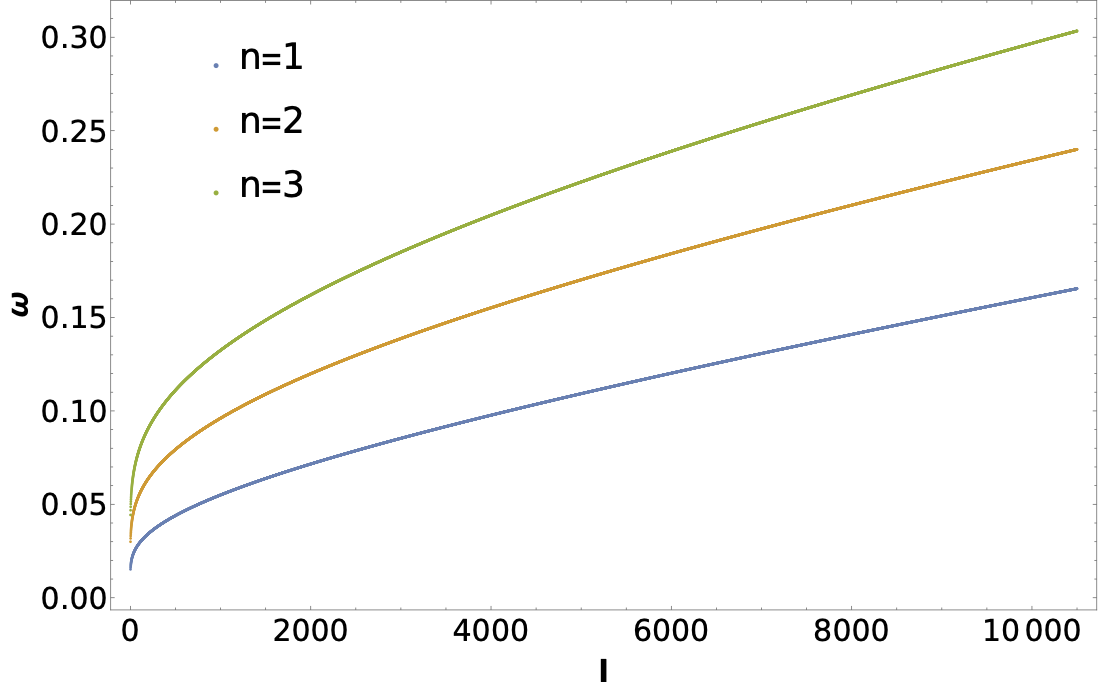}
       \caption{Plot of perturbative exact spectrum $\omega(n,l)$ for Kerr Geometry $y_o = -220$ with $r_+ = 10$ and $r_- = 1$ for $n = 1, 2$ and 3.}
       \label{fig19}
\end{figure} 

We can approximate the phase equation using the following low-$\omega$ approximations of Gamma functions
\bea \label{Kerr-phase}
   {\rm Arg} \left[\Gamma\left(\frac{2i r_{+} (r_{+}+r_{-}) \omega}{r_{+}-r_{-}}\right)\right] = -0.575 \frac{2\omega r_{+} (r_{+}+r_{-})}{r_{+}-r_{-}}-\frac{\pi}{2} \hspace{2cm} \\ 
  {\rm Arg} \left[\Gamma\left(l+1-i(r_{+}+r_{-}) \omega\right)\right]+{\rm Arg} \left[\Gamma\left(l+1-\frac{i (r_{+}+r_{-})^2 \omega}{r_{+}-r_{-}}\right)\right] = \hspace{1in} \\ =(-0.0996\ \omega - 0.982\ \omega \log(l))
   \left(\frac{2r_{+}(r_{+}+r_{-})}{r_{+}-r_{-}}\right)  \nonumber 
\eea
to find 
\begin{equation} \label{analytic-Kerr-low-lying}
    \frac{2 \omega r_{+} (r_{+}+r_{-})}{r_{+}-r_{-}} = \frac{n \pi}{-\frac{r_{+}-r_{-}}{2r_{+}(r_{+}+r_{-})}\tilde{y}_o - 0.982 \log(l)}
\end{equation}
where, $n \in \mathbb{Z}^{+}$ and $\frac{r_{+}-r_{-}}{2r_{+}(r_{+}+r_{-})}\tilde{y}_o = \frac{r_{+}-r_{-}}{2r_{+}(r_{+}+r_{-})} (y_{o}-r_{+}) + 0.675$. 
This is the {\em Kerr approximate phase equation} or the {\em Kerr analytic low-lying spectrum}. Note that this is for $m=0$. Like in Rindler or BTZ, this solution also diverges when denominator is zero for a given cutoff (negative $y_o$). This $l^{Kerr}_{max}$ is given by
\begin{equation} \label{lKerrmax}
    l^{Kerr}_{max} = e^{-0.686} e^{\frac{r_{+}-r_{-}}{1.964 r_{+}(r_{+}+r_{-})} (y_{o}-r_{+})}
\end{equation}
%Planck Length is coming as Elliptic Integrals in this geometry in mathematica, even in \theta = 0 case. 
Similarly, an expression for very low $l$'s can also be written from \eqref{analytic-Kerr-low-lying} in leading order as, 
\begin{equation} \label{ultra-low-Kerr}
    \omega = -\frac{n\pi}{y_o} + \frac{2n\pi r_{+}(r_{+}+r_{-})}{y^{2}_{o}(r_{+}-r_{-})}0.982\log (l)
\end{equation}

\subsubsection{Kerr Normal Mode Perturbation Theory}

As in the BTZ case, the solutions of the Kerr phase become unreliable around $l^{Kerr}_{max}$. But again as in BTZ, we can get past this by incorporating the perturbative corrections to the phase equation. The corrected phase equation is
\begin{equation} \label{Pert-corrected-Kerr-phase-eqn}
\begin{split}
   \tilde{R}_{1} (r_{o}/r_+ - 1)^{-\frac{iA}{r_+ - r_-}} + \tilde{S}_{1} (r_{o}/r_{+}-1)^{\frac{iA}{r_{+}-r_{-}}} = 0\\
\end{split}
\end{equation}
where, 
\begin{equation} \label{eq8.31}
\begin{split}
    E &= \frac{(A^2 + B^2 + l(l+1) (r_{+}-r_{-})^2 + i A (r_{+}-r_{-})}{(r_{+}-r_{-}-2iA)(r_{+}-r_{-})^2}\\
    \tilde{R}_{1} &= R_{1}(1+E (r_o-r_+))\\
    \tilde{S}_{1} &= S_{1}(1+E^{*} (r_o-r_+))\\
\end{split}
\end{equation}
where, $R_{1}$ and $S_{1}$ are defined in \eqref{near-horizon-Kerr-def}, with $|R_{1}| = |S_{1}| = 1$. From the form of $E$ defined here $|1+E(r_o-r_+)| = |1+E^{*} (r_o-r_+)|$. This suggests that $\tilde{R}_{1}$ and $\tilde{S}_{1}$ are also pure phases. Using these facts we can write the {\em perturbed Kerr phase equation for $m=0$}:
\begin{align} \label{perturbed-Kerr-phase-equation}
&\cos\left({\rm Arg} \left[\Gamma\left(\frac{2iA) \omega}{r_{+}-r_{-}}\right)\right]+{\rm Arg} \left[\Gamma\left(l+1-i\frac{(A-B)}{r_{+}-r_{-}}\right)\right]+ {\rm Arg} \left[\Gamma\left(l+1-\frac{i (A+B)}{r_{+}-r_{-}}\right)\right] \right. \nonumber \\
&\qquad\left. {\rm Arg}\left[1+\frac{(A^2 + B^2 + l(l+1) (r_{+}-r_{-})^2 + i A (r_{+}-r_{-})}{(r_{+}-r_{-}-2iA)(r_{+}-r_{-})}e^{\frac{(y_o-r_+)(r_{+}-r_{-})}{r_{+}(r_{+}+r_{-})}}\right]- \omega y_{o} + \omega r_{+} \right) = 0
\end{align}
The qualitative features of the various exact and approximate normal modes proceeds as in our previous discussions (Schwarzschild/Rindler and BTZ), so we settle for simply presenting the plots without further discussion.

\pagebreak

\end{document}